\documentclass[a4paper,fleqn,usenatbib]{mnras}

\usepackage[T1]{fontenc}
\usepackage{ae,aecompl}

\usepackage{graphicx}	
\usepackage{amsmath}	
\usepackage{amssymb}	
\usepackage{makecell}

\newcommand{\Msold}{M$_{\odot}$\,yr$^{-1}$}

\title[Morphology and kinematics of the gas envelope of Mira Ceti]{Morphology and kinematics of the gas envelope of Mira Ceti}
\author[P.T. Nhung et al.]{
{P. T. Nhung\thanks{E-mail: pttnhung@vnsc.org.vn}, D. T. Hoai, P. N. Diep, N. T. Phuong, N. T. Thao, P. Tuan-Anh}
\newauthor{  and P. Darriulat}
\\
Department of Astrophysics, Vietnam National Satellite Center, VAST, 18 Hoang Quoc Viet, Hanoi, Vietnam\\
}
\date{Accepted XXX. Received YYY; in original form ZZZ}

\pubyear{2016}

\begin{document}
\label{firstpage}
\pagerange{\pageref{firstpage}--\pageref{lastpage}}
\maketitle

\begin{abstract}
Observations of $^{12}$CO(3-2) emission of the circumbinary envelope of Mira Ceti, made by ALMA are analysed. The observed Doppler velocity distribution is made of three components: a blue-shifted south-eastern arc, which can be described as a ring in slow radial expansion, $\sim$1.7 km\,s$^{-1}$, making an angle of $\sim$50$^\circ$ with the plane of the sky and born some 2000 years ago; a few arcs, probably born at the same epoch as the blue-shifted arc, all sharing Doppler velocities red-shifted by approximately 3$\pm$2 km\,s$^{-1}$ with respect to the main star; the third, central region dominated by the circumbinary envelope, displaying two outflows in the south-western and north-eastern hemispheres. At short distances from the star, up to $\sim$1.5$''$, these hemispheres display very different morphologies: the south-western outflow covers a broad solid angle, expands radially at a rate between 5 and 10 km\,s$^{-1}$ and is slightly red shifted; the north-eastern outflow consists of two arms, both blue-shifted, bracketing a broad dark region where emission is suppressed. At distances between $\sim$1.5$''$ and $\sim$2.5$''$ the asymmetry between the two hemispheres is significantly smaller and detached arcs, particularly spectacular in the north-eastern hemisphere are present. Close to the stars, we observe a mass of gas surrounding Mira B, with a size of a few tens of AU, and having Doppler velocities with respect to Mira B reaching $\pm$1.5 km\,s$^{-1}$, which we interpret as gas flowing from Mira A toward Mira B.
\end{abstract}




\section{Introduction}


Mira Ceti is one of the most studied binary stars. Its distance from the Sun is estimated to be 107$\pm$10 pc 
from the Hipparcos catalogue and 91.7 pc by \citet{van Leeuwen2007}. In the present work we retain a distance 
of 100 pc, meaning that an angular distance projected on the sky of 1$''$ corresponds to 100 AU and is covered 
in $\sim$500 yr at a projected velocity of 1 km\,s$^{-1}$. The infrared environment has been recently observed 
over $\sim 5'$ using Herschel PACS by \citet{Mayer2011} at 70 and 160 $\micron$.

The main star, Mira A, is a long period variable that has been on the Asymptotic Giant Branch (AGB) long enough 
to display technetium in its spectrum \citep{Little1987, Kipper1992}. It is an M.7 type, oxygen-rich star with 
a mass loss rate of the order of 10$^{-7}$ \Msold \citep[]{Mauron1992, Knapp1998, Ryde2001}. It has a mass of 
1.2 solar masses \citep[]{Wyatt1983}, a luminosity of $\sim$9000 solar luminosities and a temperature of 
$\sim$3000 K \citep[]{Woodruff2004}. Its apparent magnitude varies between as low as $\sim$2 and as high as $\sim$10 
with a pulsation period of 333 days \citep[]{Templeton2009}. It has a very high velocity with respect to the local 
ISM, of the order of 120 km\,s$^{-1}$, oriented approximately from north to south \citep{van Leeuwen2007}.
A bow shock is visible in front of Mira A at a distance of $\sim$3$'$ from it \citep[]{Martin2007}. It generates a cometary wake visible 
over $\sim$2$^\circ$  that has been observed in the UV \citep[]{Martin2007} and the interaction of the circumstellar envelope with the interstellar medium has been seen in HI \citep[]{Matthews2008}. 
Evidence for the cometary astropause has also been obtained in the far infrared \citep[]{Ueta2008}.

\begin{figure}
  \centering
  \includegraphics[width=0.8\columnwidth]{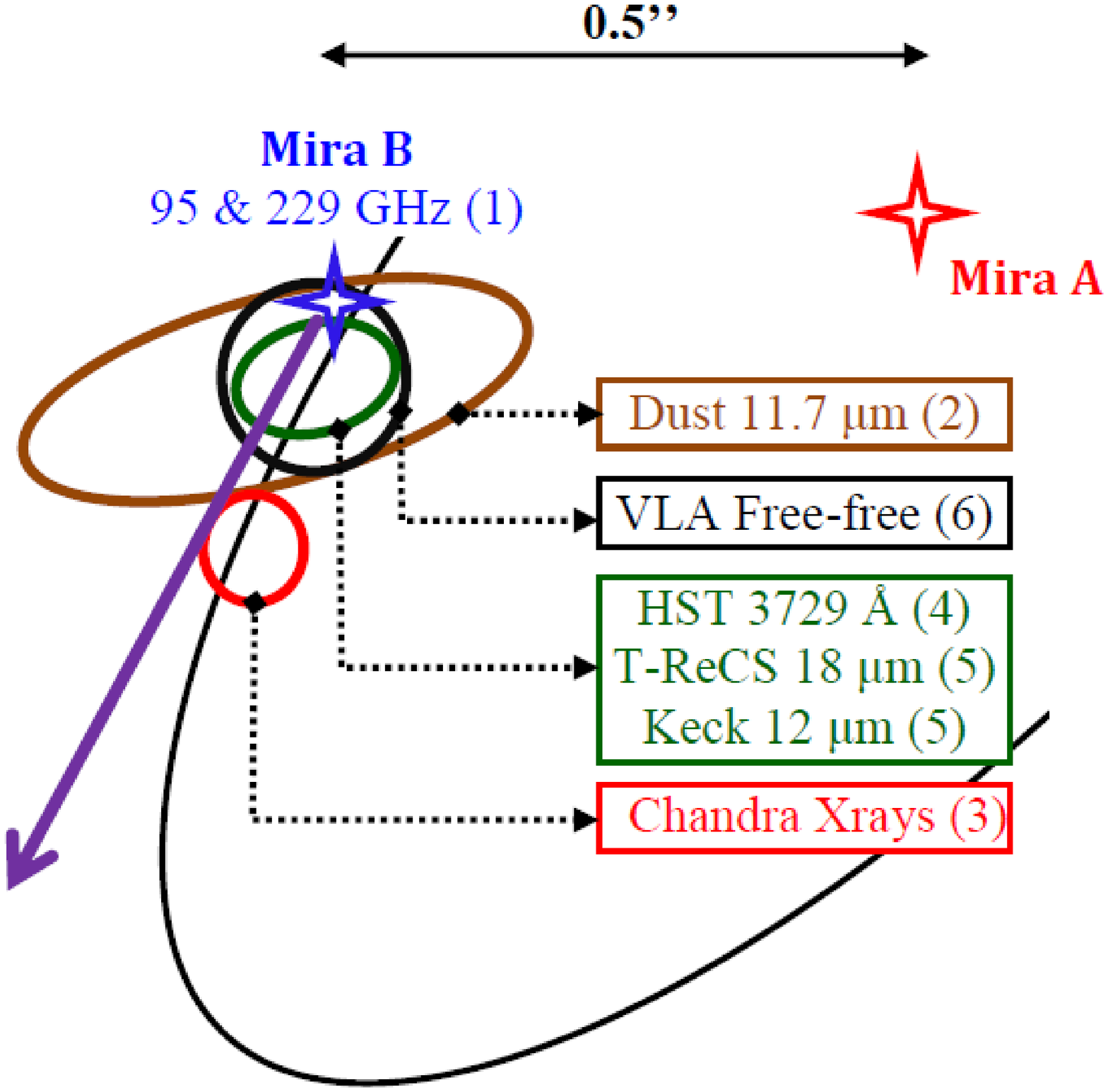}
  \caption{Schematic summary of multiwavelength emission in the Mira B environment. 
Data are from References 1 \citep[]{Ramstedt2014, Vlemmings2015}, 2 \citep[]{Marengo2001}, 
3 and 4 \citep[]{Karovska2005}, 5 \citep[]{Ireland2007}, 6 \citep[]{Matthews2006}.
The coloured ellipses are meant to give the measured position of Mira B with respect to 
Mira A (red star) with an estimate of the measurement uncertainties. We did not correct 
for the movement of Mira B with respect to Mira A during the time separating the different 
measurements, which were all made in the past ten years or so. The associated corrections 
are of the same order of magnitude as the measurement uncertainties. The magenta arrow 
displays the evolution of the observed  CO(3-2)  emission  in  the  present  analysis: 
for Doppler velocities between 42 and 44 km\,s$^{-1}$ it stays on the nominal Mira B 
position and then extends south-east along the arrow for Doppler velocities between 
44 and 46 km\,s$^{-1}$. The orbit, shown as a black ellipse, is that of Prieur et al. (2002).}
  \label{fig1}
\end{figure}

The companion, Mira B, is probably a white dwarf with a temperature of $\sim$10$^4$ K \citep[]{Sokoloski2010} at a 
projected distance on the plane of the sky \citep[measured in 2014,][]{Ramstedt2014} of 0.487$''\pm$0.006$''$ from Mira A and at a position angle of 
8$^\circ$ south of east as measured at 95 and 229 GHz on partially ionized gravitationally bound gas 
\citep{Ramstedt2014, Vlemmings2015, Planesas2016}. The orbit is poorly constrained by observations \citep[]{Prieur2002}: 
between 1923 and today, Mira B has drifted by $\sim$0.6$''$ projected on the plane of the sky toward some 
20$^\circ$ west of north (Fig.~\ref{fig1}). The orbital period is estimated to be at least 500 years. The orbit 
is nearly circular and its projected eccentricity corresponds to an angle of $\sim$22$^\circ$ with the line of 
sight. Ly$\alpha$ fluoresced H2 emission has been observed by HST from Mira B in UV spectra \citep[]{Wood2002}. 
Mira B is known to accrete matter from Mira A, with a connection between the two stars having been observed in 
near UV using the HST \citep[]{Karovska2005}, in mid infrared using the Keck telescope \citep[]{Ireland2007}, 
in soft X rays using Chandra \citep[]{Karovska2005} and in free-free emission at 8.5 GHz using the VLA 
\citep[]{Matthews2006}. The Chandra observation has revealed the occurrence of an X ray outburst and of intense 
activity in the system, also confirmed by the detection of variable OH maser emission \citep[]{Etoka2010}. 
A bipolar outflow with a wind velocity of 160$\pm$10 km\,s$^{-1}$ has been revealed in the UV \citep[]{Martin2007} 
and by the presence of knots in H$\alpha$ \citep[]{Meaburn2009}. It makes an angle of 69$^\circ\pm$15$^\circ$ 
with the plane of the sky, the northern jet being red-shifted and the southern jet blue-shifted \citep[]{Meaburn2009}.

Mira's circumstellar envelope has been observed in CO(2-1) using the Plateau de Bure interferometer 
\citep[]{Planesas1990, Josselin2000} and in CO(1-0) using the Berkely Illinois Maryland Association 
interferometer \citep[]{Fong2006}. The former see a nearly spherical envelope disturbed north and 
south by the bipolar outflow while the latter suggest a kinematics combining rotation and expansion. 
Both quote very low expansion velocities not exceeding $\pm$5 km\,s$^{-1}$.

Recently, the short distance environment of Mira A+B has been observed using the Atacama Large 
Millimetre/sub-millimetre Array (ALMA). Continuum emission has been observed at 1, 3 and 7 mm wavelength 
and its analysis \citep{Vlemmings2015, Planesas2016}, in particular in association with JVLA observations 
\citep{Matthews2015}, has resolved Mira A photosphere with unprecedented spatial resolution. 
The $^{12}$CO(3-2) line emission has been mapped with an also unprecedented spatial resolution of 
$\sim$0.5$''$ up to a distance of $\sim$10$''$ from the central star by \citet{Ramstedt2014}.  
The map reveals the presence of a bubble in the south-east quadrant and of a complex pattern of arcs. 
The authors suggest an interpretation in terms of Mira A's slow wind filling its Roche lobe and being 
accreted by Mira B in the orbital plane \citep[]{Mohamed2012}. In the present work we analyse other 
ALMA observations that have recently been open to public access and offer a slightly better resolution 
at short distance from the pair of stars. While not repeating in any way the analysis of \citet{Ramstedt2014} 
our analysis provides a useful complement to their work by using different approaches and concentrating on different issues.
In particular the better spatial resolution of the present data makes it possible for us to study 
the short distance environment of the binary in considerably more detail than available in \citet{Ramstedt2014}.

\section{Observations and data reduction}
We consider here a set of observations (ADS/JAO.ALMA/2013.1.00047.S), made on the CO(3-2) line between 
12 and 15 June 2014 with configurations including between 34 and 36 antennas and having a maximal baseline 
of 650.5 km. The data have been reduced, merged and corrected for the primary beam by the ALMA staff to clean 
maps using pixel sizes of 0.06$''\times$0.06$''$ and having a beam of angular dimensions of 
0.325$''\times$0.309$''$ FWHM at position angle of 83.2$^\circ$. The total time on source was $\sim$10.6 
minutes. The line Doppler velocity spectrum evaluated from 25 km\,s$^{-1}$ to 65 km\,s$^{-1}$ is available 
in channels of 0.4 km\,s$^{-1}$.

We define $y$ and $z$ as coordinates in the plane of the sky, $y$ pointing east and $z$ pointing north. 
The projected angular distance of pixel ($y,z$) from the origin is $R=\sqrt{y^2+z^2}$ and its position angle, 
measured clockwise from west, is $\varphi=\arctan(z/y)+\pi$. The $x$ axis is parallel to the line of sight 
with positive $x$ values being red-shifted. The measured flux density is $f(y,z,V_x)$ and its integral over 
the measured Doppler velocity $V_x$, normally between 34 km\,s$^{-1}$ and 56 km\,s$^{-1}$, is $F(y,z)=\int{f(y,z,V_x)dV_x}$. 
We define an effective emissivity $\rho=f(y,z,V_x)dV_x/dx$, that mixes information concerning both gas density and temperature.

\begin{figure}
\centering
\includegraphics[height=4.5cm,trim=2.cm 0.cm 0.cm 0.cm,clip]{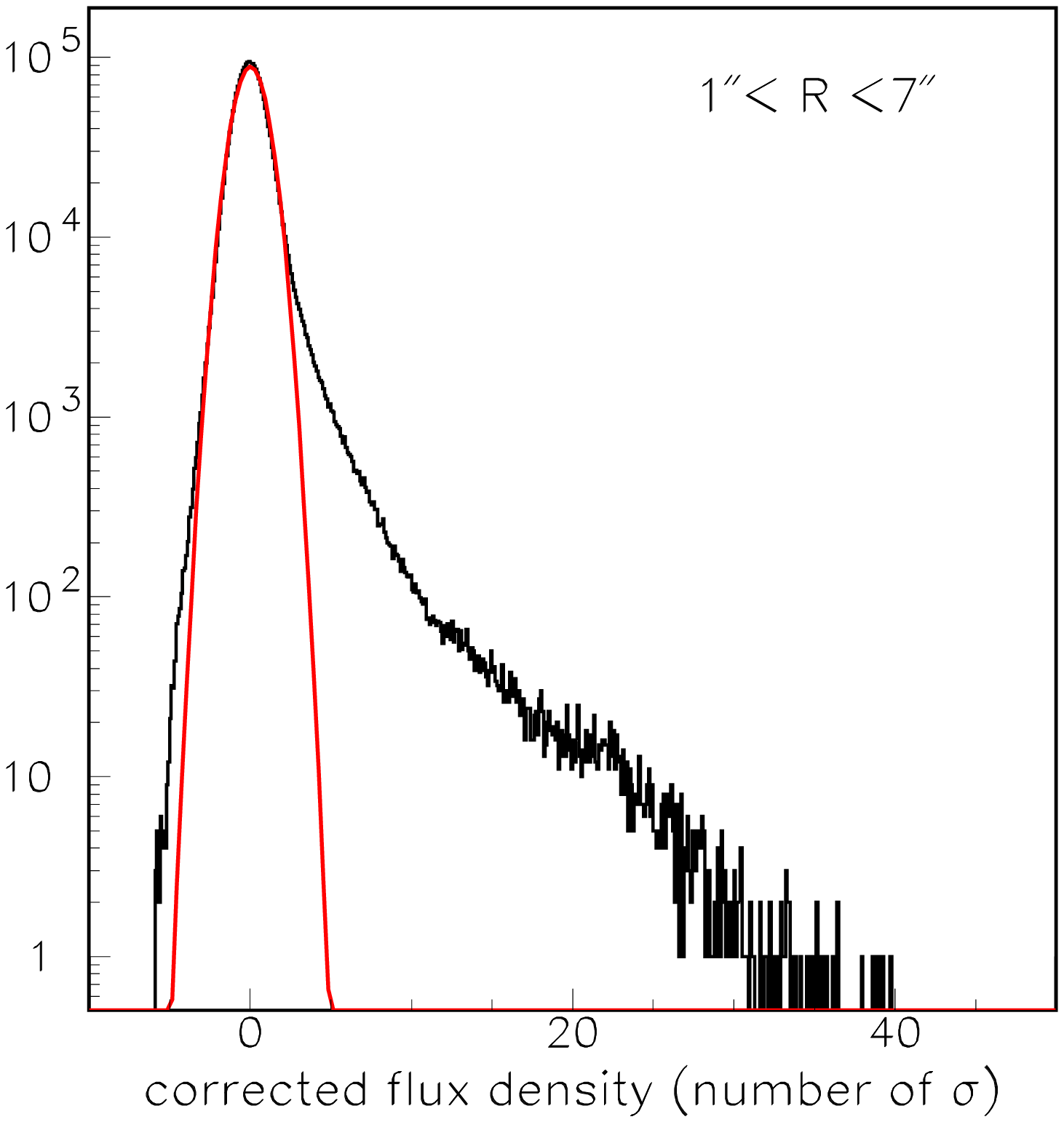}
\includegraphics[height=4.5cm,trim=2.cm 0.cm 0.cm 0.cm,clip]{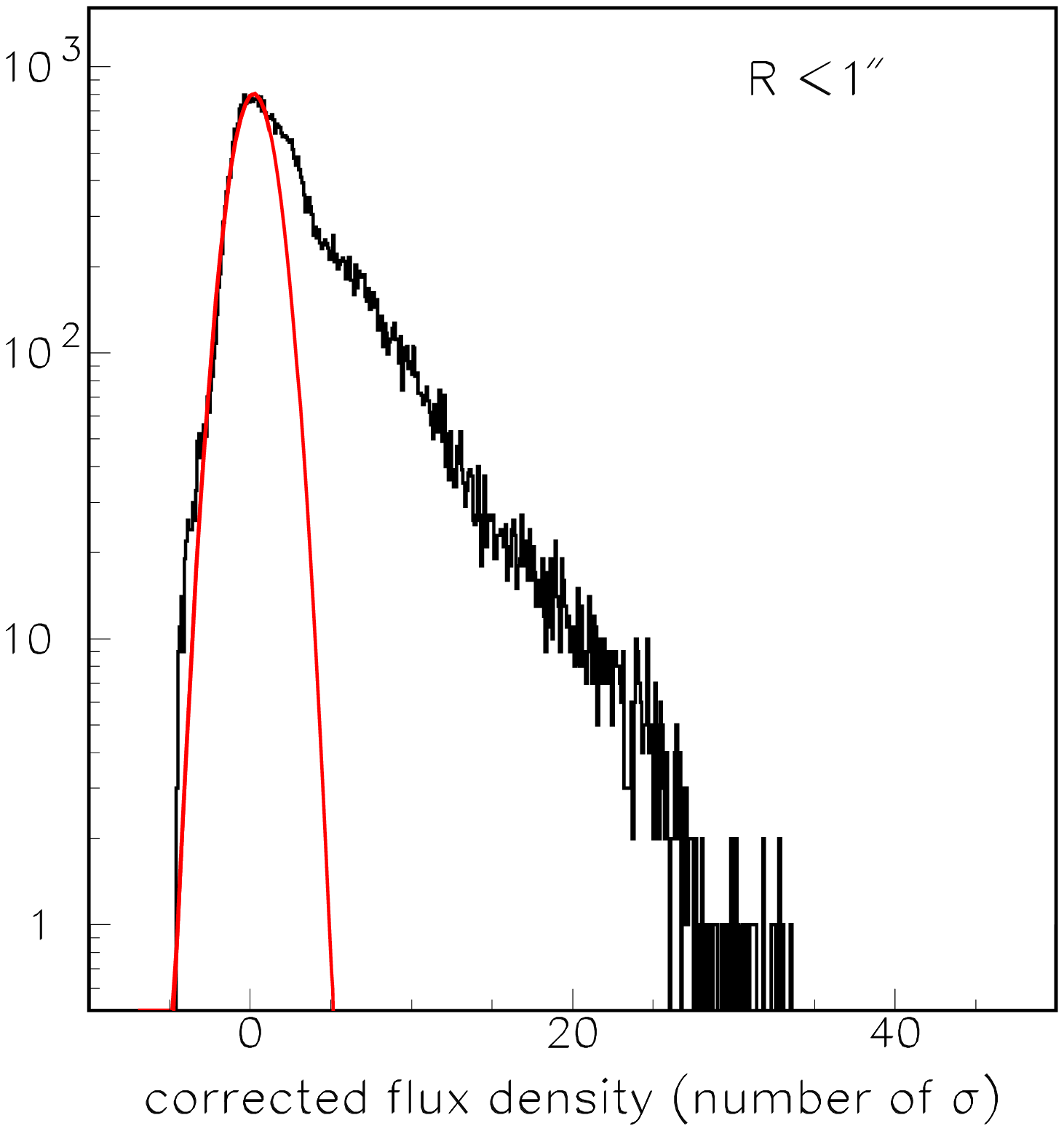}
\caption{Distributions of $[f(y,z,V_x)+\Delta{f(V_x)}]/\sigma_f(V_x)$ for 1$''$<$R$<7$''$ (left) and, after having 
increased $\Delta{f(V_x)}$ by 0.04 Jy\,arcsec$^{-2}$ and $\sigma_f(V_x)$ by 30\%, for $R$<1$''$ (right). The Gaussians 
have unit variance and are centred at the origin.}
\label{fig2}
\end{figure}

A study of the evolution of the distribution of the measured flux density when moving away from the star reveals a 
strong dependence on the velocity channel of the effective noise (meaning the peak observed at the lower flux densities). 
Indeed, as noted and explained by \citet{Ramstedt2014} using a similar set of observations, the effective noise peaks 
at different values and displays different widths in different velocity channels, the channels with strong emission 
having a limited dynamic range due to the lack of large distance coverage. We therefore evaluate, for pixels distant 
from the star by a projected angular distance 1$''$<$R$<7$''$ and for each velocity channel separately a shift $-\Delta f(V_x)$ 
and a width $\sigma_f(V_x)$ from a Gaussian fit to the effective noise peak. The resulting distribution of 
$[f(y,z,V_x)+\Delta f(V_x)]/\sigma_f(V_x)$ is shown in Fig.~\ref{fig2} (left). However, when looking at the same 
quantity in the star region ($R$<1$''$), we find that we need to increase $\sigma_f(V_x)$ by $\sim$30\% and $\Delta f(V_x)$ 
by $\sim$0.04 Jy\,arcsec$^{-2}$ in order to obtain a good description of the effective noise (Fig.~\ref{fig2} right). 
Typically, going from low to high emission, $\sigma_f(V_x)$ increases from less than 0.1 to $\sim$0.6 Jy\,arcsec$^{-2}$ 
and $\Delta f(V_x)$ from zero to $\sim$0.15 Jy\,arcsec$^{-2}$. We estimate that both quantities are obtained with an 
uncertainty of the order of 30\% of their value.  In what follows, flux densities are corrected accordingly and are 
retained only if they exceed a specified number of $\sigma_f(V_x)$, usually 2 (we speak of a cut at 2\,$\sigma$'s). 
We have checked systematically that all results presented here are not significantly affected by a change in the cut 
level, or even by the absence of cut.

\begin{figure*}
\centering
\includegraphics[height=5.5cm,trim=0.cm 0.cm 0.5cm 0.cm,clip]{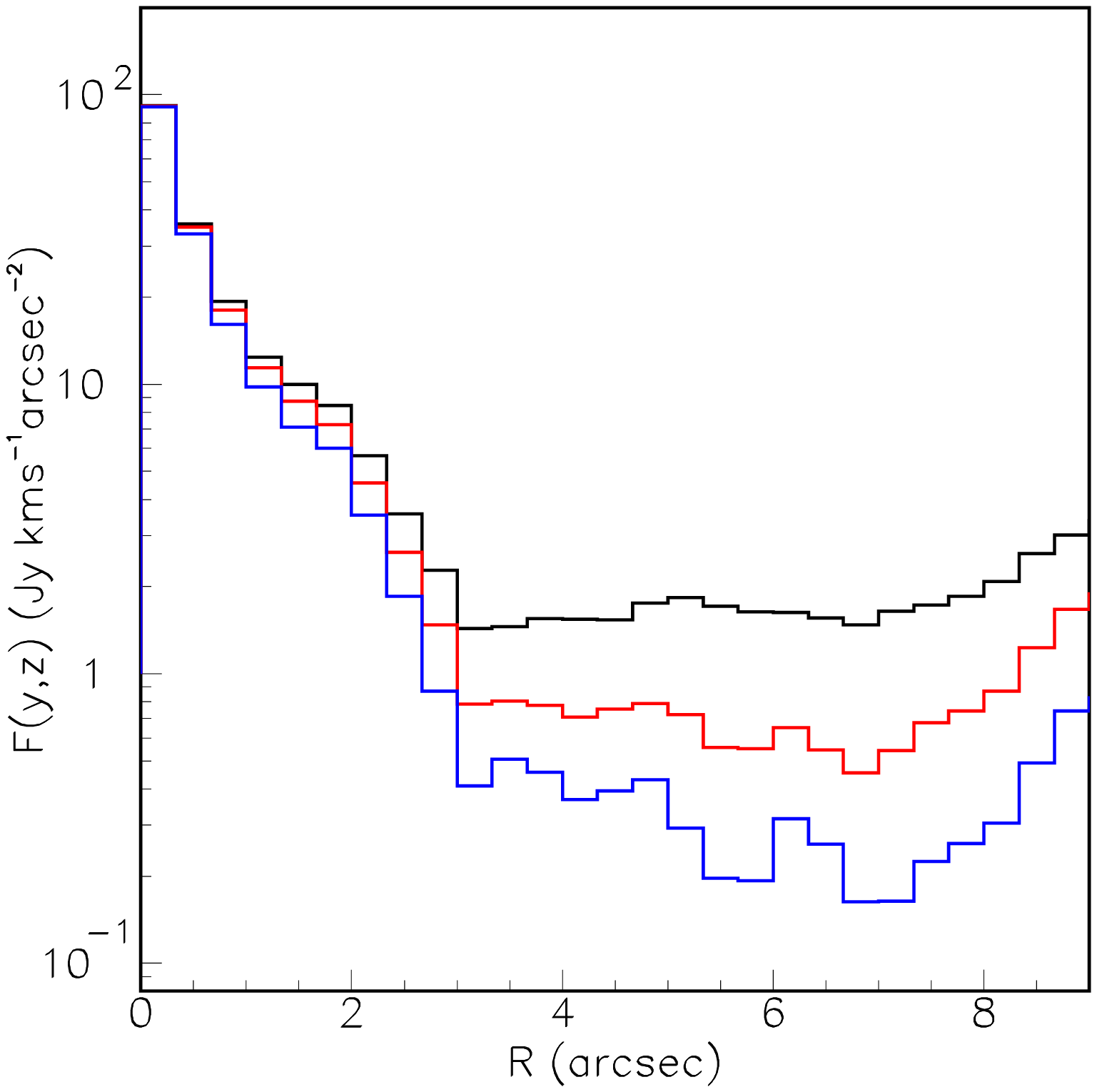}
\includegraphics[height=5.5cm,trim=0.cm 0.cm 0.5cm 0.cm,clip]{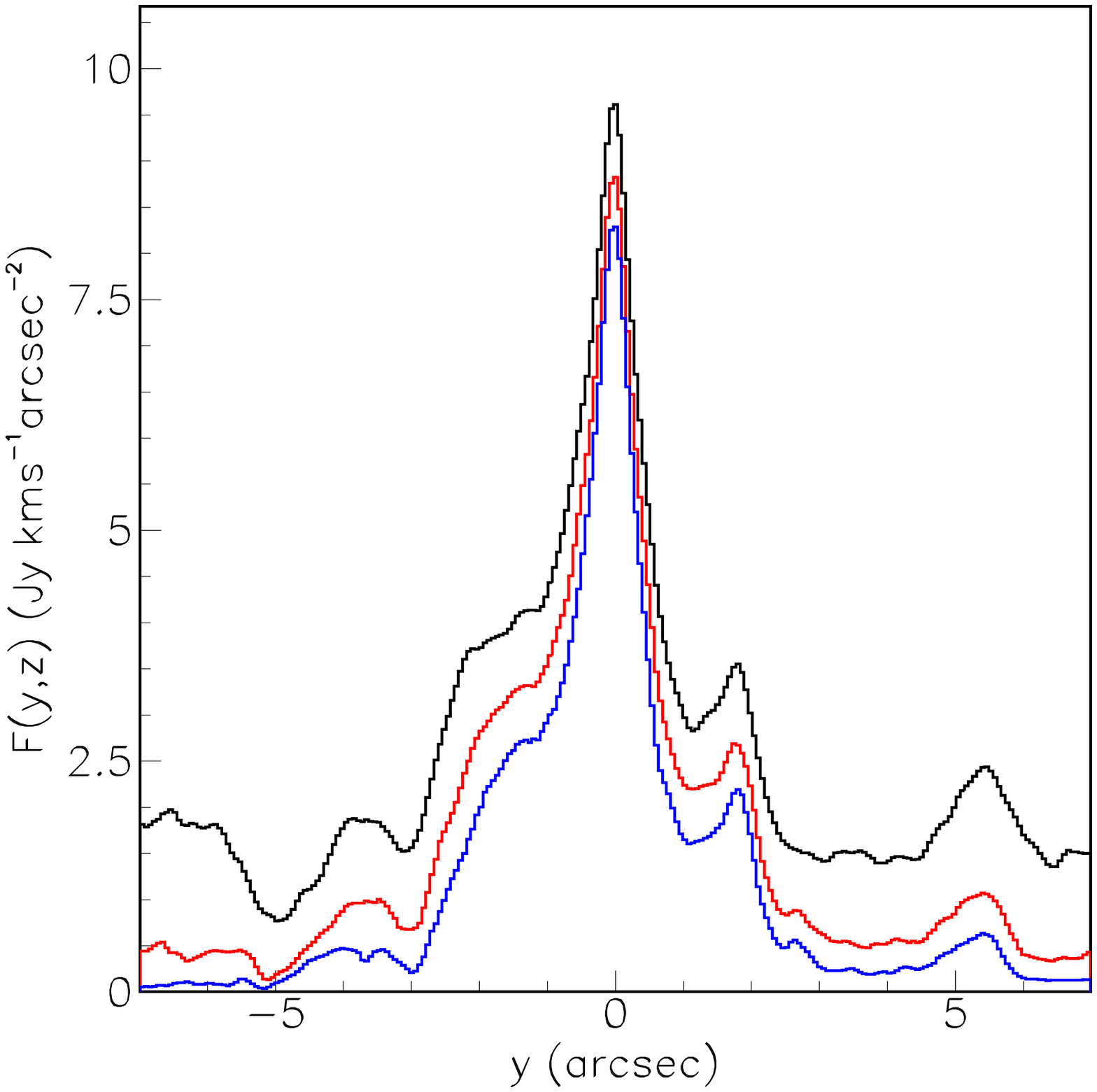}
\includegraphics[height=5.5cm,trim=0.cm 0.cm 0.5cm 0.cm,clip]{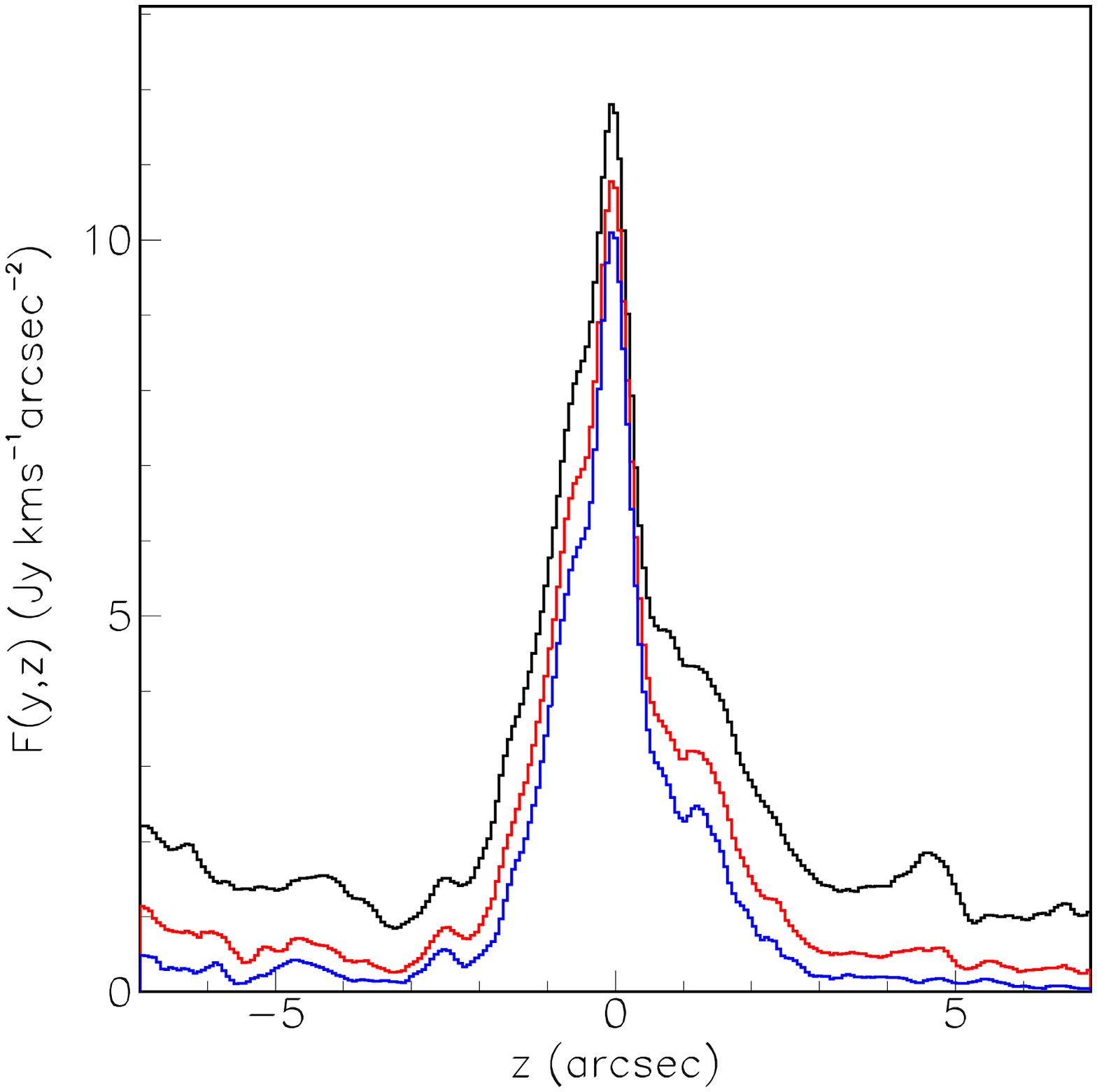}
\caption{Left: distributions of $F(y,z)$ as a function of $R$. Middle and right: distributions of $F(y,z)$ 
as a function of respectively $y$ and $z$ for $|y|$<7$''$ and $|z|$<7$''$. Colours indicate cuts at 1 (black), 2 (red) and 3 (blue) $\sigma$'s.}
\label{fig3}
\end{figure*}

The distribution of $F$ as a function of $R$, averaged over position angles and integrated between 34 km\,s$^{-1}$ 
and 56 km\,s$^{-1}$ is displayed in Fig.~\ref{fig3} (left) for cuts at 1, 2 and 3 $\sigma$'s. In what follows, we 
restrict accordingly the analysis to pixels having $R$<7$''$ in order to be free of problems related with the lack 
of large distance coverage. In practice, most of the results presented here are confined to the interior of a circle 
of radius $R$=3$''$. Distributions of $F$ as a function of $y$ and $z$ are then shown in Fig.~\ref{fig3} (middle and right) 
for cuts at 1, 2 and 3$\sigma$'s. They have been centred to have the peak of maximum emission at the origin of coordinates.

Channel maps covering $8.2''\times8.2''$ are shown in Appendix A.

\section{Morphology and kinematics of the circumbinary envelope: an overview}

\begin{figure*}
\centering
\includegraphics[height=7.cm,trim=0.cm 0.cm 0.cm 0.cm,clip]{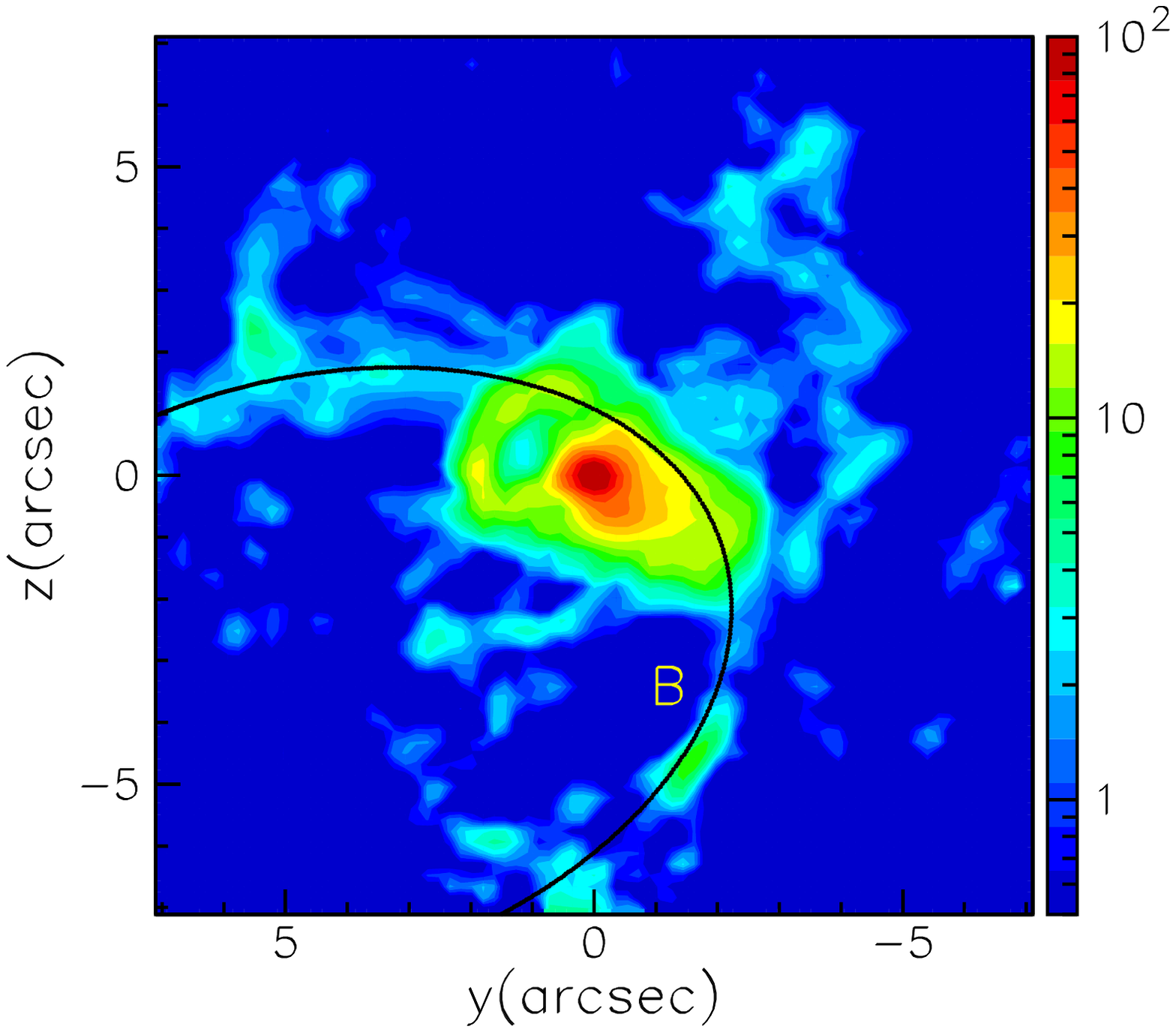}
\includegraphics[height=7.cm,trim=0.cm 0.cm 0.cm 0.cm,clip]{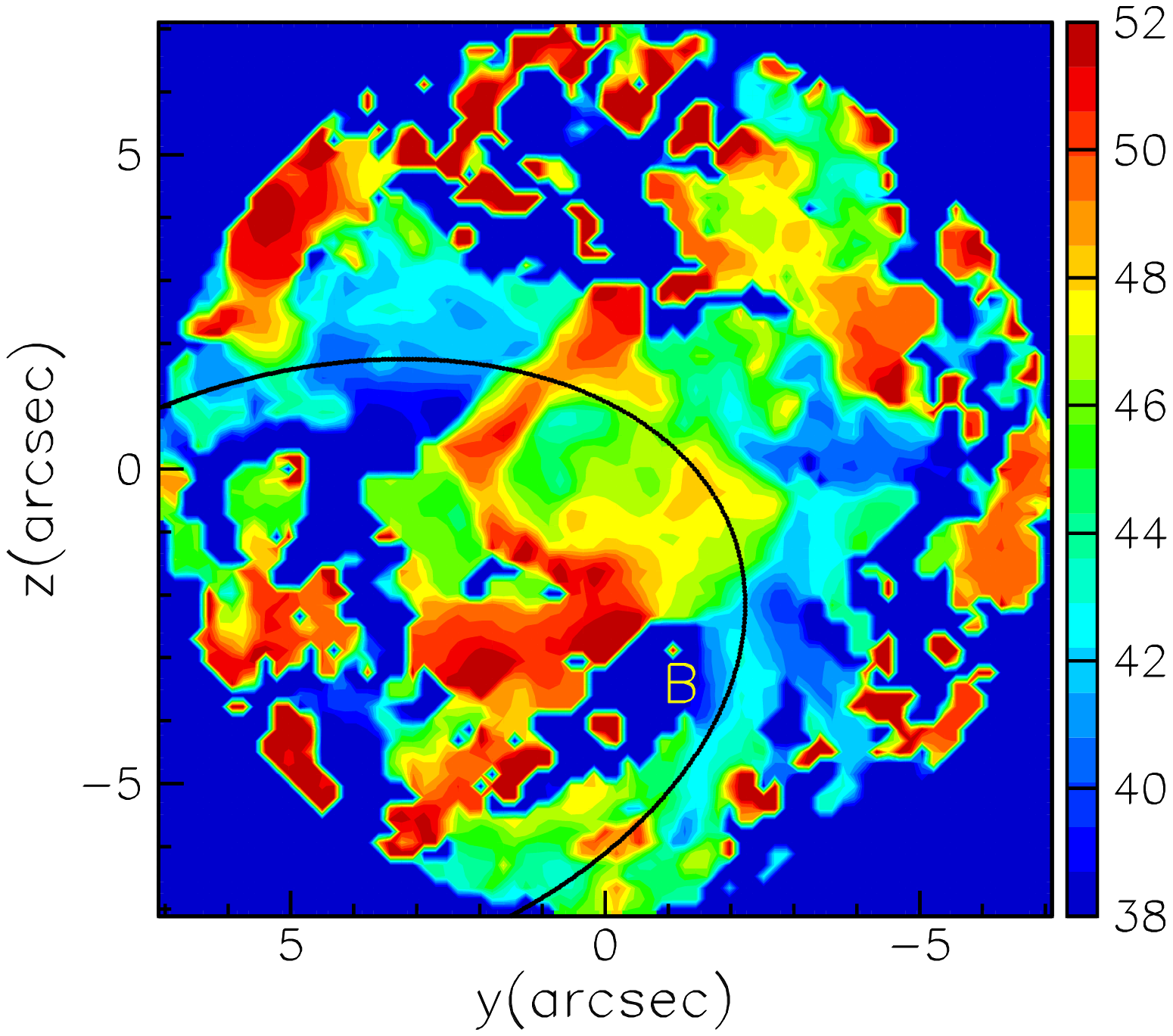}
\caption{Maps of $F$ (left, Jy\,km\,s$^{-1}$arcsec$^{-2}$) and <$V_x$> (right, km\,s$^{-1}$) for a cut at 2$\sigma$'s. 
The curve labelled B corresponds to the bubble (see text).}
\label{fig4}
\end{figure*}

\begin{figure}
\centering
\includegraphics[height=6.5cm,trim=0.cm 0.cm 0.cm 0.cm,clip]{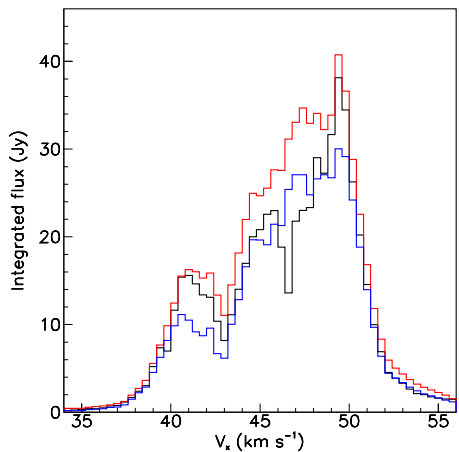}
\caption{Doppler velocity distributions in a circle of radius 7$''$ centred on Mira A 
for cuts at 2$\sigma$'s (red) and 3$\sigma$'s (blue) or without any cut (black).}
\label{fig5}
\end{figure}

Maps of the integrated flux, $F$, and of the mean Doppler velocity, <$V_x$>, are displayed in Fig.~\ref{fig4} for 
a cut at 2$\sigma$'s. It is tempting to associate regions having similar Doppler velocities and to group them to 
form simple patterns. An example is a large south-eastern arc, labelled B in Fig.~\ref{fig4}, corresponding to 
what \citet{Ramstedt2014} call a bubble. While the reality of such associations needs to be critically assessed, 
one can state that a distinctive feature of the data is the existence of regions separated by several arcseconds 
but sharing very similar Doppler velocities. Fig.~\ref{fig5} displays the Doppler velocity distribution in a 
circle of radius 7$''$ centred on Mira A. It shows the contribution of several peaks giving evidence for distinct 
families of sources, making it easier to cope with the complexity of the observed patterns, which \citet{Ramstedt2014} 
have rightly mentioned as being an essential feature of the Mira environment.

\begin{figure*}
\centering
\includegraphics[height=7.cm,trim=1.cm 0.cm 2.5cm 0.cm,clip]{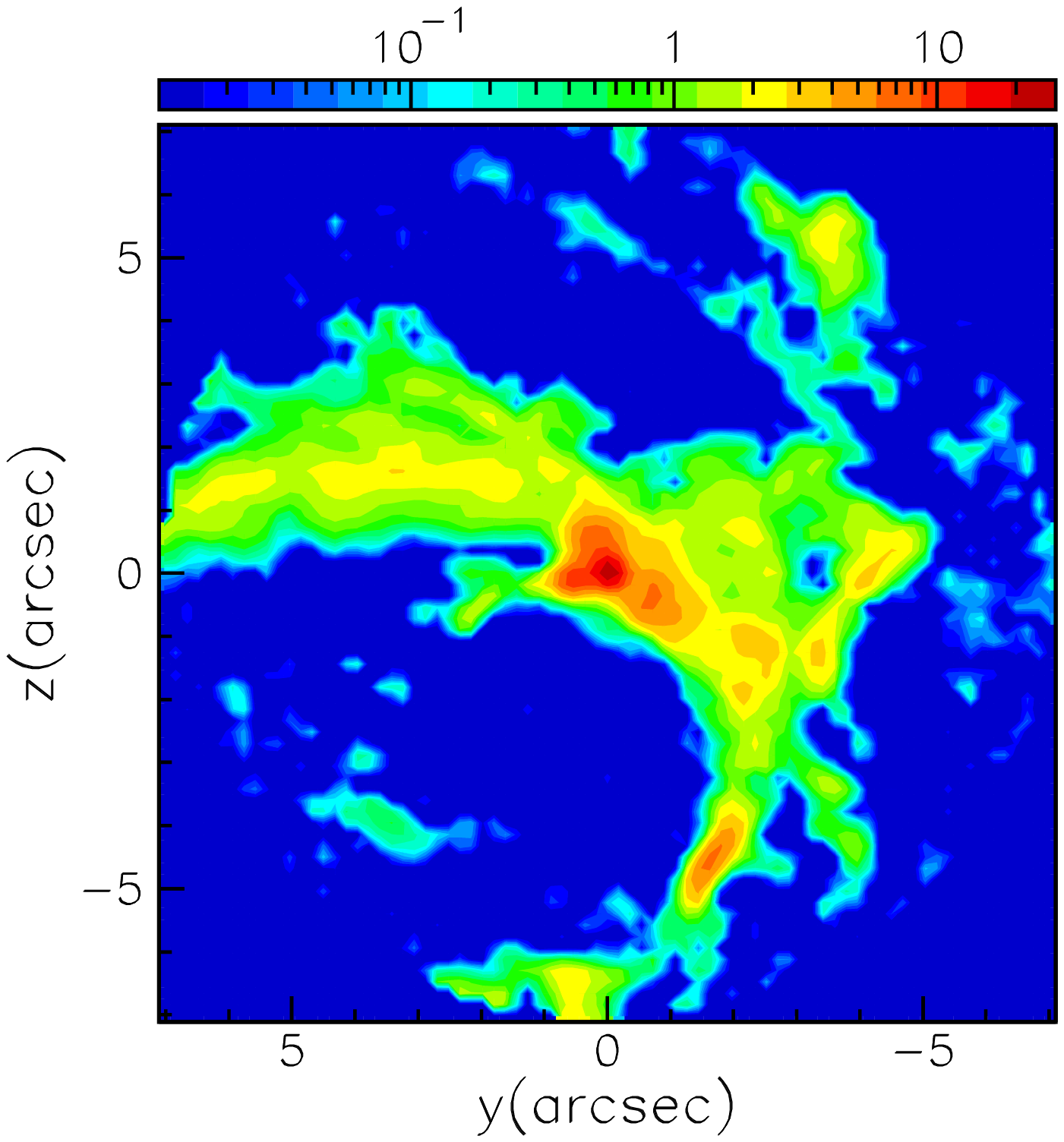}
\includegraphics[height=7.cm,trim=2.5cm 0.cm 2.5cm 0.cm,clip]{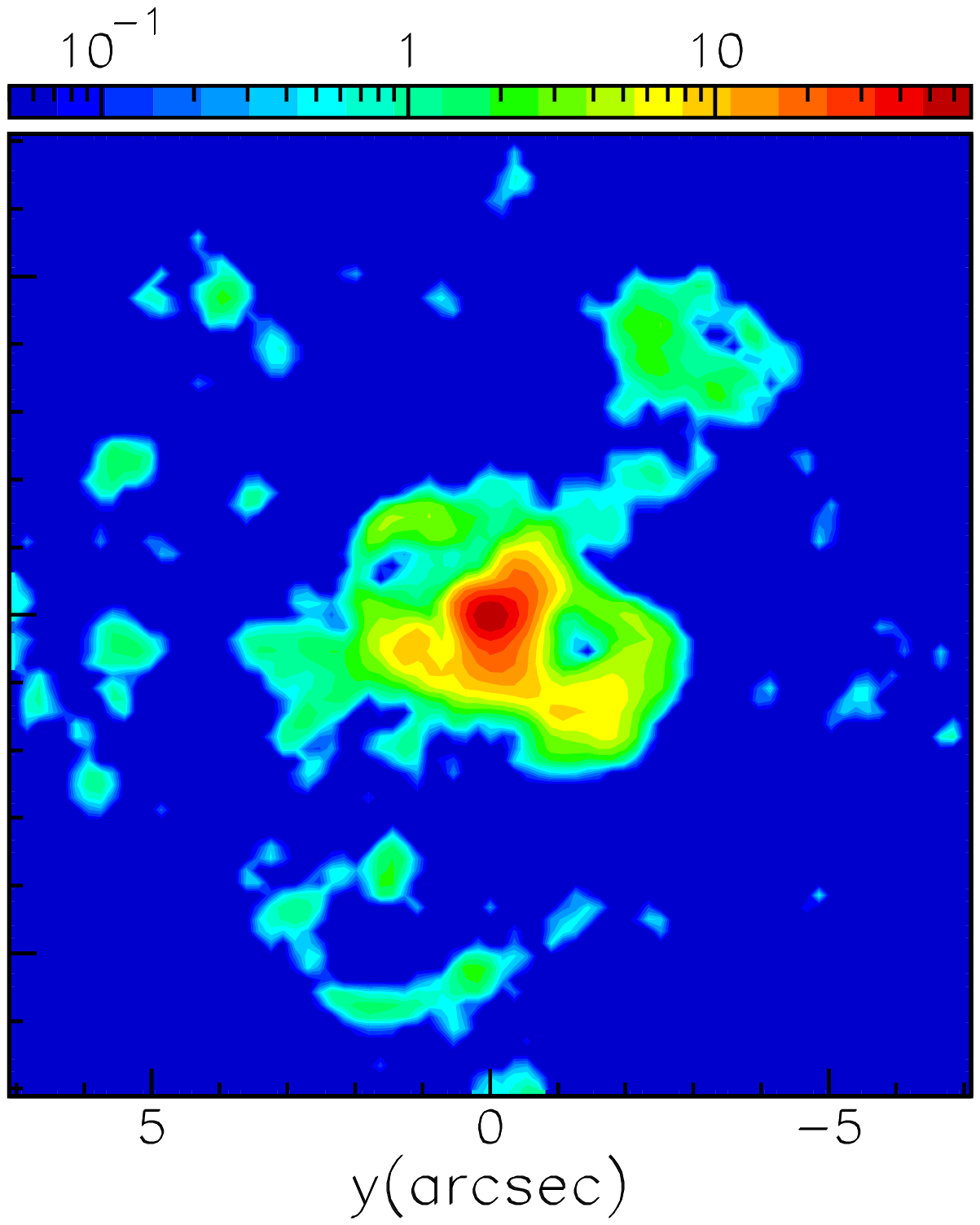}
\includegraphics[height=7.cm,trim=2.5cm 0.cm 2.5cm 0.cm,clip]{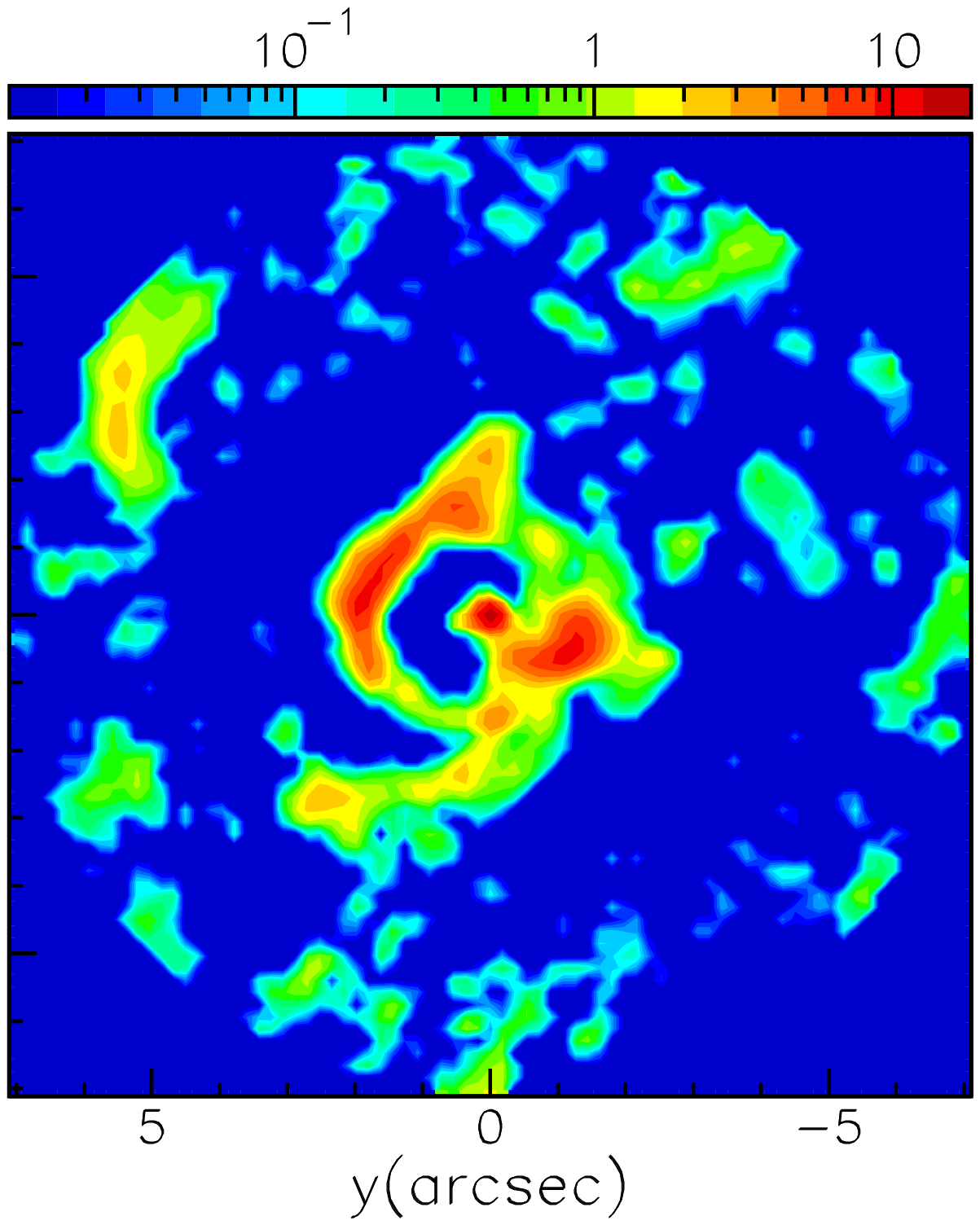}
\caption{Channel maps drawn inside a 7$''$ radius circle centred on Mira A for a 2$\sigma$ cut, 
with velocity intervals of 34-44, 45-49 and 50-56 km\,s$^{-1}$ from left to right. The colour scale is in the unit of Jy\,km\,s$^{-1}$ arcsec$^{-2}$}
\label{fig6}
\end{figure*}

Channel maps in velocity intervals representative of the structure observed in Fig.~\ref{fig5} are shown in 
Fig.~\ref{fig6}. The blue shifted region, 34 to 44 km\,s$^{-1}$, and the red-shifted region, 50 to 56 km\,s$^{-1}$, 
respectively identified by \citet{Ramstedt2014} as dominated by a bubble and by spiral arcs, will be treated separately 
from the central velocities, 44 to 50 km\,s$^{-1}$, that correspond to sources confined inside a circle of $\sim$2.5$''$ 
radius around Mira A. The three velocity intervals just mentioned are not perfectly separated, which complicates the 
analysis in the overlap regions. We shall aim at giving as objective as possible a description of what has been observed, 
without relying on a priori preconceptions, such as the existence of spiral wakes or of an accretion disk around Mira B. 
The complexity of the observed patterns is such that one can easily be induced into grouping different features in 
associations that may have no real identity. Moreover, the emission of the CO(3-2) line traces the presence of gas 
in only well-defined conditions of temperature, is strongly influenced by the presence of turbulences or of short 
wavelength radiations and is sensitive to optical thickness. In the very variable and inhomogeneous environment of 
Mira, such factors may well play an important role.

\section{The blue-shifted south-eastern bubble}

\begin{figure*}
\centering
\includegraphics[height=6.2cm,trim=1.cm 0.cm 2.cm 0.cm,clip]{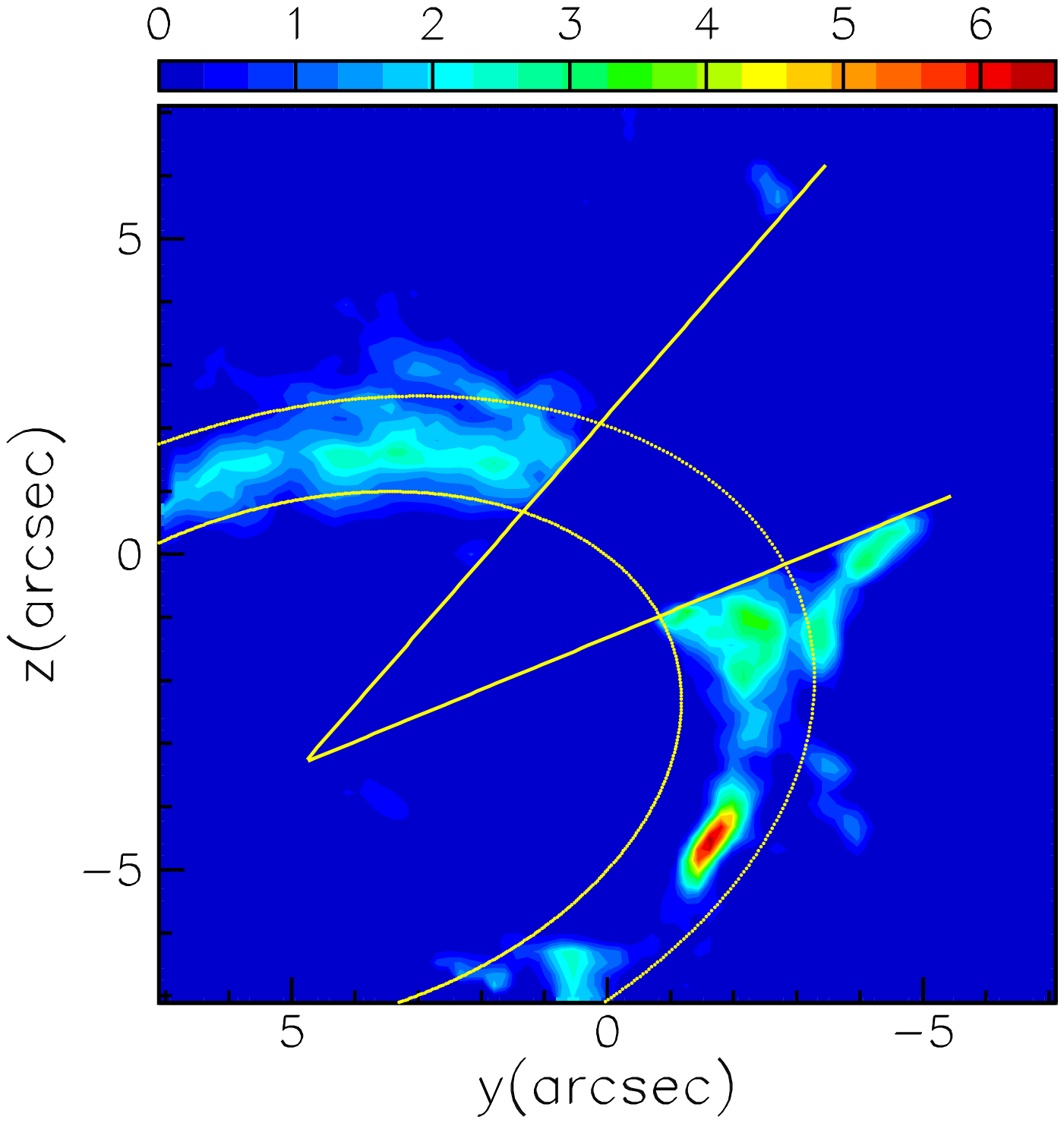}
\includegraphics[height=6.2cm,trim=.5cm 0.cm 1.5cm 0.cm,clip]{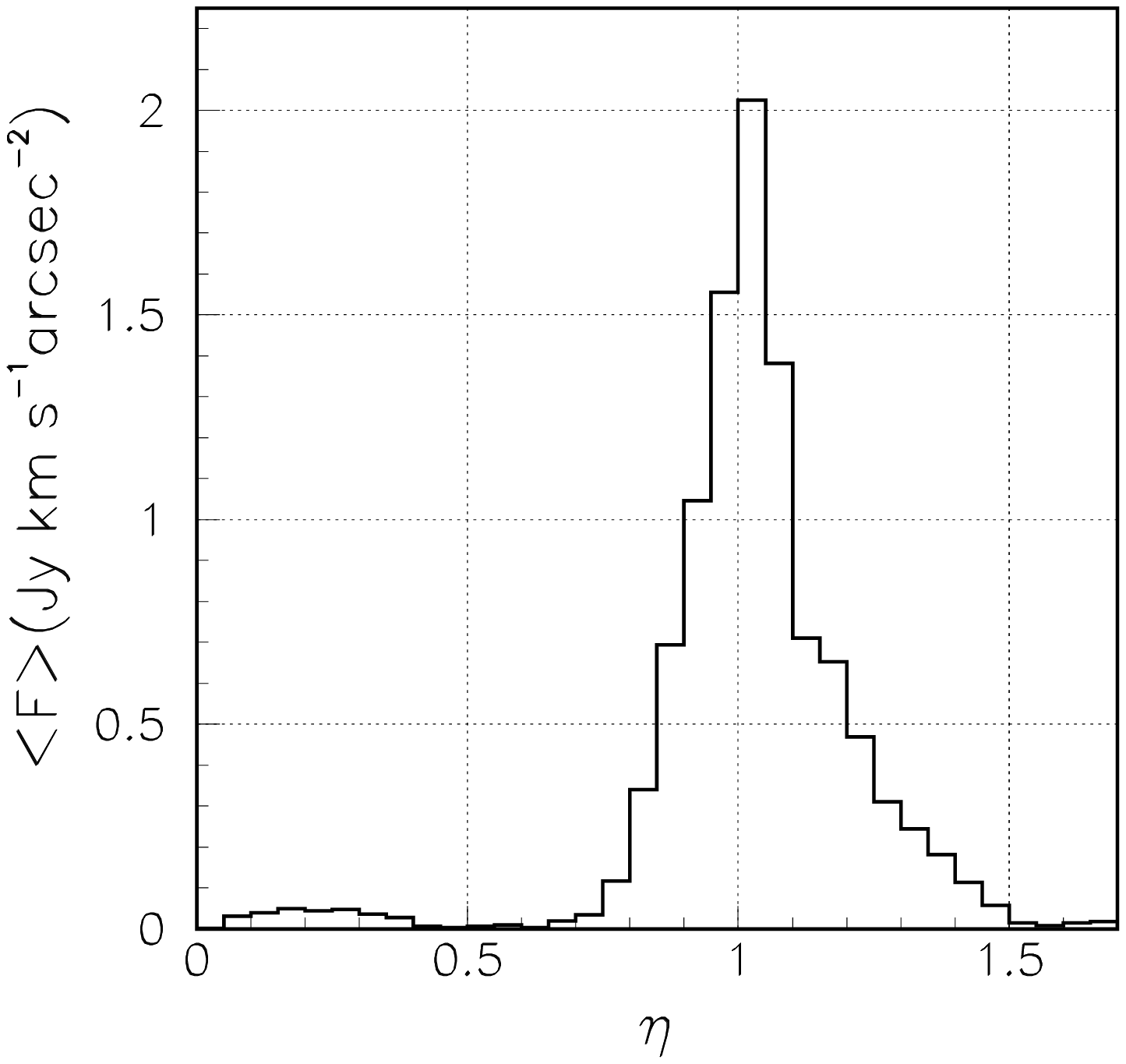}
\includegraphics[height=6.2cm,trim=.5cm 0.cm 1.5cm 0.cm,clip]{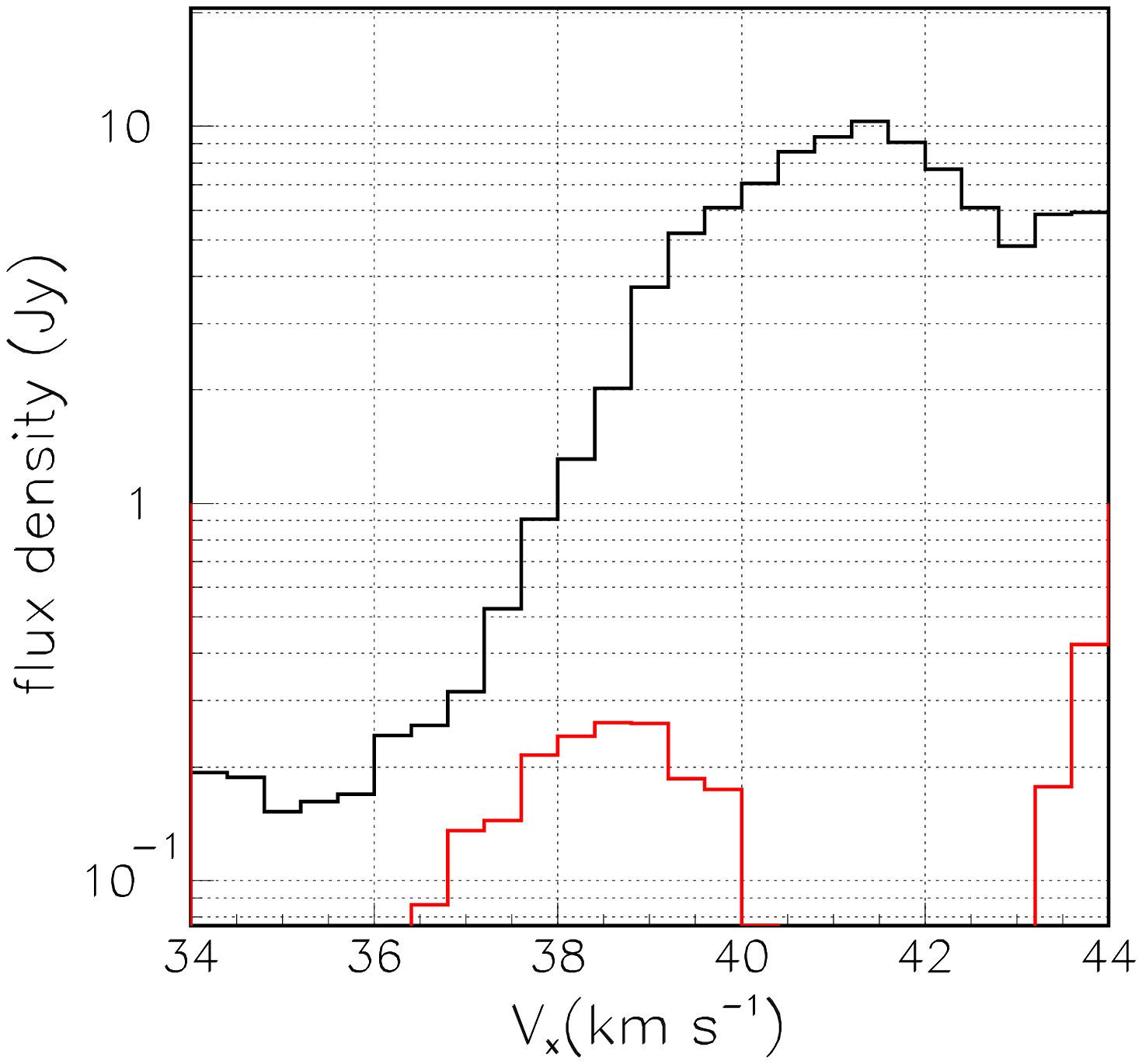}
\caption{Left: Sky map of F (Jy km\,s$^{-1}$ arcsec$^{-2}$) integrated between Doppler velocities of 34 and 44 km\,s$^{-1}$. 
The ellipses correspond to $\eta$=0.85 and 1.15, the lines to $\varphi_{bubble}$=120$^\circ$ and 155$^\circ$. 
Middle: Flux density averaged over $\varphi_{bubble}$ as a function of $\eta$. Right: Doppler velocity distribution 
for pixels having $\eta$<0.8 (red) and pixels having 0.8<$\eta$<1.4 (black) separately. }
\label{fig7}
\end{figure*}

Figure~\ref{fig7} (left) displays the sky map of $F$ integrated between Doppler velocities of 34 and 44 km\,s$^{-1}$. 
Two ellipses have been drawn by eye to roughly delineate the pattern identified by \citet{Ramstedt2014} as a bubble. 
The position angle of the major axis is 17$^\circ$ north of west, the aspect ratio (ratio of major to minor axes) is 1.5, 
the half minor and major axes measure 4.1$''$ and 5.5$''$ respectively and the coordinates of the centre are $y=4.8''$ and $z=-3.3''$. 
We define for each point ($y,z$) in the plane of the sky an elliptic radial coordinate $\eta$ cancelling at the centre 
of the ellipse and taking respective values of 0.85 and 1.15 on the two ellipses of Fig.~\ref{fig7}:
\begin{equation}
\begin{array}{c}
\xi=(y-4.8'')\cos(17^\circ)-(z+3.3'')\sin(17^\circ) \\
\zeta=(y-4.8'')\sin(17^\circ)+(z+3.3'')\cos(17^\circ) \\
\eta^2=(\frac{\xi}{7.2''})^2+(\frac{\zeta}{4.8''})^2 \\
\end{array}
\end{equation}
We also define an elliptical position angle $\varphi_{bubble}$, measured clockwise from east (rather than from west) 
such that $\tan(\varphi_{bubble}+17^\circ)=(\zeta/4.8'')/(\xi/7.2'')$. In order to keep away from contributions from 
the Mira A+B neighbourhood we require $\varphi_{bubble}$ not to be in the interval [120$^\circ$,155$^\circ$].

Figure~\ref{fig7} (middle) displays the flux per pixel and velocity bin averaged over $\varphi_{bubble}$ as a 
function of $\eta$. The bubble region sticks out very clearly. The FWHM is 0.17, meaning $\sim$1.0$''$ on average. 
The inner edge is sharper than the outer edge, while the opposite is expected from a bubble (understood as an 
ellipsoid shell of constant thickness). The inner region is strongly depleted, at the limit of sensitivity when 
using a cut at 2$\sigma$'s (as done here). The ratio of the flux integrated over the inner part of the bubble 
($\eta<$0.8) to the flux integrated over the periphery of the bubble (0.8$<\eta<$1.4) is 2.8\% with a 2$\sigma$ 
cut and 7.4\% without cut. Figure~\ref{fig7} (right) displays the Doppler velocity distribution for pixels having 
$\eta<$0.8$''$ and pixels having 0.8$''<\eta<$1.4$''$ separately.

\begin{figure}
\centering
\begin{minipage}[c][][t]{\columnwidth}
\centering
\includegraphics[height=5.cm,trim=1.2cm 0.cm 1.cm 0.cm,clip]{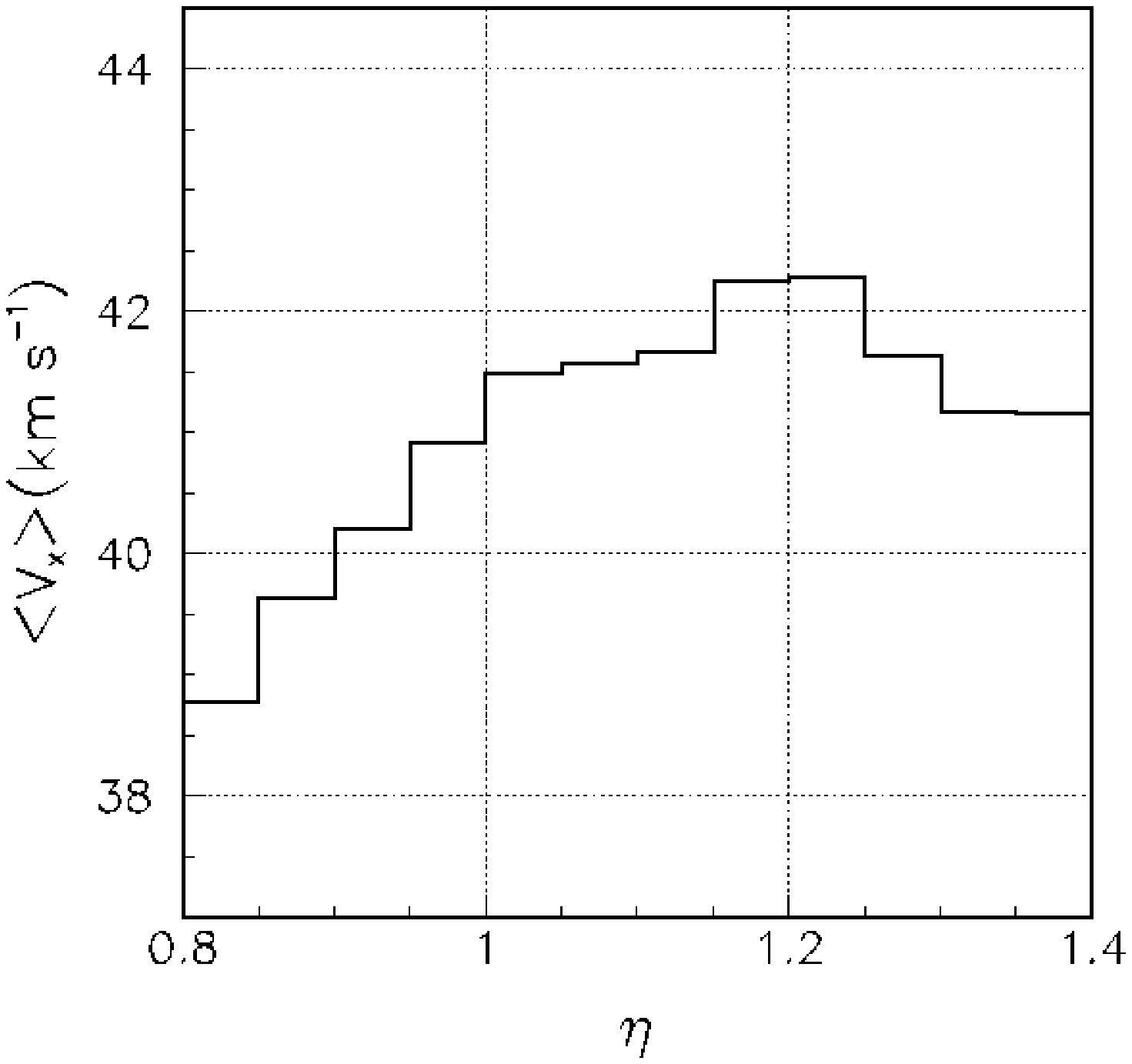}
\includegraphics[height=5.cm,trim=0.cm 0.cm 1.cm 0.cm,clip]{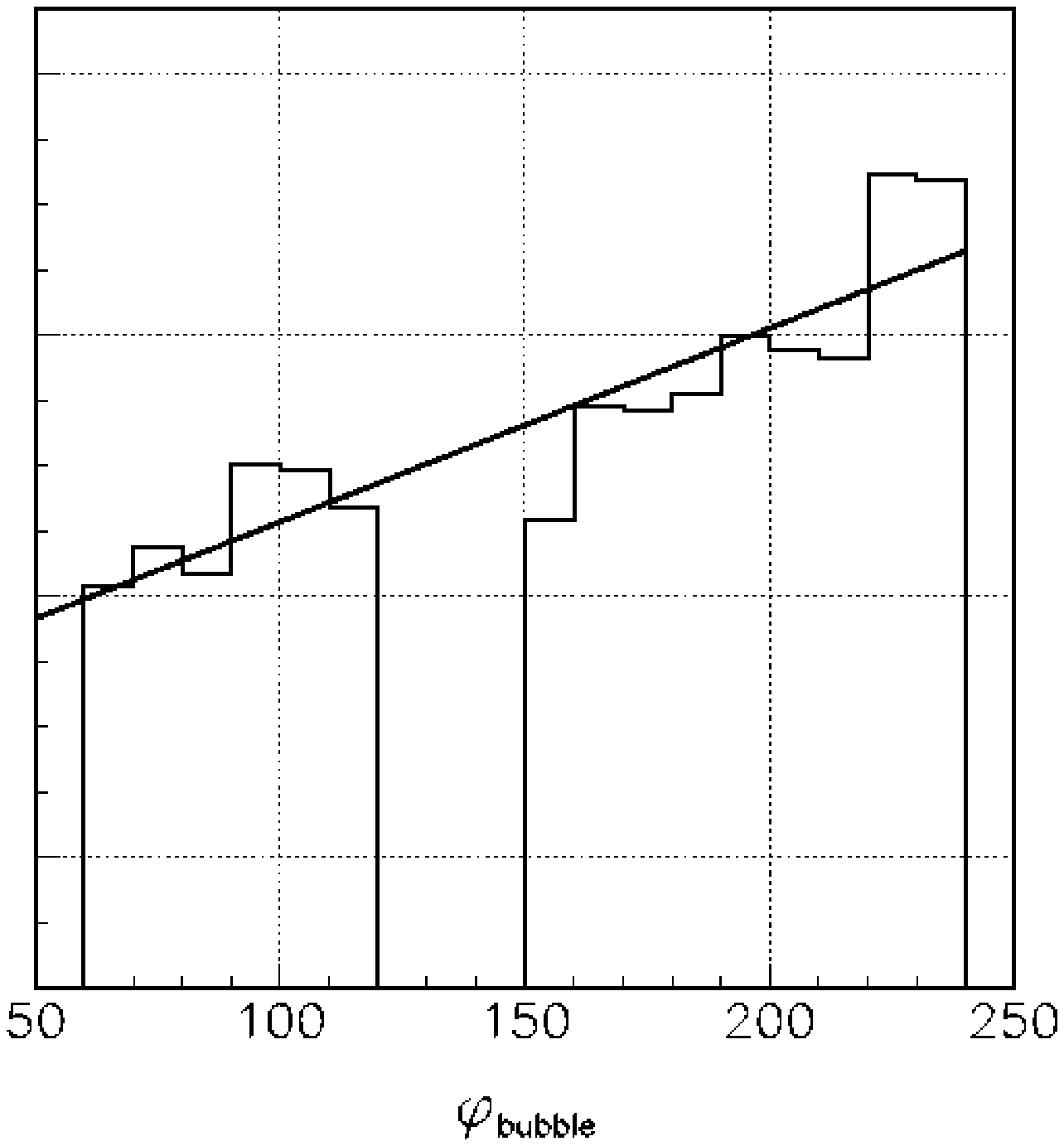}
\caption{Dependence of the mean Doppler velocity <$V_x$> on $\eta$ (left) and, for the bubble region defined 
as 0.8<$\eta$<1.4, on $\varphi_{bubble}$ (right, measured clockwise from east).}
\label{fig8}
\end{minipage}
\end{figure}

The dependence of the mean Doppler velocity <$V_x$> on $\eta$ and $\varphi_{bubble}$ is illustrated in 
Fig.~\ref{fig8} and shows small, but significant variations. The mean Doppler velocity increases 
from $\sim$39 km\,s$^{-1}$ inside the arc to between 41 and 42 km\,s$^{-1}$ on the arc and beyond. 
In the arc region, it increases from $\sim$40 km\,s$^{-1}$ at the north-eastern extremity to between 
42 and 43 km\,s$^{-1}$ at the south-western extremity. If the ellipse is interpreted as the projection 
of a circular ring in space, its angle with the plane of the sky is $\sim$50$^\circ$. Moreover, if the 
ring is expanding radially at constant velocity, as suggested by Fig.~\ref{fig8} (right), this 
constant velocity would project as $(42.5-40.0)$/2$\sim$1.3 km\,s$^{-1}$ on the line of sight, namely 
would be of the order of 1.7 km\,s$^{-1}$ in space. To span 7.2$''$, meaning $\sim$720 AU, at such 
velocity takes $\sim$2000 yr.

In summary, the blue-shifted part of the Doppler velocity distribution is dominated by a narrow arc ($\sim$1$''$ FWHM), 
referred to as ``the bubble'', that can be interpreted as the projection of a circular ring expanding 
at a velocity of less than 2 km\,s$^{-1}$, born some 2000 years ago and making an angle of $\sim$50$^\circ$ 
with the plane of the sky.

\section{Red-shifted arcs}

\begin{figure*}
\centering
\includegraphics[height=7.cm,trim=0.cm 0.cm 0.cm 0.cm,clip]{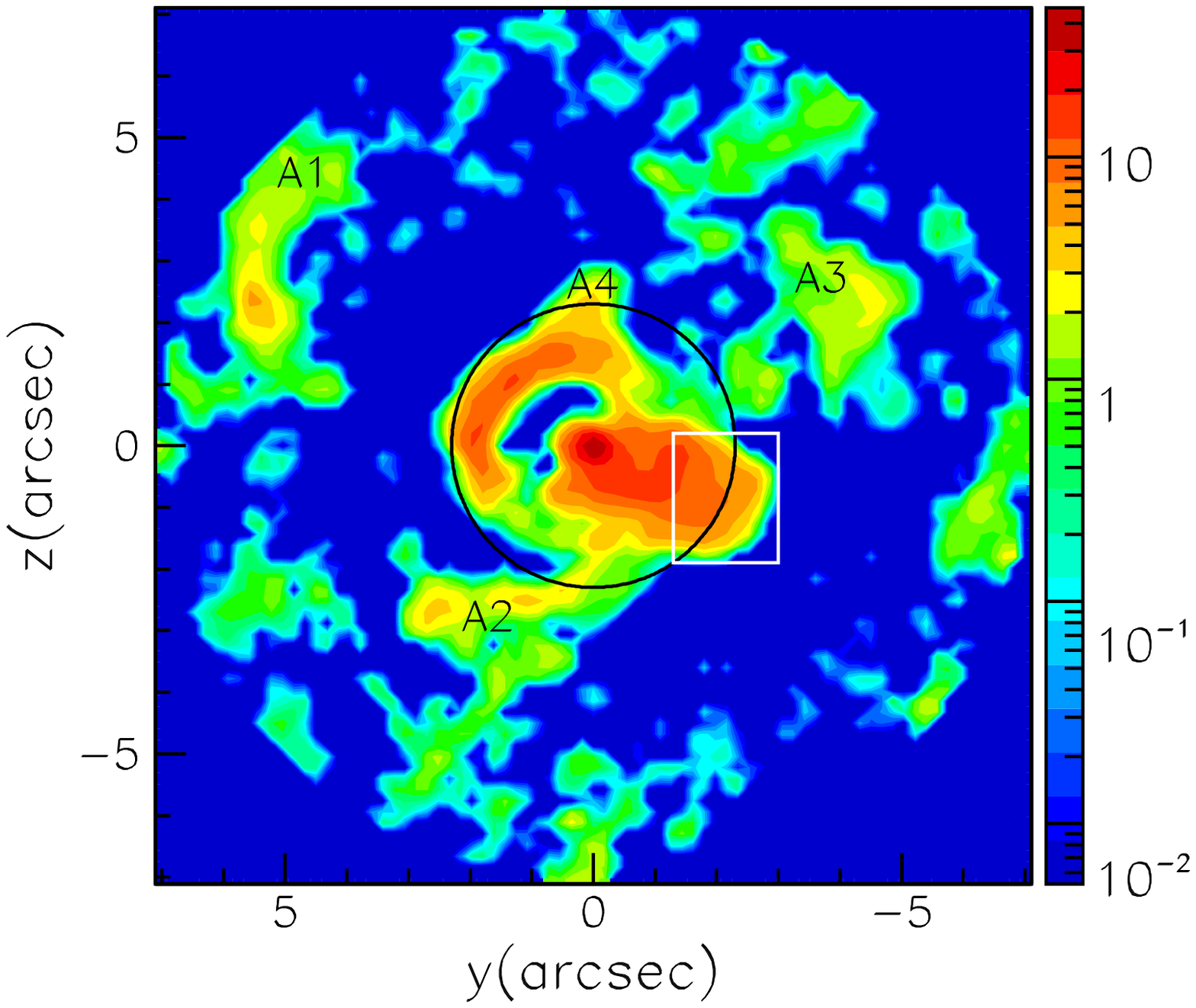}
\includegraphics[height=7.cm,trim=0.cm 0.cm 0.cm 0.cm,clip]{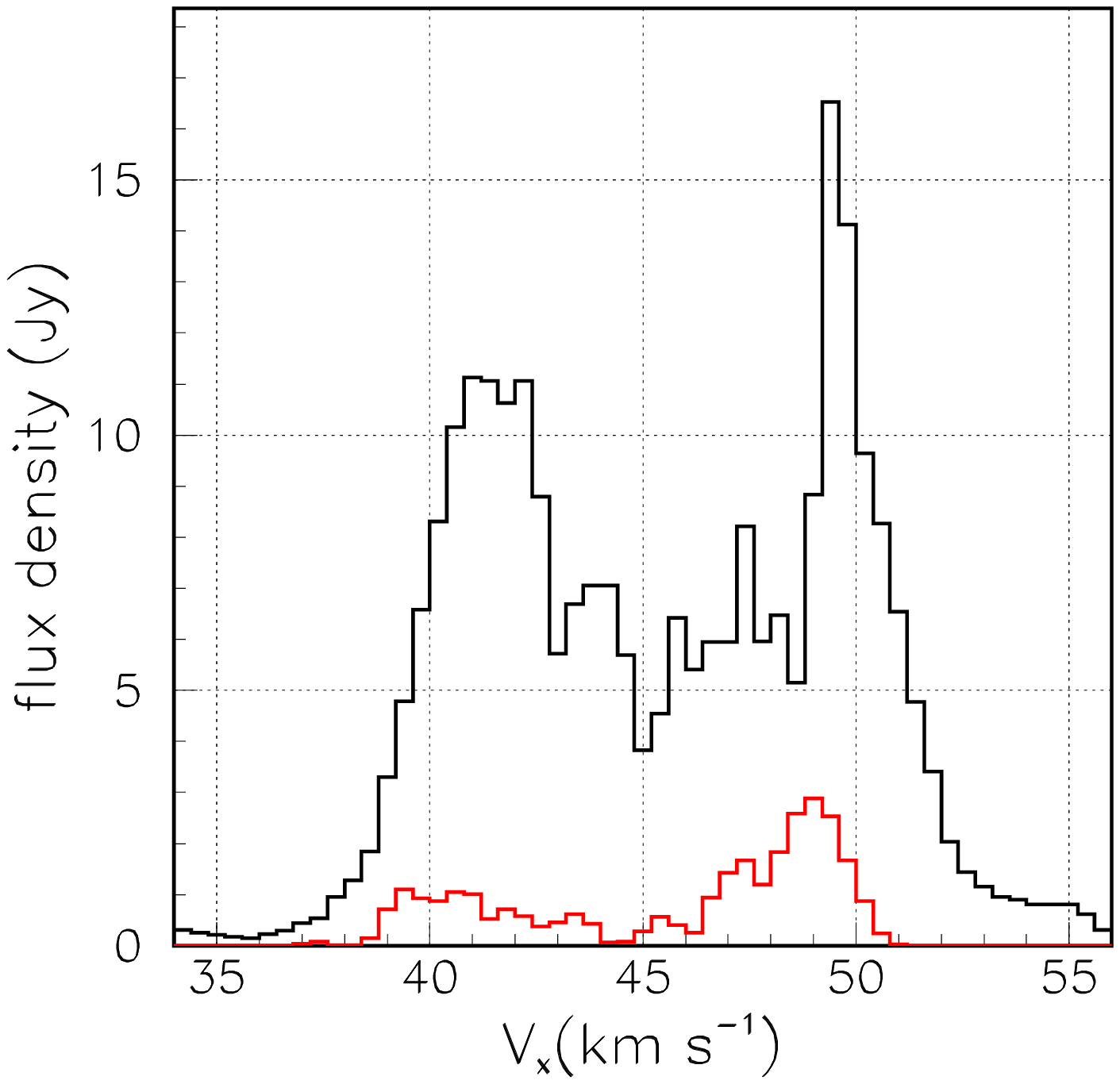}
\caption{Left: Sky map of F (Jy\,km\,s$^{-1}$ arcsec$^{-2}$) for Doppler velocities $V_x$>48 km\,s$^{-1}$. 
Right: Doppler velocity distributions for pixels located outside a circle centred on Mira A and having a radius of 2.3$''$,
 as shown on the left panel. The red histogram displays the contribution of the flux contained in the part of the rectangle
 exterior to the circle (on the left panel).}
\label{fig9}
\end{figure*}

Figure~\ref{fig9} (left) displays the sky map of $F$ integrated over Doppler velocities larger than 
48 km\,s$^{-1}$. It shows several zones of emission, the main ones being labelled A1, A2, A3 and A4 
for further reference. The region surrounding Mira A+B at distances not exceeding 2.3$''$ is treated 
in the next section.

The right panel displays the Doppler velocity distribution in the region 2.3$''$<$R$<7$''$, the red 
histogram showing the contribution of what is described as the south-western outflow in the next 
section.  Two peaks, at $\sim$41.5 and $\sim$49.5 km\,s$^{-1}$, stand out, suggesting a global expansion 
about a mean velocity of $\sim$45.5 km\,s$^{-1}$. The blue peak is easily identified as the contribution 
of the bubble, the red peak being associated with the arcs displayed in the left panel. We have 
considered various possible associations of these sources, either on spiral arms or on large circles 
or ellipses. We do not feel confident enough to give sufficiently convincing arguments about their 
physical reality and we prefer to refrain from doing so. This should not be understood as a statement 
against the existence of such associations, but we are simply unable to unravel them with sufficient 
confidence from the present data. The data used by \citet{Ramstedt2014}, having a better coverage at 
large distances are more suitable for this purpose than those used here.

\section{The close environment of the Mira A+B pair}
\subsection{Immediate environment}

\begin{figure*}
\centering
\includegraphics[height=6cm,trim=1.cm 0.cm 1.2cm 0.cm,clip]{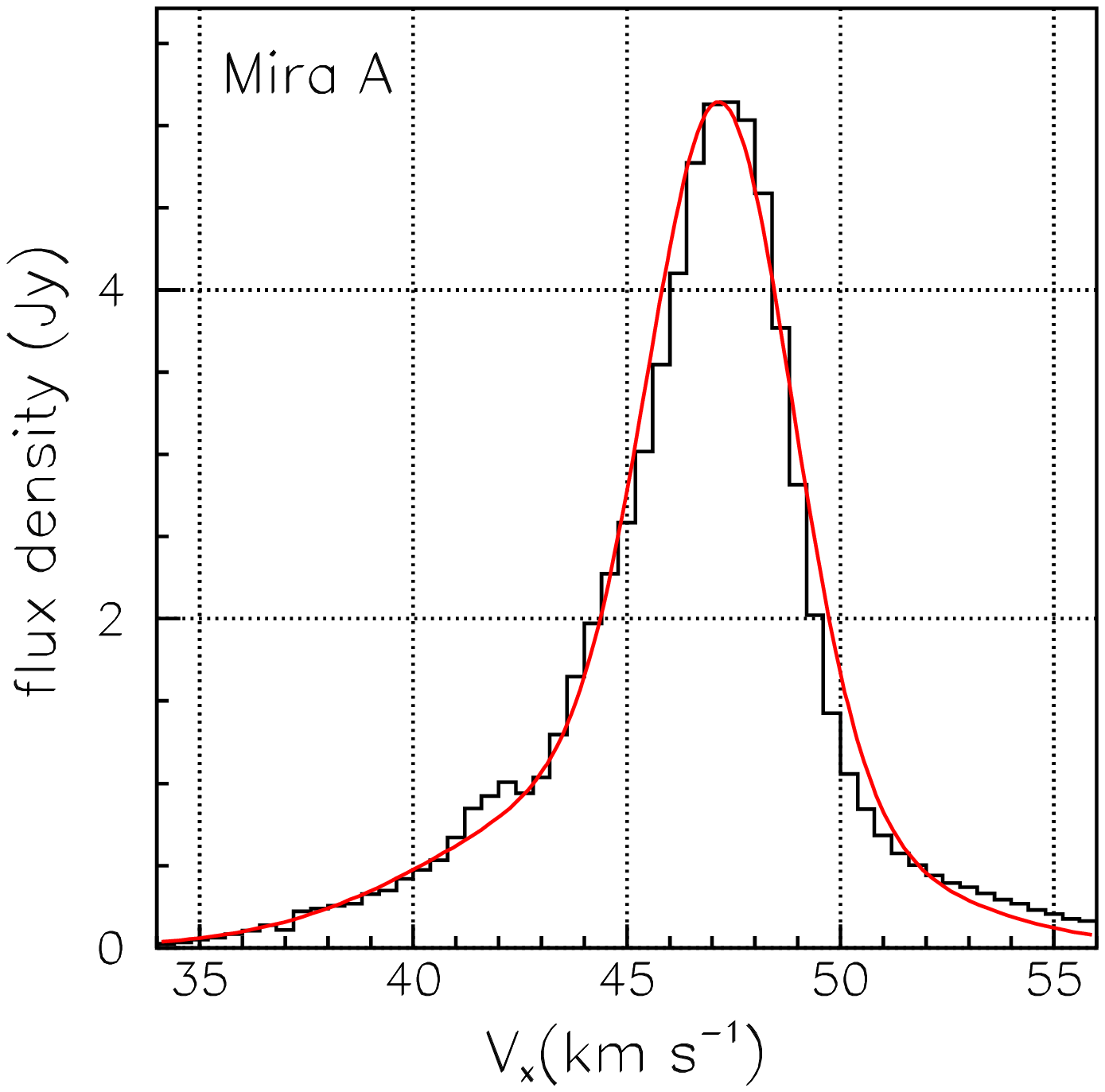}
\includegraphics[height=6cm,trim=0.5cm 0.cm 1.2cm 0.cm,clip]{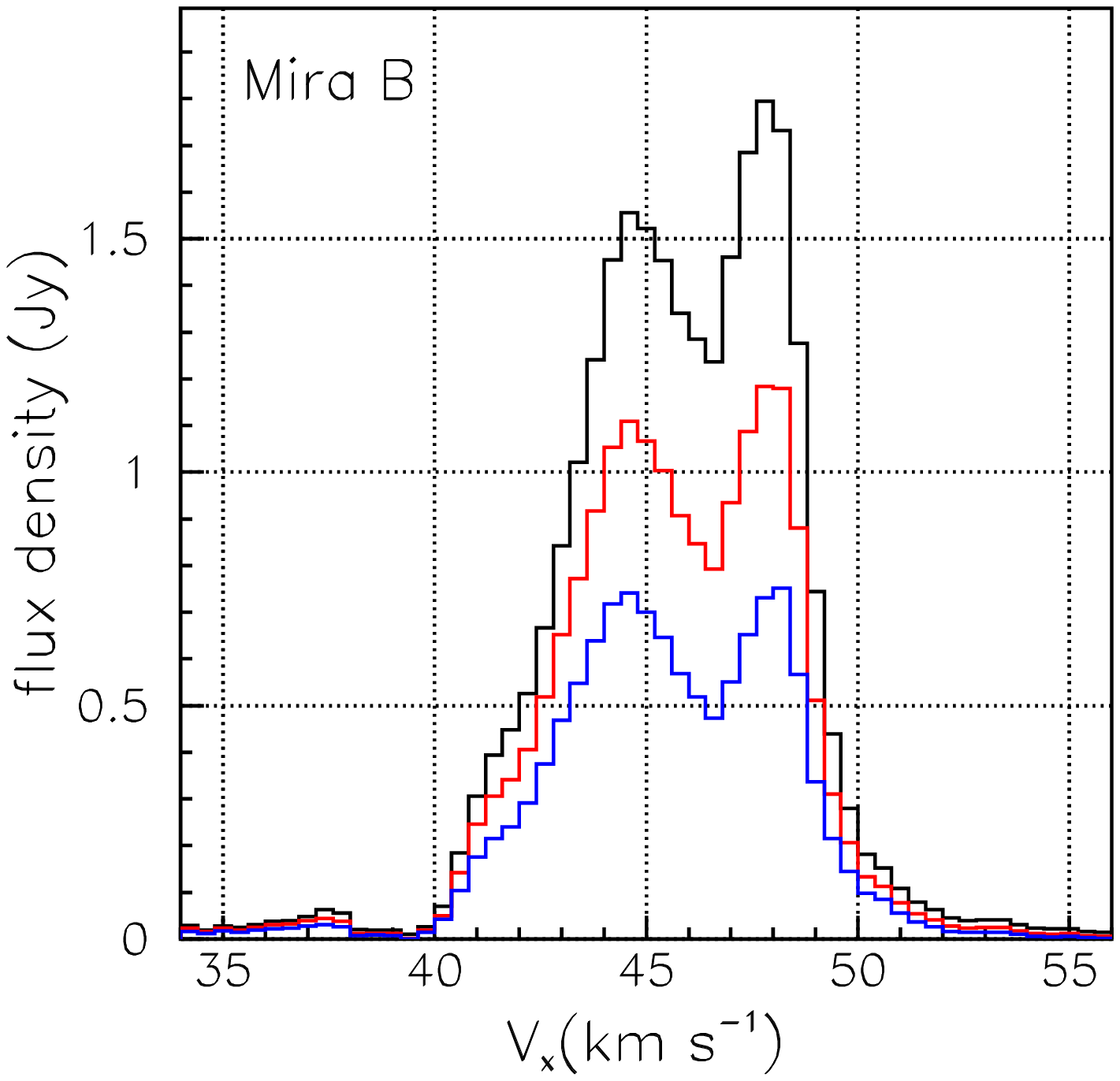}
\includegraphics[height=6cm,trim=0.5cm 0.cm 1.2cm 0.cm,clip]{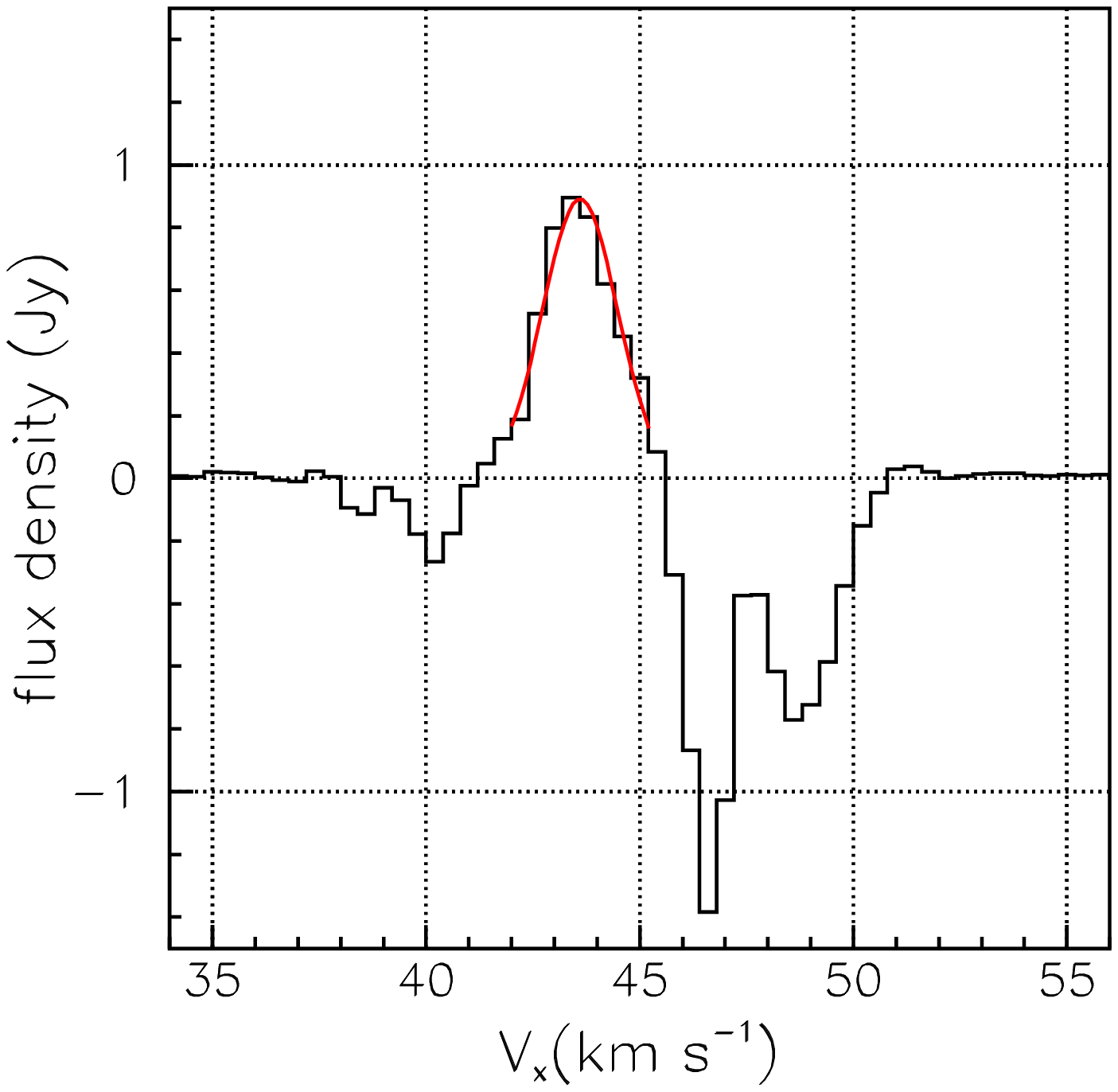}
\caption{Left: Doppler velocity distribution integrated in a circle of radius 0.3$''$ centred on Mira A. 
The fit is the sum of two Gaussians (see text). Middle: Doppler velocity distributions integrated in circles 
of radius 0.20$''$ (blue), 0.25$''$ (red) and 0.3$''$ (black) centred on Mira B. Gaussian fits to the peaks 
for the 0.20$''$ radius give mean velocities of 44.6 and 47.9 km\,s$^{-1}$, amplitudes of 0.73 and 0.74 
Jy arcsec$^{-2}$ and $\sigma$'s of 1.6 and 1.1 km\,s$^{-1}$. Right: difference between the Doppler velocity 
distributions integrated in circles of radius 0.3$''$ and centred on Mira B and on the symmetric of Mira B 
with respect to Mira A. The curve is a Gaussian fit to the peak with mean value of 43.6 km\,s$^{-1}$, FWHM 
of 2.0 km\,s$^{-1}$ and amplitude 0.89 Jy arcsec$^{-2}$.  }
\label{fig10}
\end{figure*}

From the parameters of the Mira B orbit derived by \citet{Prieur2002} we evaluate the space velocity of 
Mira B in the Mira A rest frame to be $\sim$5.0 km\,s$^{-1}$ and the Doppler shift between the two stars 
to be $\sim$1.0 km\,s$^{-1}$. Fig.~\ref{fig10} shows Doppler velocity distributions in the immediate 
environments of Mira A and B respectively.

Near Mira A, the spectrum is asymmetric, with the blue tail larger than the red tail. This cannot be 
due to the contribution of the bubble, which is quite small and indeed clearly identified at 
$\sim$42 km\,s$^{-1}$. The spectrum can be described in terms of two Gaussian components: one, 
narrow (FWHM=3.9 km\,s$^{-1}$) centred at 47.2 km\,s$^{-1}$ and a wider one (FWHM=10.5 km\,s$^{-1}$) 
centred at 45.7 km\,s$^{-1}$. Integrating over velocity, the former is 40\% larger than the latter. 
If the broader component were caused by radial expansion, it would mean an expansion velocity of 
$\sim$5 to 10 km\,s$^{-1}$, at the scale of the space velocity of Mira B in the Mira A rest frame. 
The 1.5 km\,s$^{-1}$ difference between Mira A and the centre of expansion could then be interpreted 
as the result of the centre of mass of the circumbinary envelope being blue-shifted by such an amount 
with respect to the current velocity of Mira A. If such were the case, Mira B would therefore be 
blue-shifted with respect to Mira A and closer to the Earth than Mira A.

\begin{figure*}
\centering
\includegraphics[height=7cm,trim=0.cm 0.cm 0.cm 0.cm,clip]{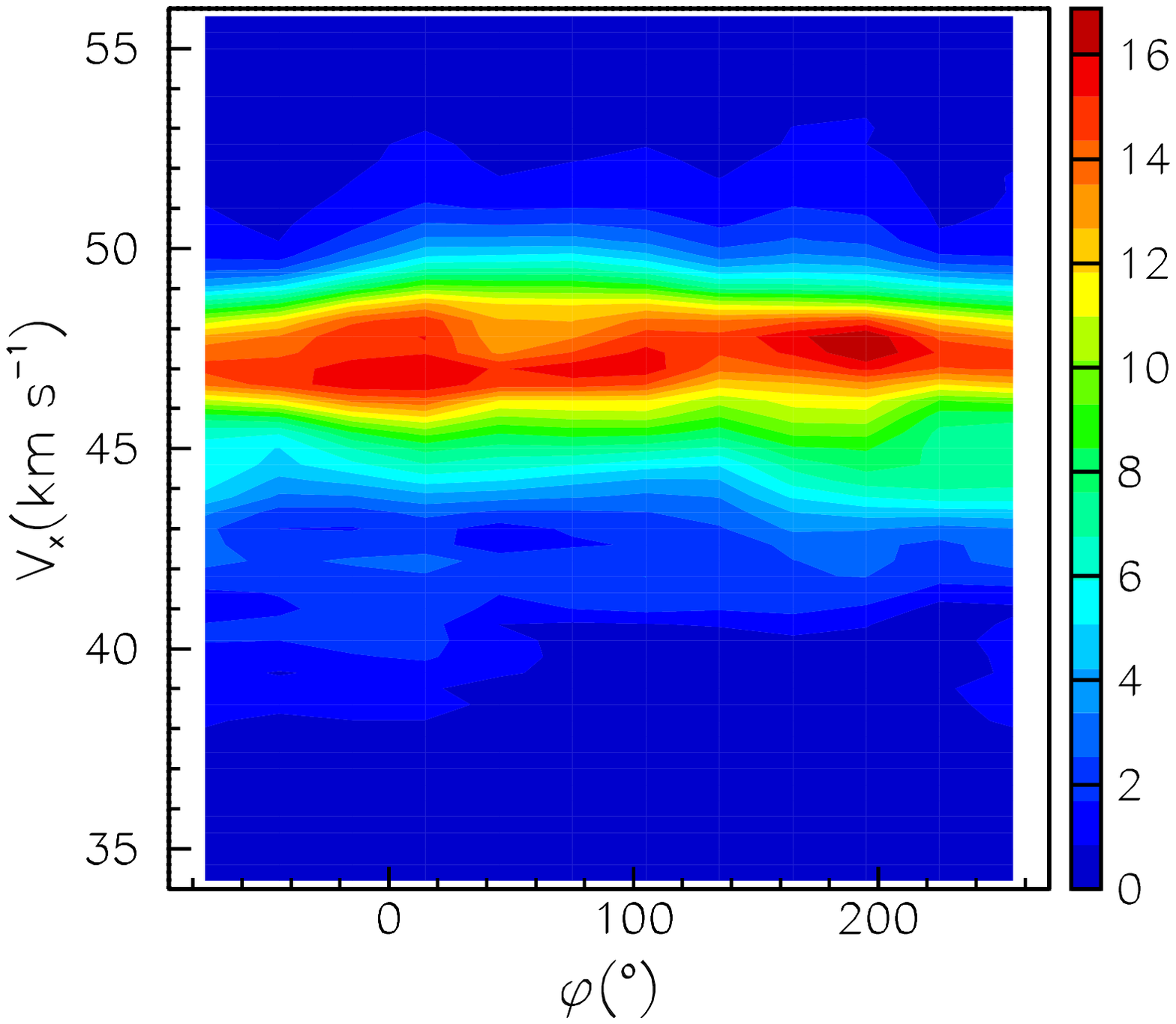}
\includegraphics[height=7cm,trim=0.cm 0.cm 0.cm 0.cm,clip]{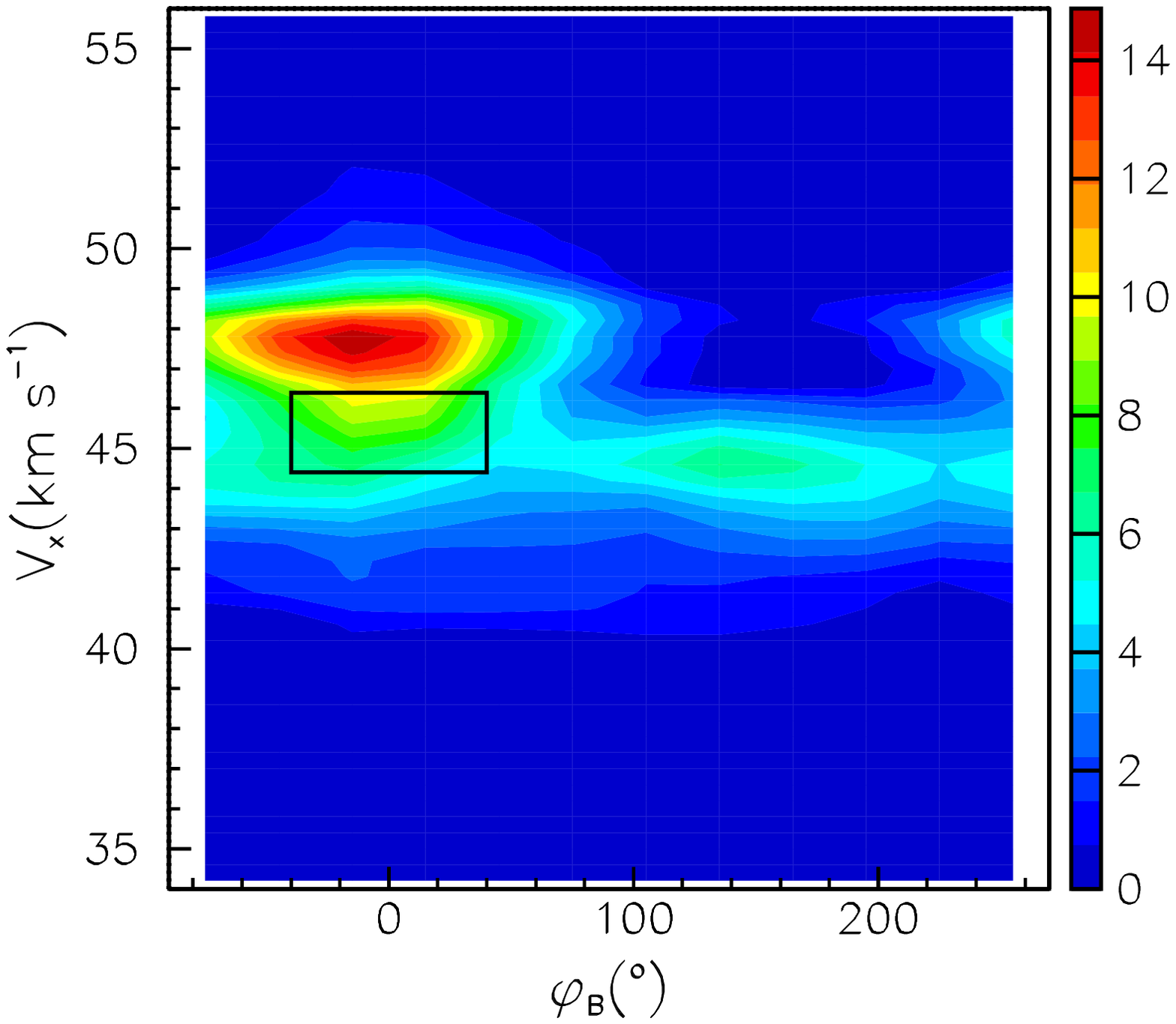}
\caption{Position-velocity diagrams centred on Mira A (left) and on Mira B (right). The abscissas are the position 
angles $\varphi$ and $\varphi_B$ measured clockwise from west around Mira A and B respectively. The Mira A diagram is 
for radii contained in the interval [0.1$''$,0.4$''$] and the Mira B diagram for radii contained in the interval [0.05$''$,0.30$''$].
The colour scale is in the unit of Jy arcsec$^{-2}$. 
The rectangle in the right panel (44.4<$V_x$<46.4 km\,s$^{-1}$ and $-$40$^\circ$<$\varphi_B$<40$^\circ$) corresponds to flux densities mapped 
in the left panel of Fig.~\ref{fig14}.}
\label{fig11}
\end{figure*}

Near Mira B (Fig.~\ref{fig10} middle) the spectrum displays two peaks, one peaking at 47.9 km\,s$^{-1}$ and 
the other at 44.6 km\,s$^{-1}$. The relative amplitude of the two peaks is not strongly dependent on the size 
of the circle over which the flux density is integrated around Mira B. This is also illustrated in 
Fig.~\ref{fig11}, which displays position-velocity diagrams confined to the immediate environment of 
Mira A and B respectively. In the present work, we call position-velocity (P-V) diagram a distribution 
in a plane spanned by the Doppler velocity and one of the sky coordinates, Cartesian or polar or else, 
of the flux density averaged over the other sky coordinate. The emission centred on Mira B covers Doppler 
velocities from $\sim$41 to $\sim$46 km\,s$^{-1}$ (here, again, the contribution from the bubble is small, 
even if relatively larger than in the preceding Mira A case). Moreover, it is spatially resolved and extends 
over several tenths of an arcsecond (the beam has a $\sigma$ of $\sim$0.13$''$).

The extension found by \citet{Vlemmings2015} in the continuum is only 0.026$''$ FWHM, significantly smaller 
than what is seen here. It is worth recalling on this occasion that such dimensions are much larger than what 
is expected from the accretion disk, which is far from being detectable. The emission detected in the present 
study, is from molecular carbon monoxide, implying temperatures at the 100 K scale, and a low density of short 
wavelength radiations that dissociate molecules. The emission detected in the continuum by \citet{Vlemmings2015} 
at $\sim$95 and $\sim$229 GHz is from partially ionized, gravitationally bound gas, namely much closer to Mira B 
than in the present case (and still farther away than the accretion disk).
  
The difference between the fluxes integrated in circles of 0.3$''$ radius around Mira B and its symmetric with 
respect to Mira A ($y$=$-$0.5$''$, $z$=0.07$''$) is displayed in the right panel of Fig.~\ref{fig10}. If the peak 
at 44.5 km\,s$^{-1}$ seen in the middle panel were due exclusively to gas gravitationally bound to Mira B, the 
difference spectrum would also peak at 44.5 km\,s$^{-1}$ above a zero baseline and reveal the emission associated 
with the bound gas. Instead, it peaks at 43.6 km\,s$^{-1}$, 1 km\,s$^{-1}$ lower, and becomes negative between 
$\sim$45.5 and 50 km\,s$^{-1}$. This implies that rather than seeing gas gravitationally bound to Mira B, we 
rather see an asymmetry of the gas flowing from Mira A, with an enhancement in the direction of Mira B, namely 
eastward and toward the Earth.

\begin{figure*}
\centering
\includegraphics[height=11cm,trim=0.cm 0.cm 0.cm 0.cm,clip]{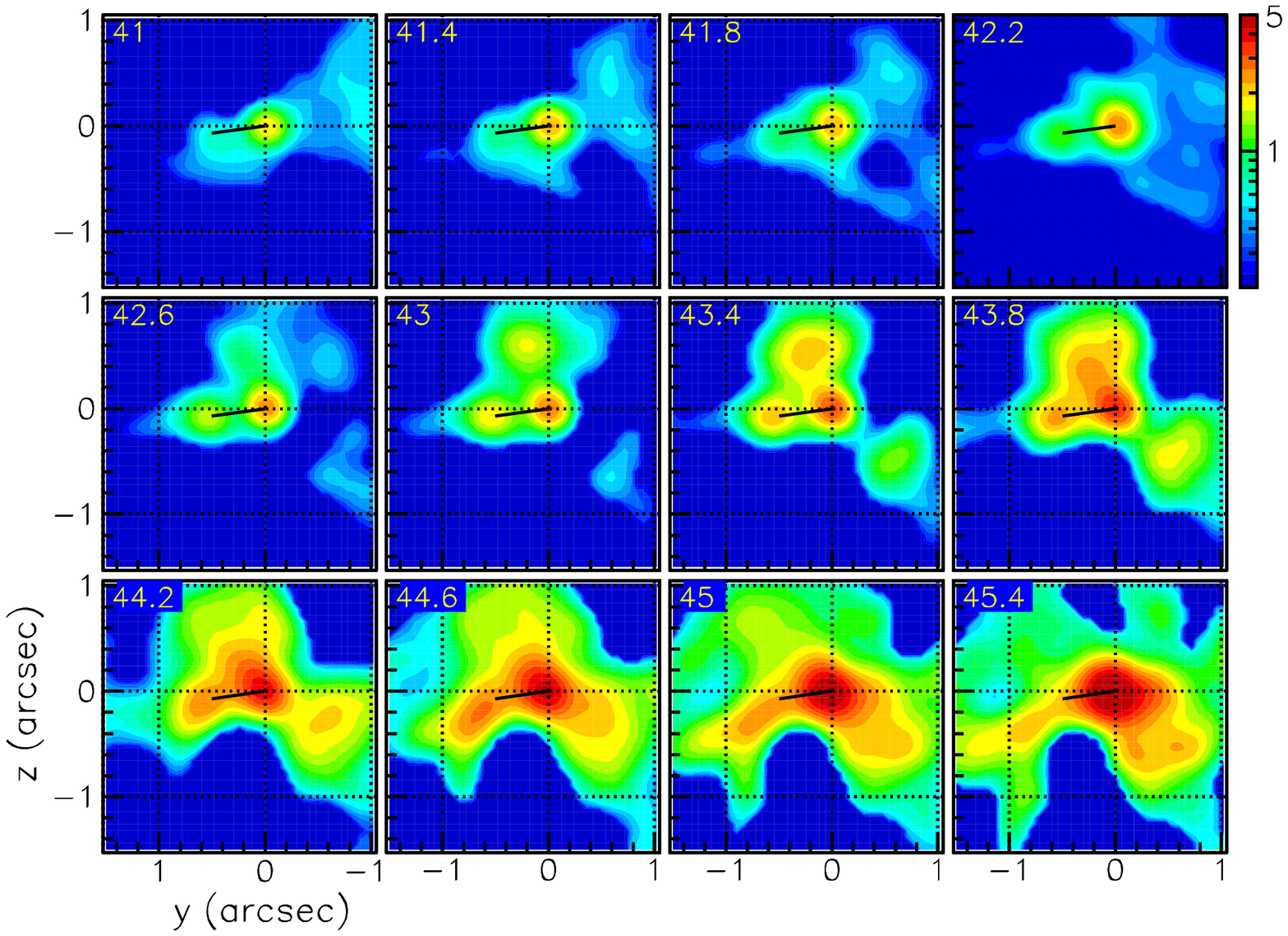}
\caption{Channel maps starting at 40.8 km\,s$^{-1}$ and ending at 45.6 km\,s$^{-1}$ in successive 0.4 km\,s$^{-1}$ wide channels.
 Each panel is labelled with the central velocity (km\,s$^{-1}$). The colour scale is in the unit of Jy\,km\,s$^{-1}$ arcsec$^{-2}$.
Black lines show the relative positions of Mira A and Mira B.}
\label{fig12}
\end{figure*}

A close look at the immediate environment of Mira A+B between 41 and 46 km\,s$^{-1}$ is given by the channel 
maps displayed in Fig.~\ref{fig12}. Up to 44 km\,s$^{-1}$, the emission is approximately centred on Mira B 
but for higher Doppler velocities it evolves into a south-eastern arm that may be interpreted as an outflow 
having its source at Mira A and flowing toward negative $x$ values, being eventually focused by Mira B. At 
the same time, two other outflows seem to originate from Mira A, one aiming south-west and the other aiming 
some 20$^\circ$ east of north. They will be discussed in the following sections.

\begin{figure*}
\centering
\includegraphics[height=11cm,trim=0.cm 0.cm 0.cm 0.cm,clip]{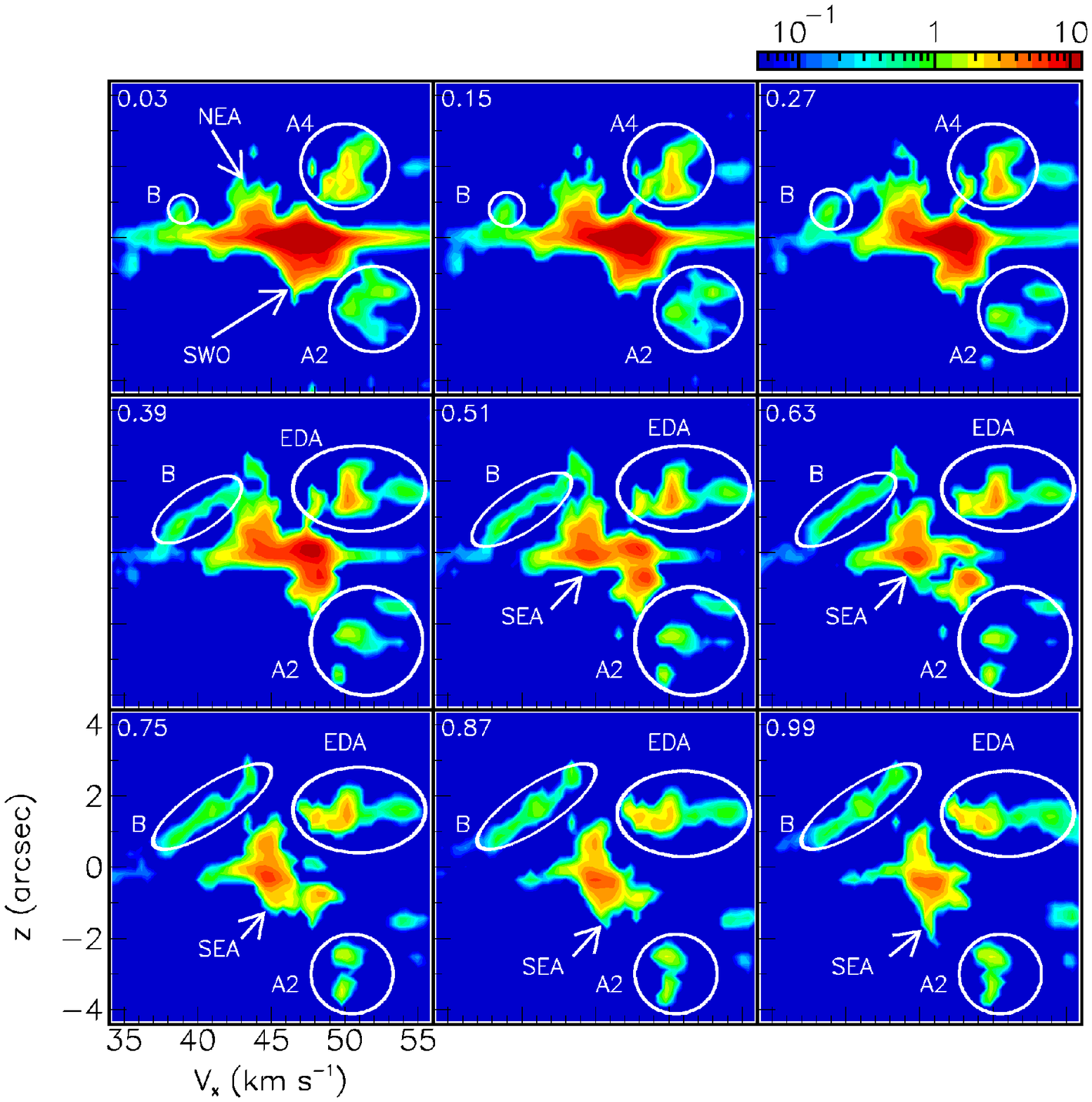}
\caption{P-V diagrams over adjacent $y$ strips, each two pixels wide (0.12$''$), $z$ being used as space coordinate. 
Each panel is labelled with the central value of $y$. The contributions of the bubble (B) and of red-shifted arcs 
(A2, A4, see Fig.~\ref{fig9}, and Eastern Detached Arm, see section 6.2 and Fig.~\ref{fig20}) are indicated. Also shown, on the first panel, 
are the contributions of the north-eastern arm (NEA) and of the south-western outflow (SWO). Starting from panel 5, 
yellow arrows point to the developing south-eastern arm (SEA).}
\label{fig13}
\end{figure*}

Another illustration is given in Fig.~\ref{fig13}, which displays P-V diagrams over adjacent $y$ strips, 
each two pixels wide (0.12$''$), $z$ being used as space coordinate. In addition to easily identifiable 
contributions from the bubble and from red-shifted arcs, they show the south-eastern arm and the north-eastern 
arm just mentioned. The north-eastern arm covers Doppler velocities between $\sim$42.5 and $\sim$45.5 km\,s$^{-1}$ 
in the north-eastern quadrant of the sky map, with both $y$ and $z$ reaching to $\sim$1$''$ from Mira A. 
On average, it is blue-shifted by 3.5 km\,s$^{-1}$ with respect to Mira A; if assigned to a wind having 
an expansion velocity of $\sim$7 km\,s$^{-1}$, this would imply an angle of $\sim$60$^\circ$ with the line of sight.

In the southern hemisphere, the P-V diagrams having $y$<0.5$''$ are dominated by the south-western outflow; 
however, from Mira B onward, one sees the south-eastern arm starting to develop, covering Doppler velocities 
between $\sim$43.5 and $\sim$46.5 km\,s$^{-1}$. At the same time, as $y$ increases, the contribution of the 
south-western outflow fades away. The projected velocity on the plane of the sky of Mira B with respect to 
Mira A is $\sim$3 km\,s$^{-1}$. The mean Doppler velocity of the south-eastern arm is $\sim$45 km\,s$^{-1}$, 
blue-shifted by 2.5 km\,s$^{-1}$ with respect to Mira A; again, if assigned to a wind having an expansion 
velocity of $\sim$7 km\,s$^{-1}$, this would imply an angle of $\sim$70$^\circ$ with the line of sight and 
a velocity of $\sim$6 km\,s$^{-1}$ in the plane of the sky. If the interpretation of the south-eastern arm 
as an outflow having its source at Mira A and being focused by Mira B is correct, the combined movements 
of the expanding gas and of the orbiting Mira B would generate a small southward curvature of the south-eastern 
arm compatible with what is observed.

In summary, the observation between Doppler velocities of 42 and 44 km\,s$^{-1}$ of an enhanced emission at 
the precise location of Mira B (Fig.~\ref{fig12}) gives evidence for the white dwarf to play an important 
role in shaping the morphology and kinematics of the gas flow in its environment. The evolution of the P-V 
diagrams displayed in Fig.~\ref{fig13} east of Mira B provides another illustration adding to this evidence. 
At first sight, one might think that we are observing a mass of gas surrounding Mira B, with a size of a few 
tens of AU, and having Doppler velocities with respect to Mira B reaching 1.5 km\,s$^{-1}$ in each of the 
red and blue directions. In this case, the velocity dispersion would be the result of rotation or/and accretion 
or/and thermal agitation. The \mbox{P-V} diagram of Fig.~\ref{fig11} (right) could reveal a rotation only if its 
axis were significantly inclined with respect to the line of sight and a bipolar flow only if its axis were 
significantly inclined with respect to the plane of the sky. It does not show any clear sign of such effects 
but better spatial and spectral resolutions would be necessary to place sensible limits on the corresponding 
velocities. Moreover, if we were observing gas gravitationally bound to Mira B, one would need to assume that 
the orbit of \citet{Prieur2002} predicts a Doppler shift between Mira A and B three times smaller than what it really is.

It seems therefore more plausible to assume that we are seeing gas flowing from Mira A toward Mira B, being 
eventually focused or even partly trapped by Mira B. 
More precisely, the gas flow emitted by Mira A in a solid angle facing Mira B, therefore blue-shifted, concentrates toward Mira B very much as predicted by Model 1 of \citet{Mohamed2012} in the framework of their Wind Roche Lobe Overflow (WRLOF) picture.
The velocity distribution displayed in the right panel of Fig.~\ref{fig10} gives strong support to this 
interpretation. While it is sensible to assume that the gas ejected from Mira A, and attracted and eventually 
trapped by Mira B, will ultimately be accreted, none of this can be seen from the present data (and is not 
expected to). The evidence for the gas flow to be enhanced in the neighbourhood of Mira B does not exclude 
a possible small contribution of gravitationally bound gas, but the present observations prevent to quantify 
it reliably. A hydrodynamic model of the gravitational attraction and focalization produced by Mira B would 
help a better understanding of the dynamics at play.

\begin{figure*}
\centering
\includegraphics[height=6.cm,trim=1.cm 0.cm 0.cm 0.cm,clip]{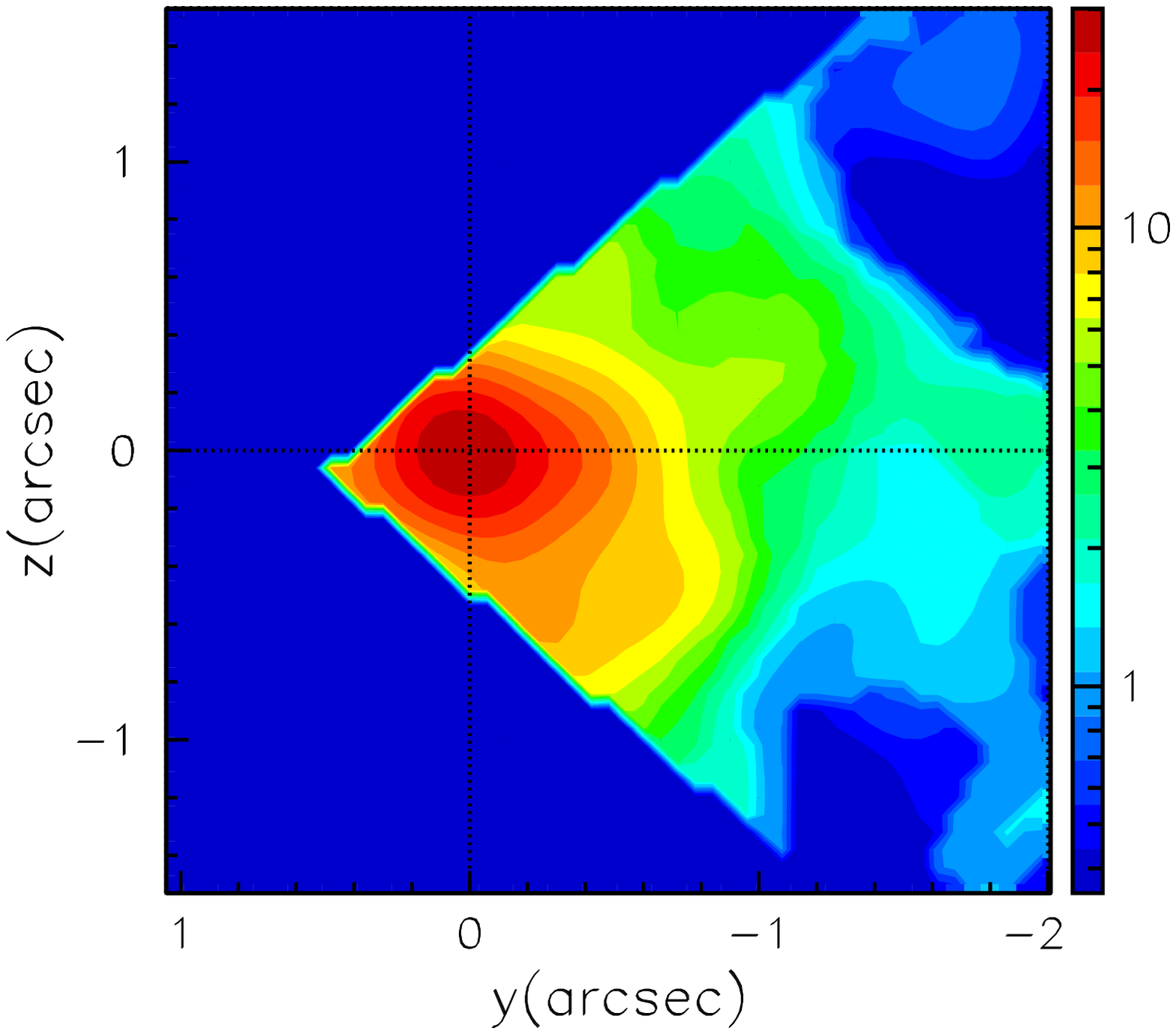}
\includegraphics[height=6.cm,trim=.5cm 0.cm 1.5cm 0.cm,clip]{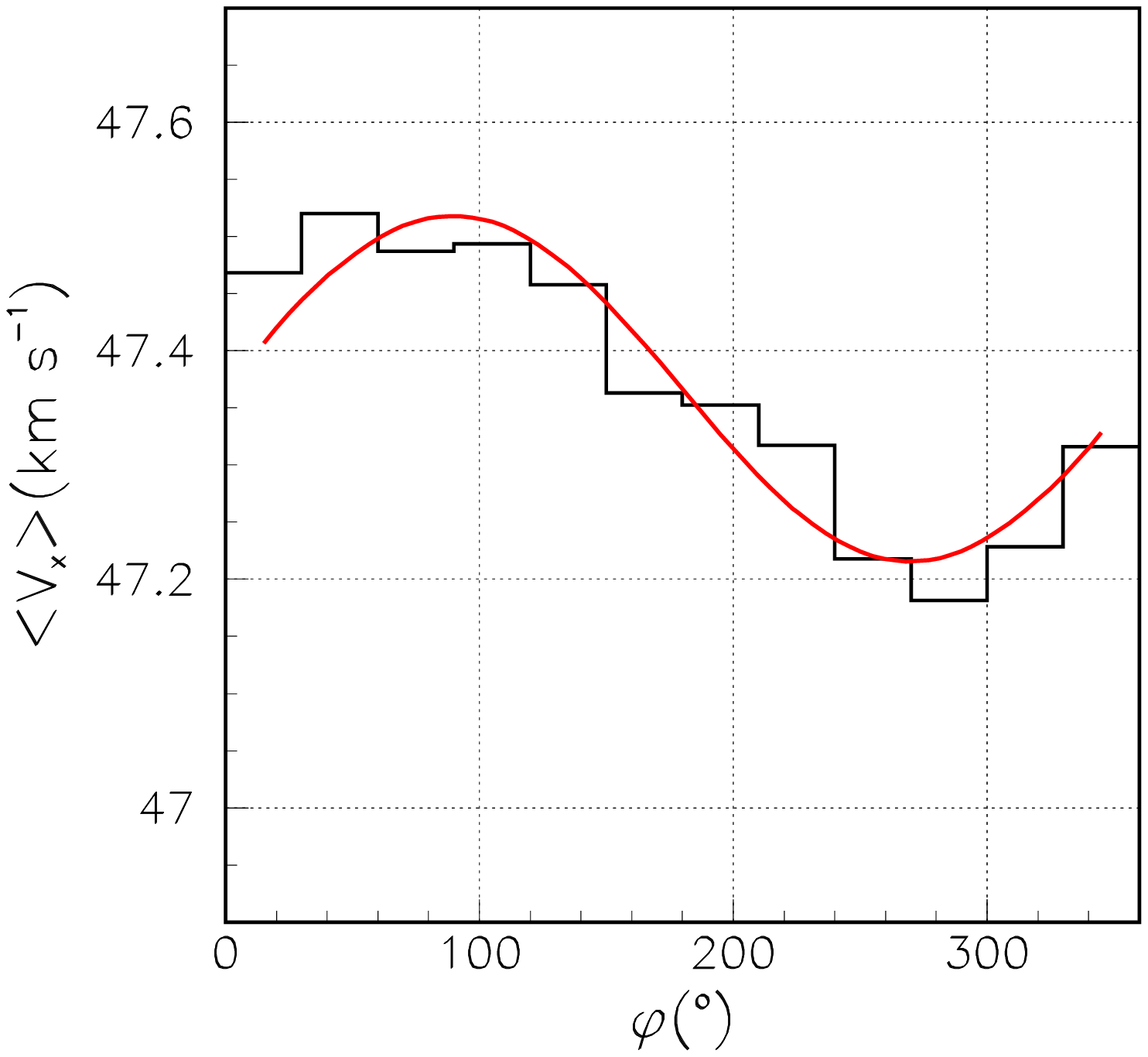}
\includegraphics[height=6.cm,trim=.5cm 0.cm 1.5cm 0.cm,clip]{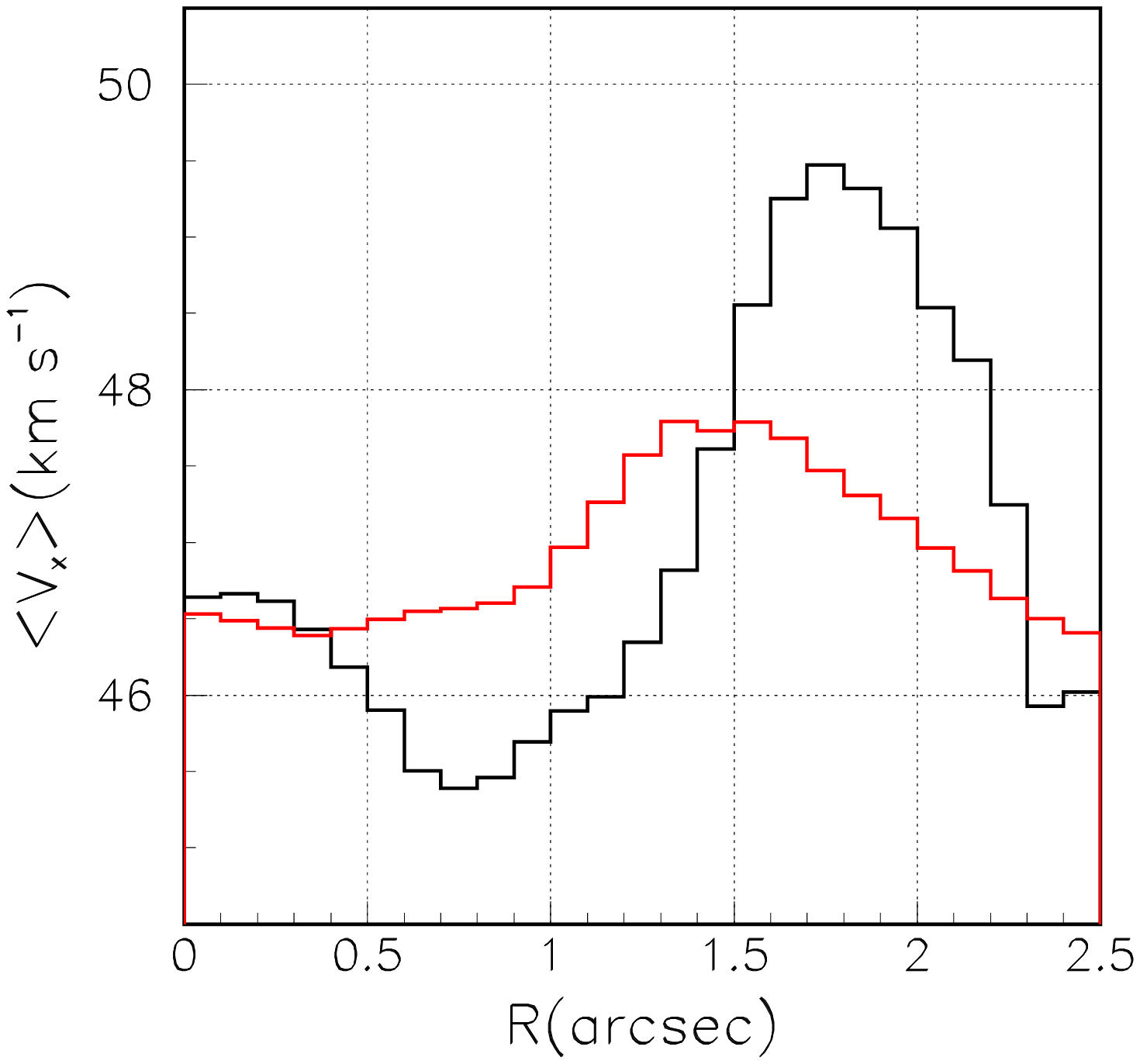}
\caption{Left: Sky map of the flux densities contained inside the rectangle drawn in the right panel of Fig.~\ref{fig11}.
 Middle: Dependence on position angle measured with respect to Mira A of the Doppler velocity averaged in a circle 
centred on Mira A for distances between 0.1$''$ and 0.6$''$ and 45<$V_x$<50 km\,s$^{-1}$. The sine wave fit corresponds 
to a bipolar outflow of $\pm$0.15 km\,s$^{-1}$ amplitude pointing south, with a mean value of 47.4 km\,s$^{-1}$. 
Right: Distribution of <$V_x$> on $R$, averaged over east (black) and west (red) position angles separately.}
\label{fig14}
\end{figure*}

Finally, Fig.~\ref{fig14} provides additional information related to the P-V diagrams of Fig.~\ref{fig11}. 
The left panel maps the flux in the rectangle displayed in the right panel of Fig.~\ref{fig11}, at Doppler 
velocities intermediate between the Mira A and Mira B maxima, showing that emission is dominated by Mira A 
but slightly shifted toward Mira B. The middle panel displays the dependence on position angle measured with 
respect to Mira A of the Doppler velocity averaged in a circle centred on Mira A for distances between 
0.1$''$ and 0.6$''$ and 45<$V_x$<50 km\,s$^{-1}$. The sine wave fit corresponds to a bipolar outflow of 
$\pm$0.15 km\,s$^{-1}$ amplitude pointing north and south, with a mean value of 47.4 km\,s$^{-1}$. When 
retaining only pixels closer to the star, the amplitude of the oscillation decreases but the mean velocity 
and the phase shift are unaffected. The right panel displays the dependence on $R$ of the mean Doppler 
velocity <$V_x$> averaged over position angle in the eastern hemisphere (toward Mira B) and the western 
hemisphere separately. While starting together at $\sim$47 km\,s$^{-1}$ they quickly depart from each other: 
the eastern part is first blue-shifted with respect to the western part but, beyond $\sim$1.5$''$, the reverse applies.

\subsection{An overview at larger distances}

\begin{figure*}
\centering
\includegraphics[height=6.cm,trim=1.cm 0.cm 0.cm 0.cm,clip]{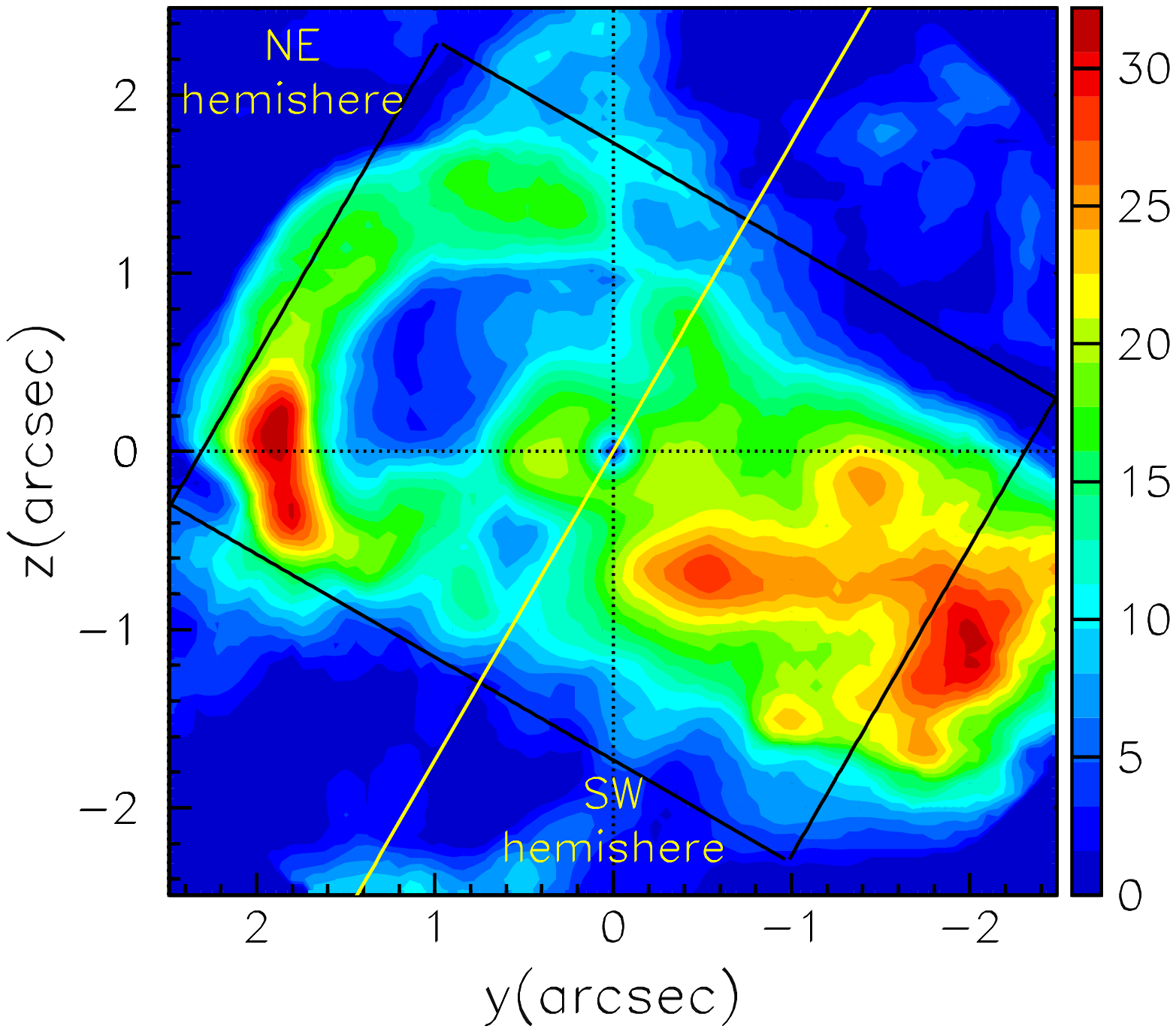}
\includegraphics[height=6.cm,trim=.5cm 0.cm 1.5cm 0.cm,clip]{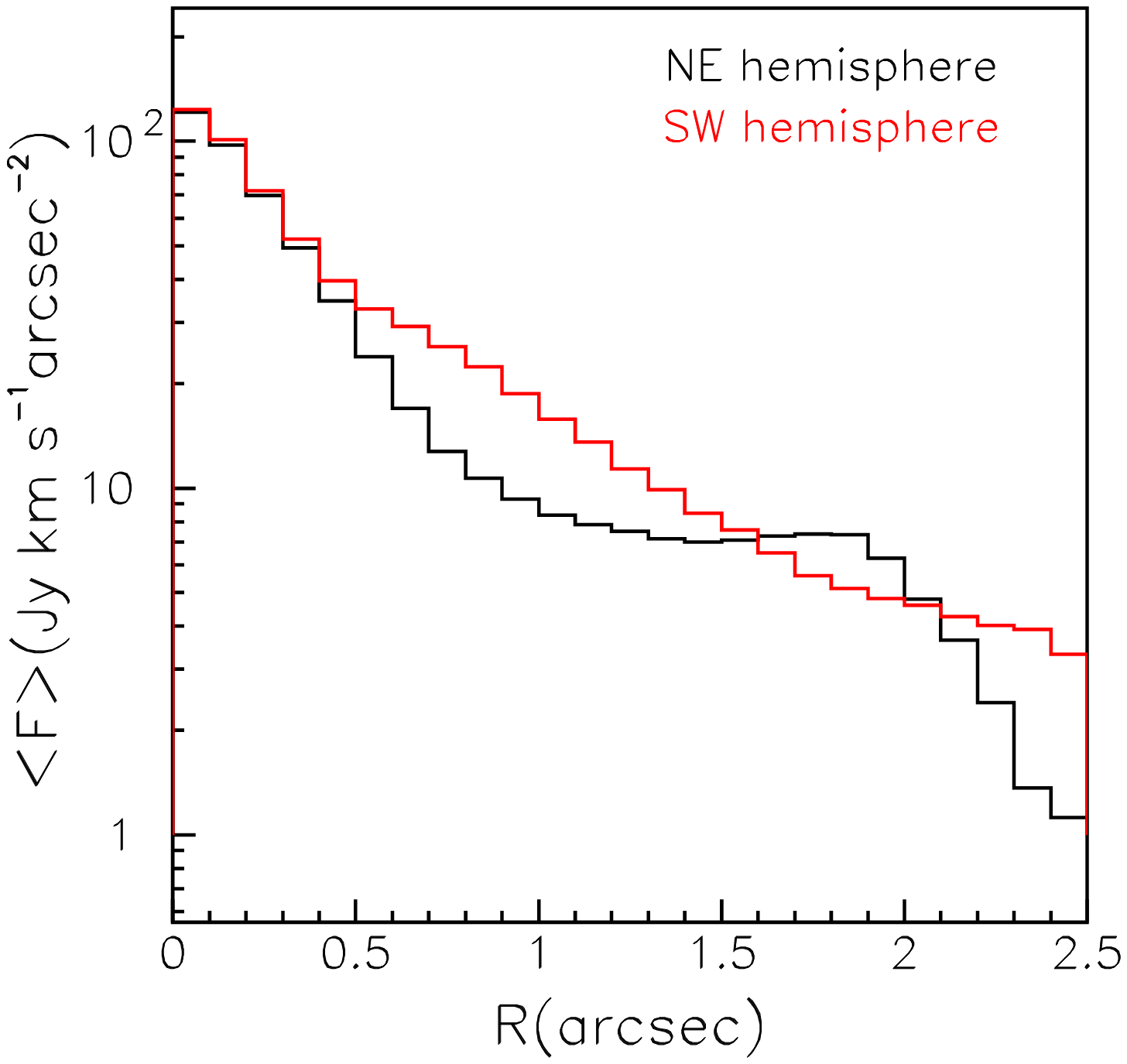}
\includegraphics[height=6.cm,trim=.5cm 0.cm 1.5cm 0.cm,clip]{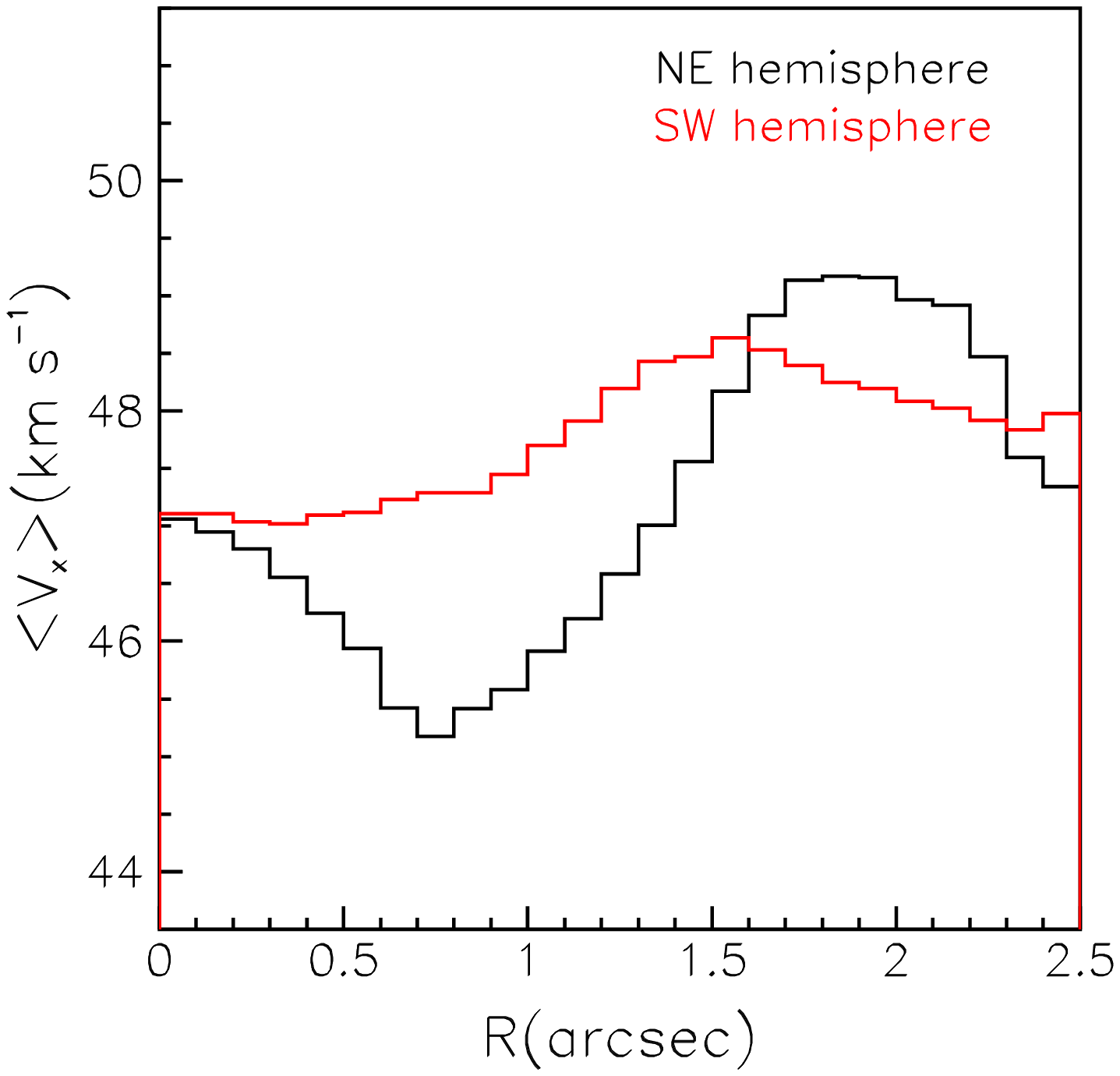}
\caption{Left: sky map of the product $F \times R$ integrated over Doppler velocities in the range $R$<3$''$ and 42<$V_x$<53.2 km\,s$^{-1}$. 
The rectangle is the region within which rotation is being looked for in the right panel of Fig.~\ref{fig16}. Middle: radial distribution 
of $F$ averaged over position angle in the same range, separately in the north-eastern (black) and south-western (red) hemispheres 
(defined with respect to an axis $y'$ pointing 30$^\circ$ east of south, see left panel). Right: radial distribution of <$V_x$> averaged 
over position angle in the same range and with the same convention. }
\label{fig15}
\end{figure*}
 
\begin{figure*}
\centering
\includegraphics[height=6.cm,trim=1.cm 0.cm 1.cm 0.cm,clip]{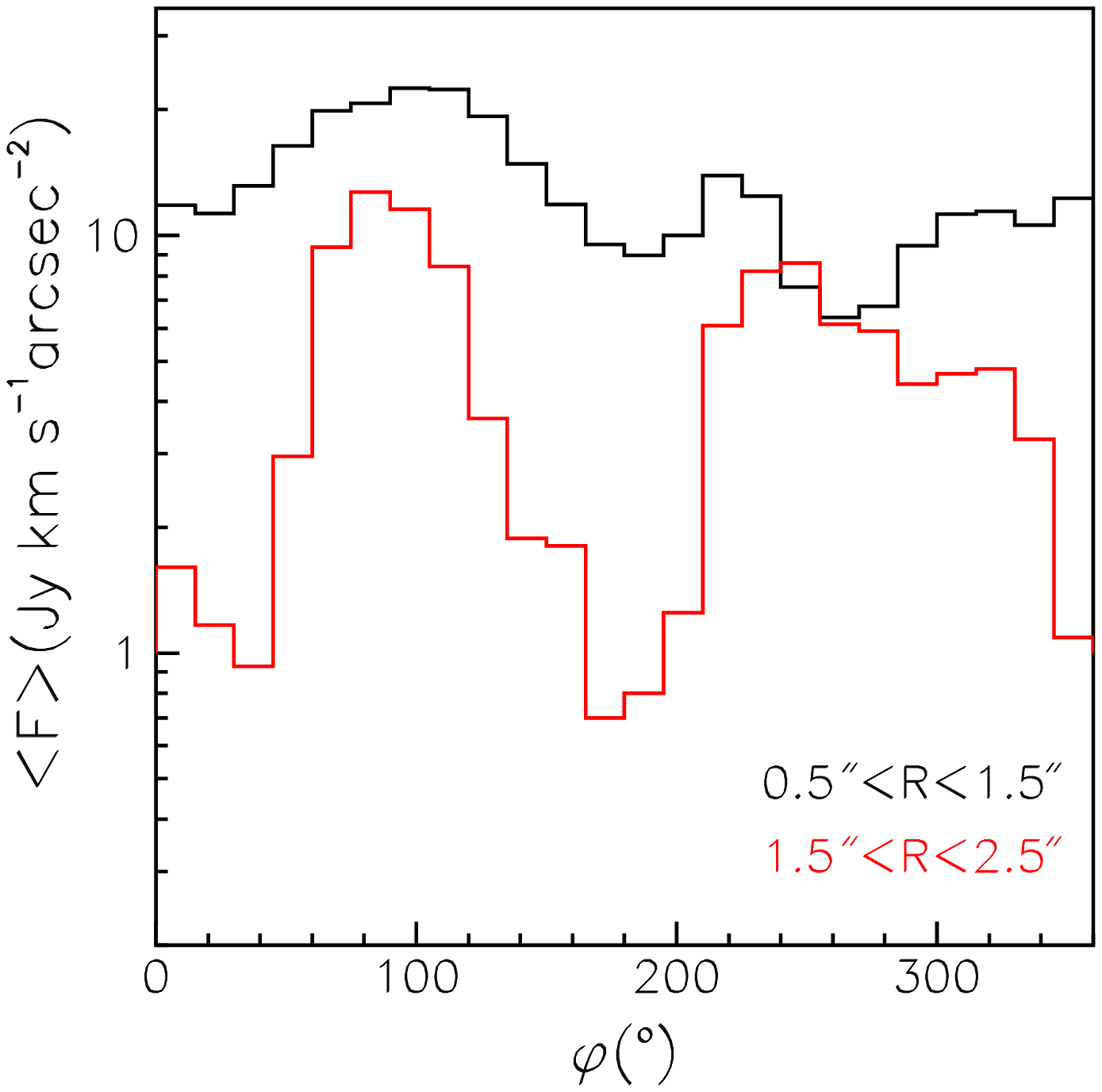}
\includegraphics[height=6.cm,trim=.5cm 0.cm 1.cm 0.cm,clip]{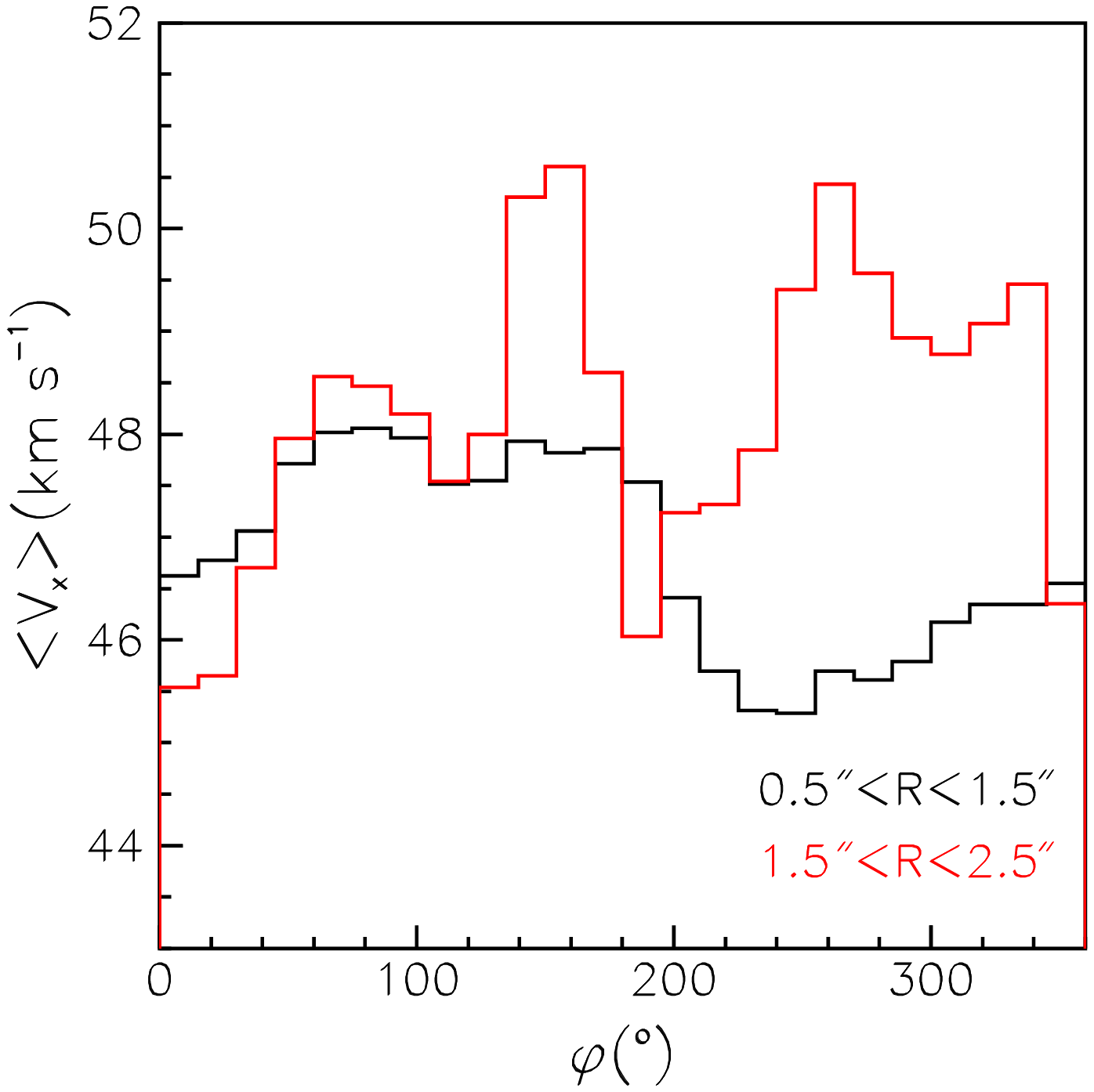}
\includegraphics[height=6.cm,trim=.5cm 0.cm 1.5cm 0.cm,clip]{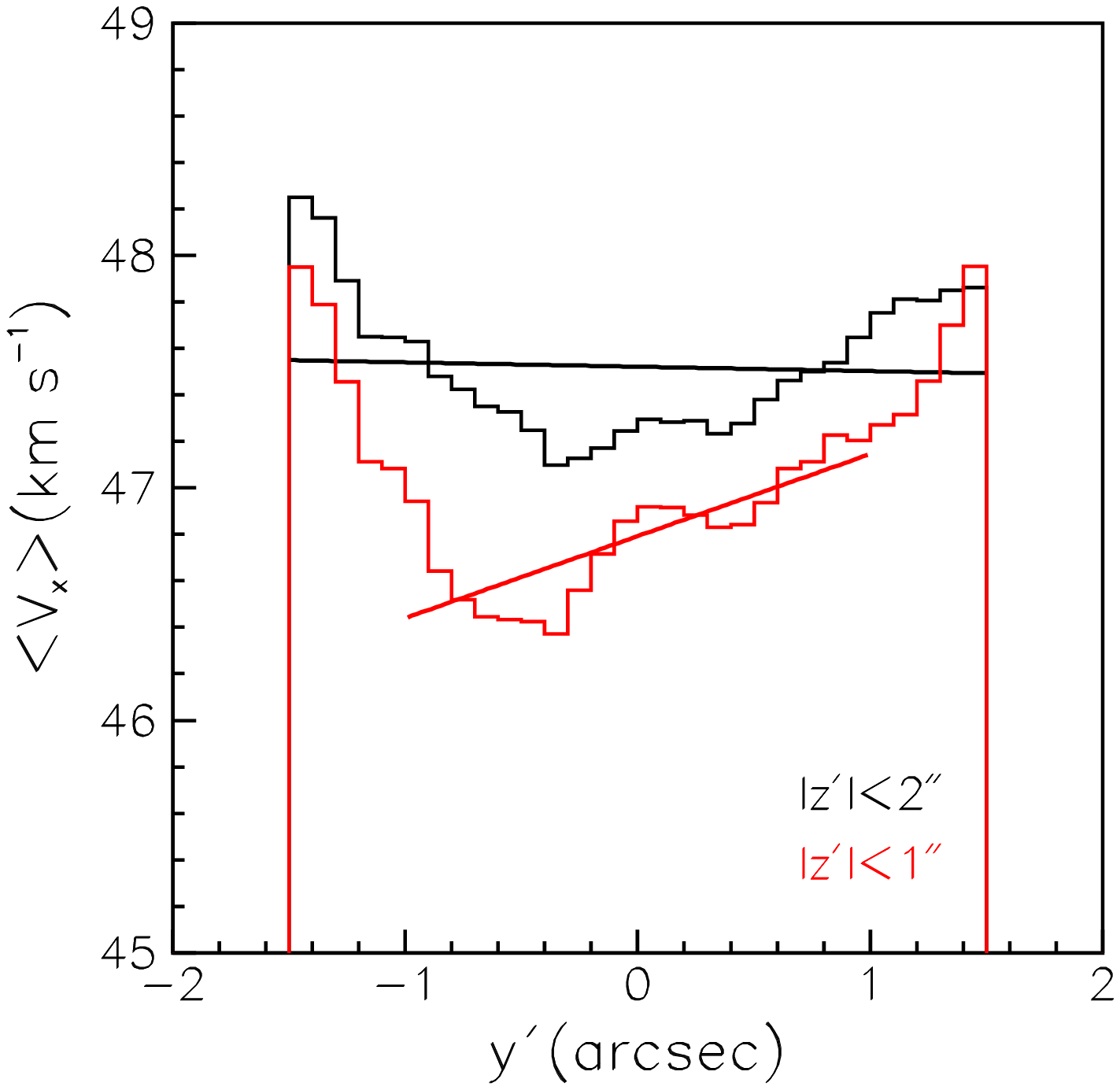}
\caption{Left: distribution on position angle (measured clockwise from west) of the flux integrated over 42<$V_x$<53.2 km\,s$^{-1}$ 
and averaged over 0.5$''$<$R$<1.5$''$ (black) or 1.5$''$<$R$<2.5$''$ (red); Middle: distribution of <$V_x$> averaged over $R$ in the same 
ranges and with the same convention. The position angle is measured clockwise from 60$^\circ$ north of west. Right: distribution of 
the mean Doppler velocity as a function of $y'=-z\cos(30^\circ)+y\sin(30^\circ)$ averaged over $|z'|$<2$''$ (black, rectangle shown in the 
left panel of Fig. \ref{fig15}) or $|z'|$<1$''$ (red) with $z'=y\cos(30^\circ)+z\sin(30^\circ)$. The lines are linear fits to $y'$ intervals 
$|y'|$<1.5$''$ (black) or $|y'|$<1$''$ (red) respectively.}
\label{fig16}
\end{figure*}

When looking on a larger scale (Fig.~\ref{fig15} left), in a radius of 3$''$ around Mira A and excluding the blue-shifted 
and red-shifted ends of the Doppler velocity spectrum, emission is enhanced along a direction pointing some 30$^\circ$ south 
of west over a broad range of position angles. We refer to this enhancement as the south-western outflow. Facing it, pointing 
to some 30$^\circ$ north of east (defined as $z$' axis), another enhancement displays a strong central depression, reaching up 
to $\sim$1.5$''$ projected distance from Mira A. We refer to it as the north-eastern outflow. When averaging over position 
angles, the integrated flux decreases smoothly with radius in the south-western outflow but is steeper at short distances 
and strongly enhanced between 1.5$''$ and 2$''$ in the north-eastern outflow (Fig.~\ref{fig15} middle). Similarly, the 
mean Doppler velocity remains confined between 47 and 48 km\,s$^{-1}$ in the south-western outflow but is strongly blue-shifted 
at short distances, around 1$''$, and strongly red-shifted at large distances, around 2$''$ in the north-eastern outflow 
(Fig.~\ref{fig15} right). When averaging instead over $R$ the position angle distribution displays a south-west to north-east 
ratio in excess of 2 for 0.5$''$<$R$<1.5$''$, but is better equilibrated between south-west and north-east for 1.5$''$<$R$<2.5$''$ 
(Fig.~\ref{fig16} left). Similarly, the mean Doppler velocity displays a strong north-east to south-west asymmetry for 
0.5$''$<$R$<1.5$''$, decreasing abruptly from $\sim$48 km\,s$^{-1}$ west to $\sim$46 km\,s$^{-1}$ east, but is better equilibrated 
between south-west and north-east for 1.5$''$<$R$<2.5$''$ (Fig.~\ref{fig16} middle).

In order to reveal a possible rotation about the approximate axis of symmetry $z'$, we display in Fig.~\ref{fig16} 
(right) the distribution on the perpendicular axis, $y'$, of the mean Doppler velocity averaged over $-2''$<$z'$<$2''$ or  $-1''$<$z'$<$1''$ 
and 42<$V_x$<53.2 km\,s$^{-1}$. In both $z'$ intervals we find no clear evidence for rotation. However, in the central region 
($|y'|$<1$''$,$|z'|$<1$''$) the data can accommodate a rotation of $\sim \pm$0.5 km\,s$^{-1}$ arcsec$^{-1}$.

\begin{figure}
\centering
\includegraphics[height=7cm,trim=.5cm 0.cm 0.cm 0.cm,clip]{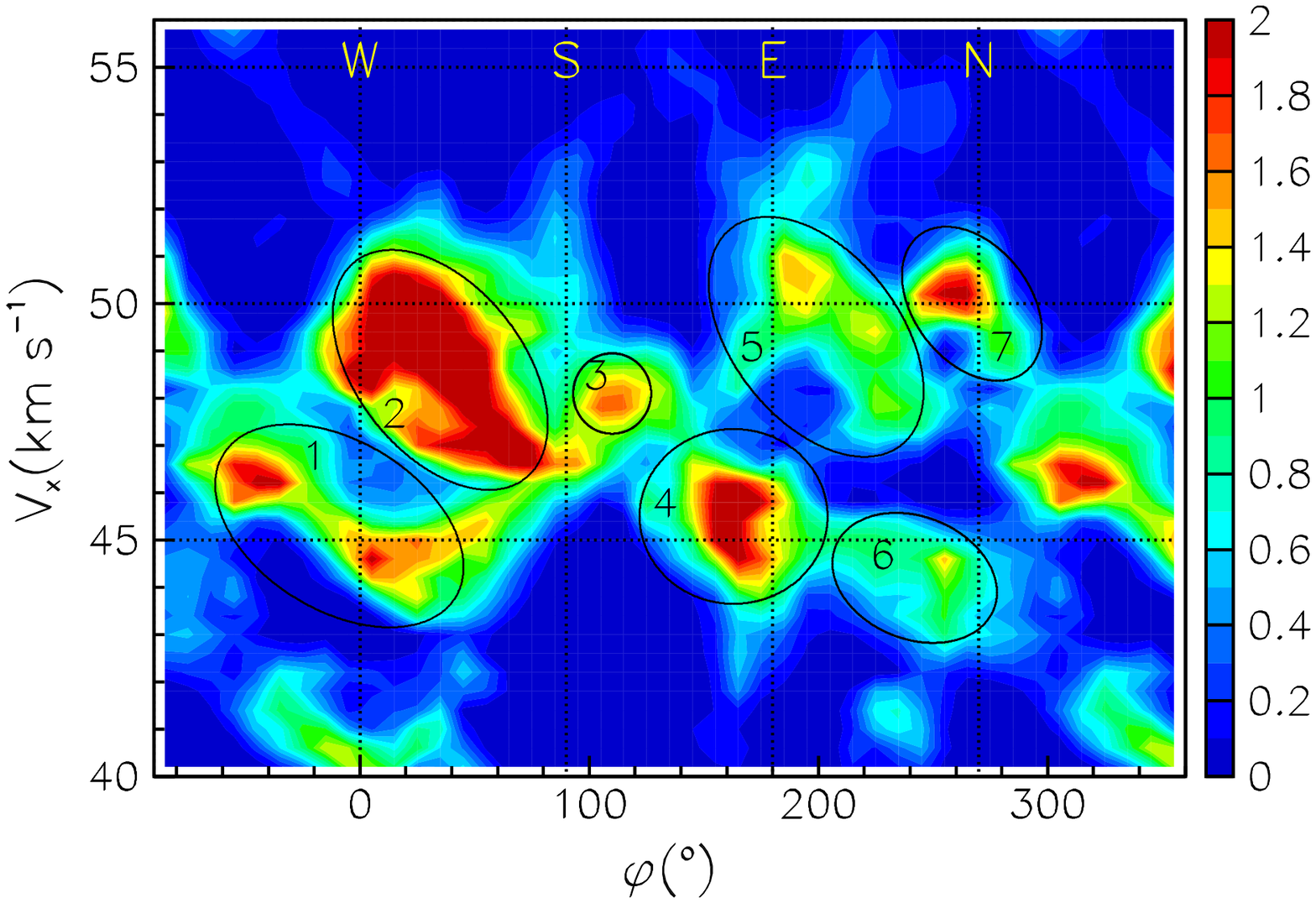}
\caption{Position-velocity diagram in a circle centred on Mira A for distances to Mira A between 0.5$''$ and 2.5$''$.  
The ellipses delineate area associated with the sky maps displayed in Fig. \ref{fig18}.}
\label{fig17}
\end{figure}

Figure~\ref{fig17} displays a P-V diagram averaged over 0.5$''$<$R$<2.5$''$ using the position angle $\varphi$ as space 
coordinate. The western outflow covers from $\sim$60$^\circ$ north of west to $\sim$30$^\circ$ east of south, where its 
limbs are seen to share similar Doppler velocities around 47 km\,s$^{-1}$, and has an axis close to the plane of the sky, 
pointing $\sim$30$^\circ$ south of west, where Doppler velocities span from $\sim$43 to $\sim$52 km\,s$^{-1}$, meaning a 
radial expansion velocity of at least 5 km\,s$^{-1}$. The eastern outflow, displays an atypical pattern with two nearly 
separated arms, one blue-shifted and pointing between east and south-east, the other red-shifted covering the north-eastern quadrant.

\begin{figure*}
\centering
\includegraphics[height=5.5cm,trim=1.cm 0.cm 0.cm 0.cm,clip]{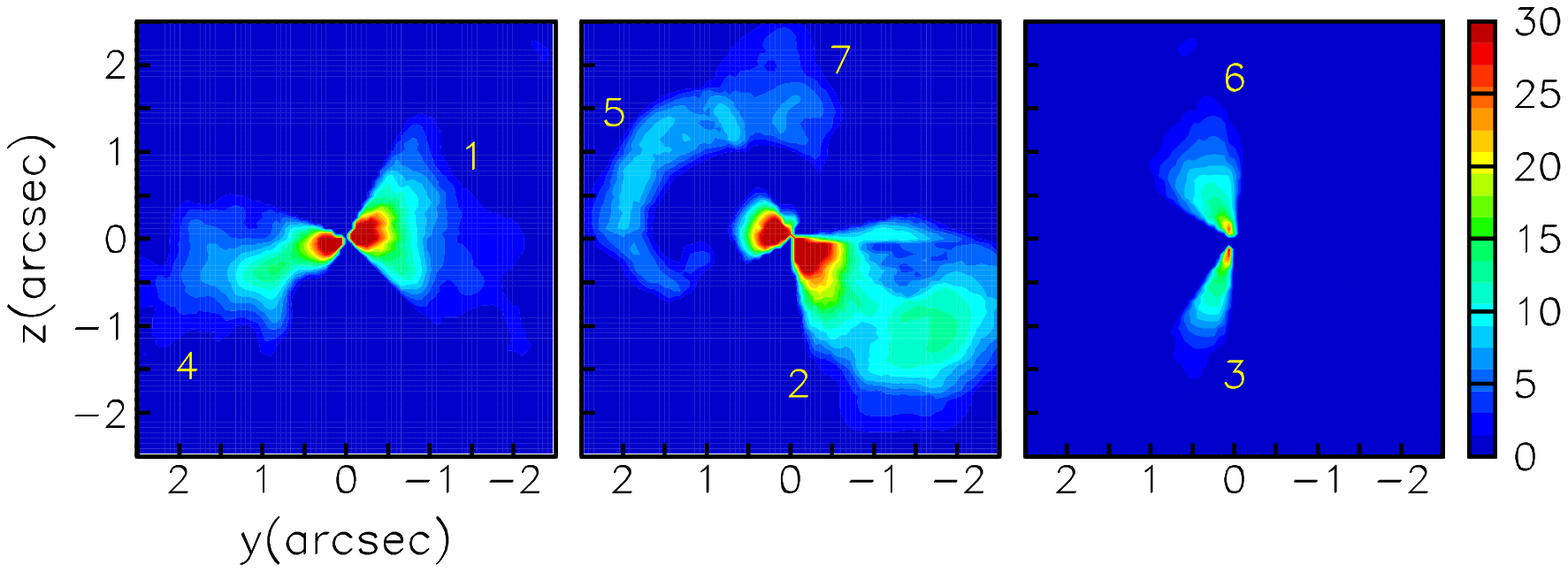}
\caption{Maps of F (Jy\,km\,s$^{-1}$ arcsec$^{-2}$) for the regions of the P-V diagram delineated by ellipses in Fig. \ref{fig17}.}
\label{fig18}
\end{figure*}

Sky maps associated with distinct regions of emission identified in the P-V diagram of Fig.~\ref{fig17} are 
displayed in Fig.~\ref{fig18}. Regions 1 and 2 cover the south-western outflow but region 2 extends about 
twice as far from Mira A than region 1 does, emission being enhanced at larger distances from the star. While 
region 4 faces region 1 and displays a similar morphology, regions 5 and 7 face region 2 but populate only large 
distances from the star, in excess of $\sim$1.5$''$, at variance with region 2. The Doppler velocity distribution 
in the part of region 2 having $R$>1.3$''$ is precisely the same as in the part having $R$<1.3$''$. Region 4 
corresponds to the south-eastern arm and region 6 to the north-eastern arm that were identified in Fig.~\ref{fig12}, 
the former flowing across Mira B and the latter connecting to Mira A at Doppler velocities of the order of 44 km\,s$^{-1}$.

\begin{figure*}
\centering
\includegraphics[height=5.5cm,trim=1.cm 0.cm 0.cm 0.cm,clip]{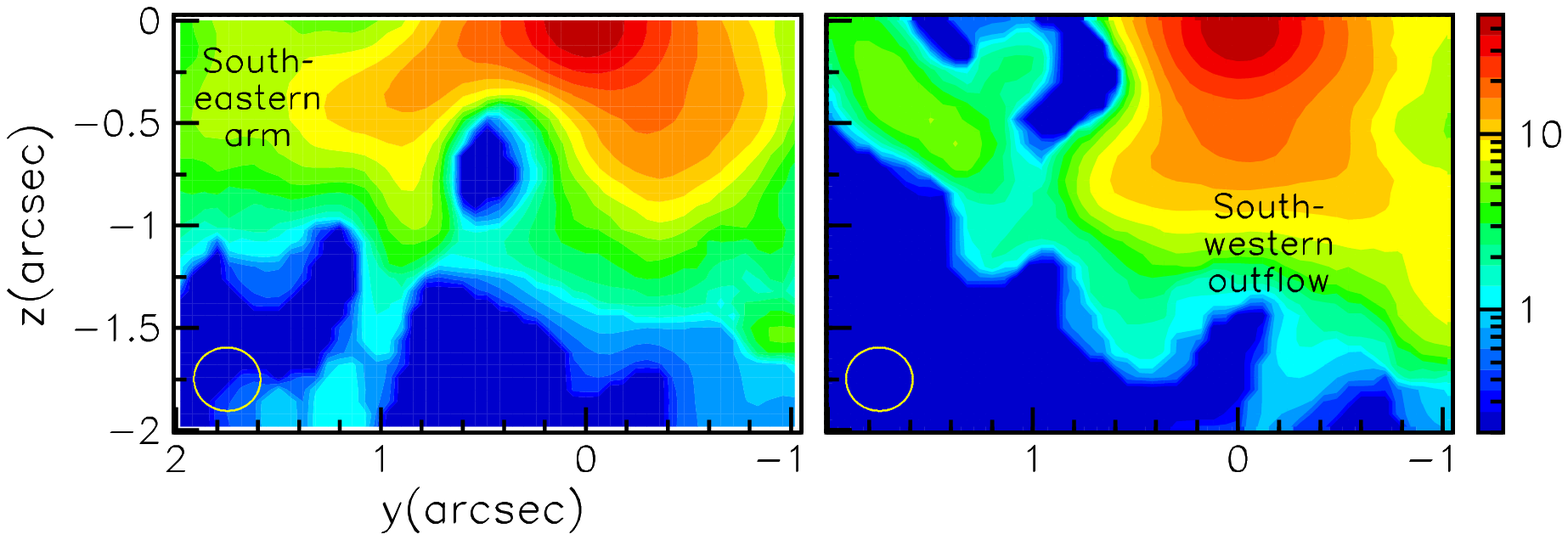}
\caption{Sky maps of F over 43<$V_x$<47 km\,s$^{-1}$ (left) and 47<$V_x$<50 km\,s$^{-1}$ (right) displaying the south-eastern arm 
(left) and the southern limb of the south-western outflow (right). The beam is shown in the lower left corner of each panel.}
\label{fig19}
\end{figure*}

It is not clear whether region 3 has an identity of its own or is simply the southern limb of the south-western 
outflow. To answer this question we show in Fig.~\ref{fig19} maps of the relevant region for blue-shifted and 
red-shifted Doppler velocities separately. The southern limb is seen to point south and to significantly overflow 
in the eastern hemisphere, making the case for treating region 3 separately from region 2 very weak.

\begin{figure*}
\centering
\includegraphics[height=10.cm,trim=1.cm 0.cm 0.cm 0.cm,clip]{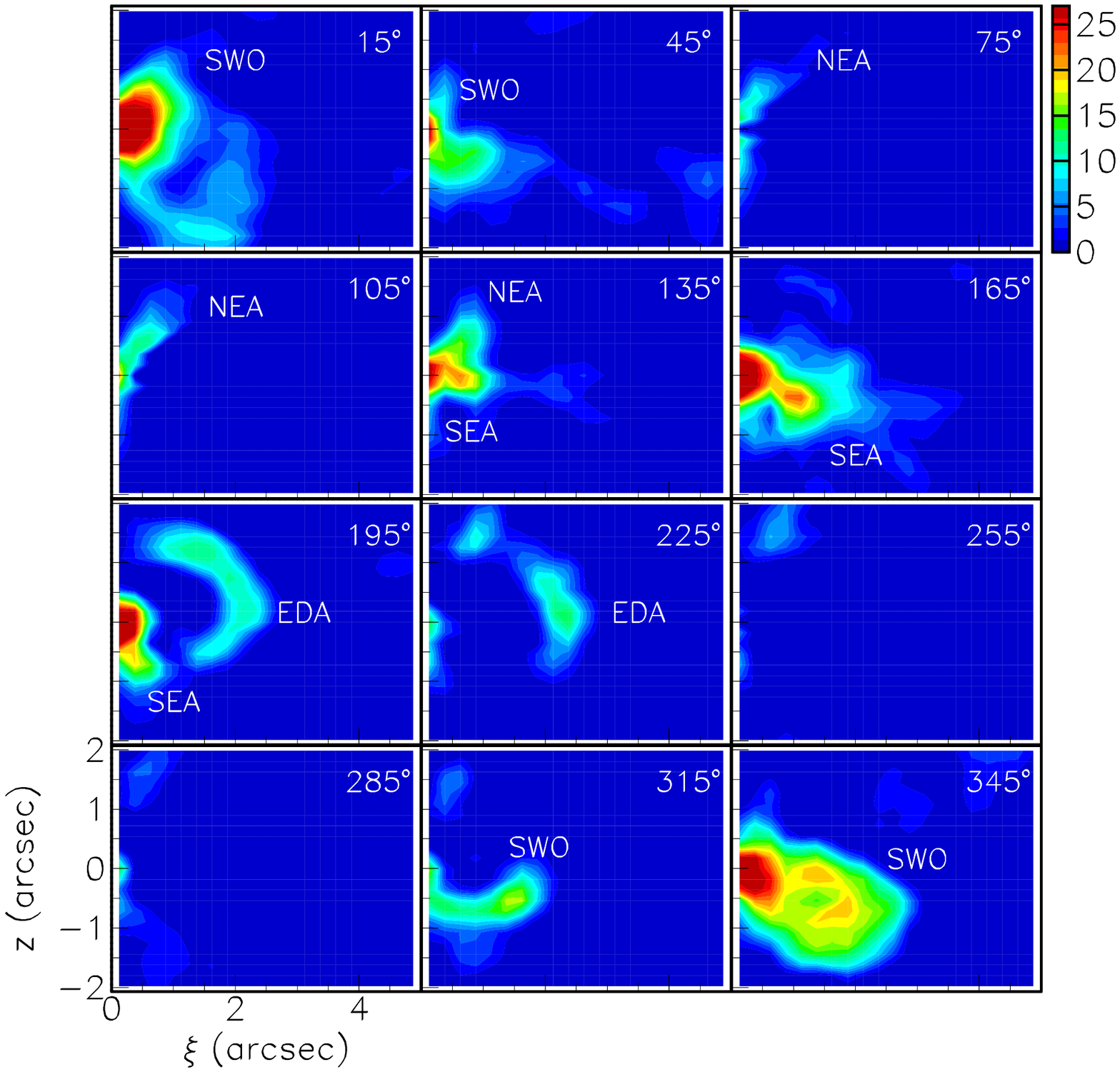}
\caption{Maps of the effective emissivity reconstructed in space under the assumption of a pure radial expansion at constant velocity of
 7 km\,s$^{-1}$. The ordinate is $z$ and the abscissa is $\xi=\sqrt{x^2+y^2}$. Each panel is for a 30$^\circ$ wide interval of $\omega$ measured
 clockwise from 0 on the $y$ axis. The label in each panel gives the value of at the centre of the interval. The upper and lower rows are west, 
the two middle rows are east. The labels stand for south-western outflow (SWO), north-eastern arm (NEA), south-eastern arm (SEA), eastern detached arc (EDA).}
\label{fig20}
\end{figure*}

P-V diagrams have the advantage of offering a visual representation of the information contained in the Doppler velocity 
and therefore helping with the conception of a 3-D representation of the morphology. Another approach in the same spirit 
is illustrated in Fig.~\ref{fig20}, displaying a 3-D representation of the detected emission assuming radial expansion 
of the circumbinary envelope at constant velocity $V_{rad}$. In such a case, at any point ($x,y,z$) the relation 
$V_x/V_{rad}=x/\sqrt{x^2+R^2}$ applies, meaning $x=R\,V_x/\sqrt{V_{rad}^2-V_x^2}$. Writing 
$\xi=\sqrt{x^2+y^2}$ one can then image the effective emissivity in any ($z,\xi$) plane making an angle $\omega$ with 
the $y$ axis by multiplying the measured flux density by $dV_x/dx=V_{rad}R^2(x^2+R^2)^{-3/2}$ and 
$dx\,dy/(d\xi\,d\omega)=\xi$. In Fig.~\ref{fig20}, $V_{rad}$=7 km\,s$^{-1}$, all Doppler velocities being 
measured with respect to a mean velocity $V_0$=47.6 km\,s$^{-1}$. Other reasonable choices of $V_{rad}$ and $V_0$ have 
been tried and the results have been found not to differ strongly from those presented here. Contributions from the 
bubble and from the red-shifted arcs are minimized by requiring 43<$V_x$<52 km\,s$^{-1}$.

While Fig.~\ref{fig20} gives a perfectly legitimate model of the observations, it must be clear that it may have 
nothing to do with reality. Indeed, such a model can be constructed for any data cube, whether expanding or not 
radially, a result of the under-determination of the problem. However, it provides a useful visualisation of the 
morphology, in particular of its topology, as long as it is interpreted with care. In particular, it is clear that 
it is of little value when $V_x$ is close to $V_0$: in such a case, there is essentially no Doppler shift to deal 
with, and therefore no hope to obtain a meaningful evaluation of $x$ from the Doppler velocity.

The north-eastern and south-eastern arms are clearly identified in the fifth and sixth panels of Fig.~\ref{fig20} 
respectively. The south-western and north-eastern outflows cover panels ($-$60$^\circ$<$\omega$<60$^\circ$) and 
panels (120$^\circ$<$\omega$<240$^\circ$) respectively. Detached shells are seen not only east, as expected, 
but also west, in the first panel. Detailed studies of the north-eastern and south-western outflows are presented in the next sub-section.

In summary, the sky map of the circumbinary envelope at distances up to 2.5$''$ from Mira A displays some 
symmetry about an axis pointing $\sim$30$^\circ$ south of west but a significant dissymmetry between the 
associated hemispheres at distances not exceeding 1.5$''$: the south-western hemisphere is dominated by a 
broad conical outflow, referred to as the south-western outflow, while the north-eastern hemisphere is dominated 
by two arms, the south-eastern arm clearly related to Mira B, and the north-eastern arm. At larger distances, 
the two hemispheres are more similar. In particular, in both hemispheres cavities are visible, with a clearly 
detached shell in the north-eastern hemisphere. It looks as if at short distances a large gas volume, red-shifted 
and covering the north-eastern quadrant, had been swept away or prevented to emit in CO(3-2) because of 
temperature or turbulences or else. Obviously, in both hemispheres, winds cannot be assumed to be stationary 
and the observed morphology bears the marks of history. While the kinematics show clear signs of radial expansion, 
with typical velocities between 5 and 10 km\,s$^{-1}$, no sign of significant rotation has been detected at the 
level of 0.5 km\,s$^{-1}$\,arcsec$^{-1}$ or more. Finally, the effective emissivity is concentrated in the vicinity 
of the plane of the sky rather than displaying rotational invariance about the approximate axis of symmetry observed on the sky map.

\subsection{South-western and north-eastern outflows}

\begin{figure*}
\centering
\includegraphics[height=8.cm,trim=1.cm 0.cm 0.cm 0.cm,clip]{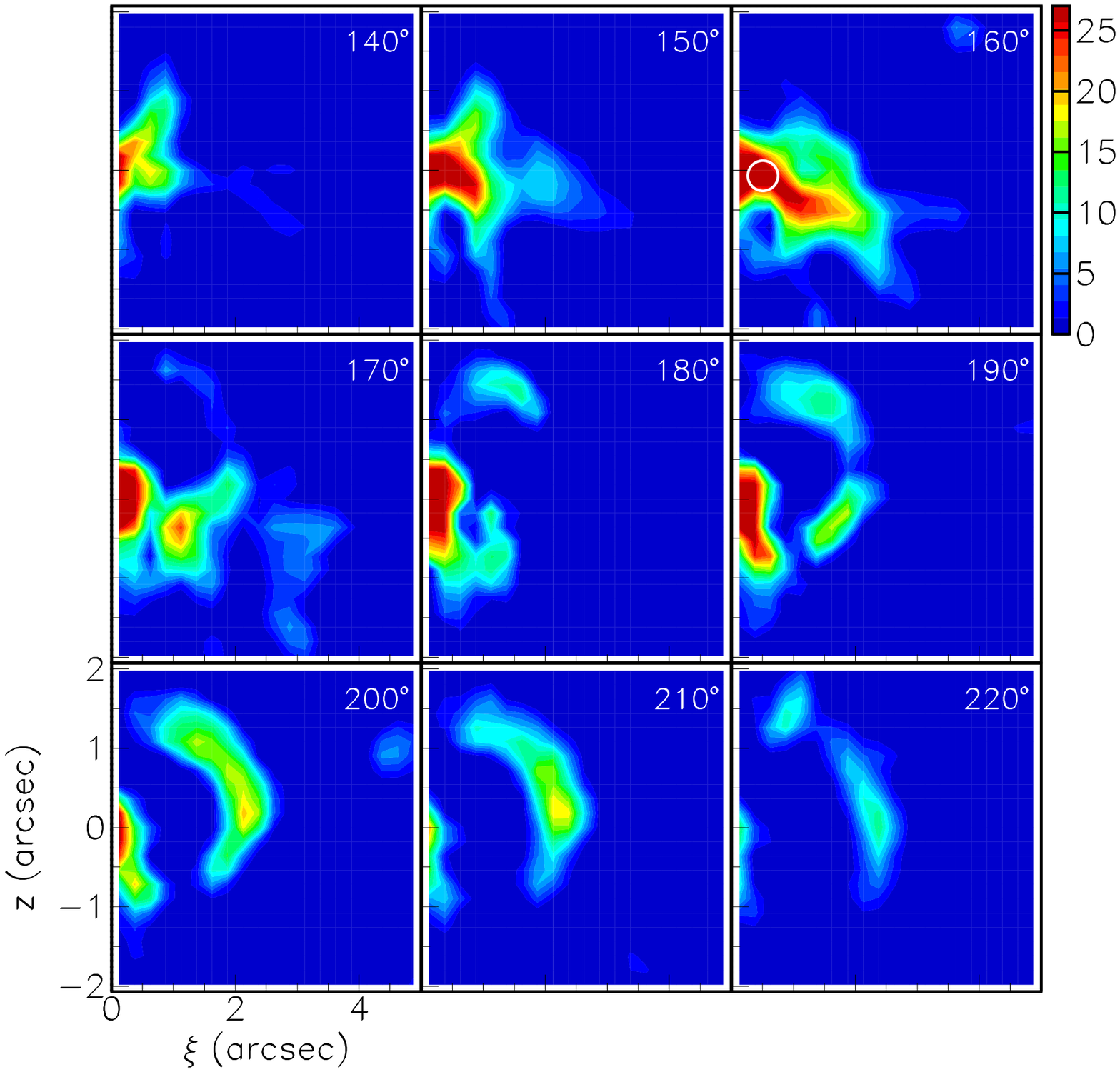}
\includegraphics[height=8.cm,trim=0.cm 0.cm 0.cm 0.cm,clip]{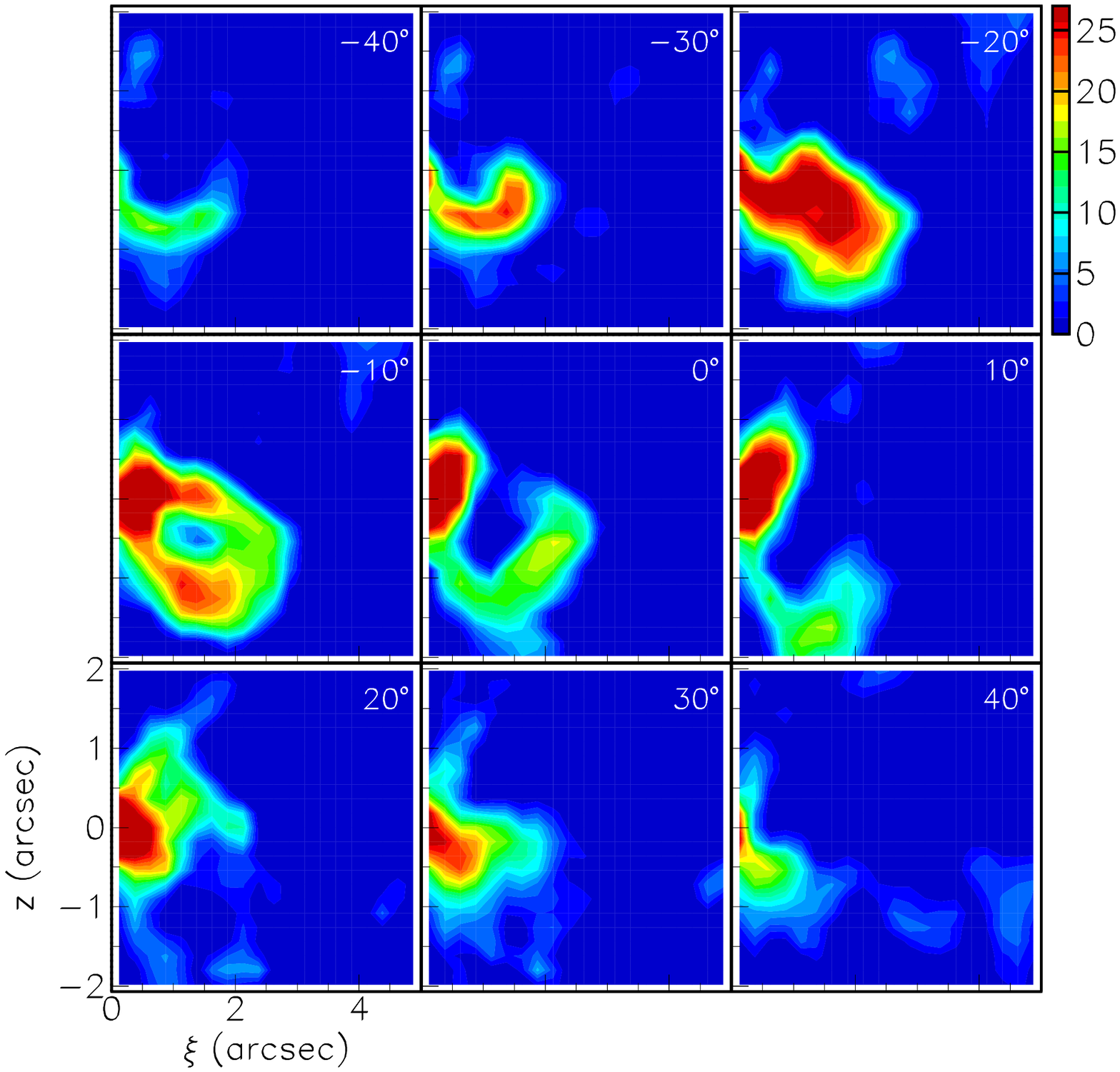}
\caption{South-western (left) and north-eastern (right) outflows: maps of the effective emissivity reconstructed in space 
under the assumption of a pure radial expansion at constant velocity of 7 km\,s$^{-1}$. The ordinate is $z$ and the abscissa 
is $\xi=\sqrt{x^2+y^2}$. Each panel is for a 10$^\circ$ wide interval of $\omega$ measured clockwise from 0 on the $y$ axis. 
The label in each panel gives the value of $\omega$ at the centre of the interval. The white circle in the $\omega=160^\circ$ panel shows
 the position of Mira B for $V_x=45$ km\,s$^{-1}$. }
\label{fig21}
\end{figure*}

Figure~\ref{fig21} shows maps of the effective emissivity in $z$ vs $\xi$ planes scanning across the south-western 
and north-eastern outflows in 10$^\circ$ steps in $\omega$ from $-$45$^\circ$ to 45$^\circ$ and 135$^\circ$ to 
225$^\circ$ respectively. The south-western outflow and north-eastern arm are particularly important at 
$\omega=-20^\circ$ and 160$^\circ$ respectively, namely back-to-back in $\omega$.  However, they both point 
south by some 30$^\circ$ and are not back-to-back in space, the north-eastern arm being the main outflow 
component in the northern hemisphere (one can just see part of it at $\omega=140^\circ$ and 150$^\circ$).

In both maps, the emission is seen to be enhanced at large distances. This is spectacular in the eastern outflow 
where an arc covering close to a steradian contributes to all panels above $\omega$=170$^\circ$. Its contribution 
to the middle panel of Fig.~\ref{fig15}, its clear separation from the bulk of the outflow and its narrow width, 
less than 100 AU, are reminiscent of what is observed in detached shells around some carbon stars 
\citep[]{Maercker2014, Olofsson2010, Mattsson2007, Schoier2005}. Such detached shells are usually observed on 
the dust rather than on the gas but there is evidence for dust and gas having similar morphologies in these cases. 
Also, they cover much larger solid angles, up to 4$\pi$, and are not seen in oxygen-rich M stars. The present data 
do not allow, without confronting them to the predictions of hydrodynamic models, to make definitive statements 
about the physical nature of these arcs, in particular to evaluate the respective contributions of gas density 
and temperature to the depression separating the arcs from the bulk. Yet, one can safely state that some violent 
event must have occurred some centuries ago, similar to the ejections of matter associated with thermal pulses 
and/or helium flashes in the case of carbon stars. In this context, the presence of a similar arc in the south-western 
outflow at $\omega$=0$^\circ$ and possibly 10$^\circ$, and of a small cavity at $\omega=-10^\circ$,  suggest a 
common origin for all these arcs, all distant from Mira A by some 2$''$, meaning 150 years or so at 7 km\,s$^{-1}$.

An interesting feature, a small apparent cavity, is seen in the north-eastern outflow at $\omega=160^\circ$, possibly 
related to Mira B which sits in the middle of the outflow for $V_x$=45 km\,s$^{-1}$. It is also seen in Fig.~\ref{fig12} 
between 45 and 46 km\,s$^{-1}$.

\section{Discussion}

At distances from Mira A not exceeding 2.5$''$, emission is dominated by two outflows aiming respectively at 
$\sim$30$^\circ$ north of east and $\sim$30$^\circ$ south of west, along an axis with respect to which the sky map is 
approximately symmetric. However, the effective emissivity is not rotation-invariant with respect to this axis, being 
rather concentrated near the plane of the sky, nor is it symmetric with respect to the equator associated with this 
axis, defining a north-eastern and a south-western hemispheres. Moreover, we detected no rotation velocity about this 
axis at a level of 0.5 km\,s$^{-1}$\,arcsec$^{-1}$ or more. We have tentatively identified different sources, which 
are listed in Table \ref{tab1}, together with the associated CO(3-2) flux, each region being defined by approximate intervals 
$\Delta{R}$, $\Delta{\varphi}$ and $\Delta{V_x}$ respectively.

\begin{table*}
\centering  
\caption{Properties of some of the main sources}    
\label{tab1}        
\renewcommand{\arraystretch}{1.1}
\begin{tabular}{|c|c|c|c|c|}
\hline 
Name & \makecell{$R$ \\($''$)} & \makecell{$\varphi$ \\($^\circ$)} & \makecell{${V_x}$ \\(km\,s$^{-1}$)} & \makecell{$F$ \\(Jy km\,s$^{-1}$)} \\
\hline 
Mira A & 0-0.5 & 0-360 & 44-52 & 4.15 \\
\hline 
South-western outflow & 0.5-2.5 & $-$60-120 & 43-52 & 7.90 \\
\hline 
North-eastern arm & 0.5-2.5 & 220-280 & 43-45.5 & 0.62 \\
\hline 
South-eastern arm & 0.5-2.5 & 120-190 & 43-47 & 1.49 \\
\hline 
Eastern detached arc & 1.5-2.5 & 180-290 & 47-52 & 1.55 \\
\hline 
Dark eastern sector & 0.5-1.5 & \makecell{170-210 \\ 200-280} & \makecell{47-48.5 \\ 45.5-47} & 0.15\\
\hline
\end{tabular} 
\end{table*}
At distances from Mira A not exceeding $\sim$1.5$''$, the south-western emission consists of a slightly red-shifted 
outflow covering a broad solid angle at the scale of a steradian and expanding radially at a velocity between 5 and 
10 km\,s$^{-1}$. The north-eastern emission consists instead of two slightly blue-shifted arms separated by a dark 
region covering the north-eastern quadrant and identified by \citet{Ramstedt2014} as a wake of Mira B. However, as 
Mira B is currently moving nearly north, we do not see how this dark region could be associated with its wake. 
Beyond $\sim$1.5$''$ and up to $\sim$2.5$''$ projected distance from Mira A, evidence has been presented for detached 
arcs that are reminiscent of the detached shells observed in some carbon stars and associated with thermal pulses 
and/or helium flashes. This is particularly spectacular in the north-eastern hemisphere, where a detached arc covers 
the whole quadrant, but seems also to apply, although less clearly, in the south-western hemisphere.

It is remarkable that the above description ignores the presence of Mira B. It is only when looking at the south-eastern 
arm in the immediate environment of the A+B pair that evidence has been found for enhanced emission from a blue-shifted 
gas flow having its source at Mira A and concentrated in the direction of Mira B, with a possible small contribution 
of gas gravitationally bound to Mira B as detected at much shorter distances by \citet{Vlemmings2015} from the continuum 
emission of partially ionized gas. We have shown that a gas flow emitted by Mira A and concentrating in the direction of Mira B is clearly prefered to a static gas volume surrounding Mira B. The WRLOF model of \citet{Mohamed2012}, and in particular its Model 1 version, offers a natural framework in which to draw such a picture. It explains the enhancement around Mira B as the result of the Mira A wind leaking through the Lagrange point to Mira B,
thereby implying that a large fraction of it would be channelled to Mira B and eventually be accreted 
onto it. This is at variance with models in which the wind escaping Mira A is isotropic, giving a chance of being 
trapped by Mira B to only a very small fraction of it. What is observed here favours clearly the former hypothesis over 
the latter but gives also evidence for two additional outflows escaping Mira A in directions far away from the Lagrange 
point: the north-eastern arm and the abundant south-western outflow, which are of course absent from the simple version 
of the WRLOF model. Assuming that the WRLOF mechanism applies, at least qualitatively, to the south-eastern arm, one would 
then expect a significant fraction of its flux to be accreted by Mira B and the downstream gas flux to be smaller than 
the upstream one. It is not clear whether the depression seen in the $\omega=160^\circ$ panel of Fig.~\ref{fig21} 
downstream of Mira B can reliably be taken as evidence for such an effect.

The interpretation in terms of the WRLOF model was also favoured by \citet{Ramstedt2014}, in particular 
in the context of the possible presence of spriral arcs at larger distance.

The precise role played by Mira B in shaping the circumbinary envelope raises some unanswered questions. The elongation 
along an approximate symmetry axis pointing 30$^\circ$ south of west observed in the present data is similar to that 
observed at  94 GHz by \citet{Vlemmings2015} who find a major axis of 0.042$''$, an aspect ratio of $\sim$1.2 and a 
position angle 36$^\circ$ south of west. However, the continuum was observed four months later than the CO line and 
earlier observations of the star atmosphere suggest that the position angle of the elongation is very strongly dependent 
on the phase of the Mira A pulsation \citep[]{Karovska1991, Haniff1992, Wilson1992, Quirrenbach1992}. If this were true, 
the similarity between the elongations observed by ALMA in the continuum and on the line would be pure coincidence. 
One might nevertheless still suppose that the elongation observed in the present CO data is caused by anisotropic 
pulsation amplitudes, which would then be responsible for shaping the outflow, causing the south-west to north-east 
enhancement observed here and producing the south-western outflow. In this process, dominated by the pulsation of Mira 
A atmosphere and its asymmetry, Mira B would play no obvious role. In the north-eastern hemisphere, however, additional 
mechanisms need to be invoked to explain the separation in two distinct arms. While focusing by Mira B must play a role 
in shaping the south-eastern arm, it is difficult to imagine how it could generate the broad dark region referred to as 
the dark eastern sector in Table \ref{tab1}. This sector suggests a deficit of some 4 Jy\,km\,s$^{-1}$ of north-eastern emission 
with respect to south-western emission: how could this be related to Mira B? It seems rather related to whatever violent 
event produced the eastern detached arc. Of course, the emission deficit may be due to conditions of temperature preventing 
the population of the J=3 level of the CO molecule, the mass deficit associated with the dark eastern sector being then 
much smaller or even inexistent. Both the accretion disk of the white dwarf and the occurrence of thermal pulses are known 
to release important amounts of energy, but their relative importance in shaping the circumbinary envelope is not clear 
to us. To answer such questions, the present observations are but one of the factors to be taken into account, the whole 
multiwavelength spectrum having to be taken into consideration. In this context, observations of the emission of another 
molecular line of carbon monoxide would bring very useful information on the temperature distribution and allow for 
disentangling, at least partly, the respective roles of temperature and density.

\section{Summary}
We have presented a description of the morphology and kinematics of the gas envelope of Mira Ceti using high spatial 
resolution observations of $^{12}$CO(3-2) emission made by ALMA. We have tried to keep away as much as possible from 
preconceptions that might have biased the analysis. Yet, much is known of this source from multiwavelength observations, 
and this knowledge must be taken into account when trying to make sense of what has been observed. A distinctive 
feature of the observations analysed here is the strong imprint of the mass loss history on the morphology of the 
circumbinary envelope. Evidence has been presented for the distinct contributions of remnants of events that occurred 
a few thousand years ago and events that occurred a few hundred years ago. The former include a south-eastern 
blue-shifted bubble that can be described as a circular ring expanding radially at a velocity of less than 2 km\,s$^{-1}$ 
and making an angle of $\sim$50$^\circ$ with the plane of the sky, together with a number of red-shifted arcs. The present 
data do not allow for a reliable association of these arcs in spirals, as proposed by \citet{Ramstedt2014} using ALMA 
observations having better large distance coverage.

The central region is dominated by the circumbinary envelope, displaying two outflows in the north-eastern and south-western hemispheres. At short distances from the star, up to $\sim$1.5$''$, they display very different morphologies. The south-western outflow covers a broad solid angle, expands radially at a rate between 5 and 10 km\,s$^{-1}$ and is slightly red shifted. The north-eastern outflow consists of two arms, with a separation of $\sim$90$^\circ$ on the plane of the sky, both blue-shifted, bracketing a broad dark region where emission is suppressed. At larger distances from the star, between $\sim$1.5$''$ and $\sim$2.5$''$ the dissymmetry between the two hemispheres is significantly smaller and both hemispheres display evidence for detached arcs, particularly spectacular in the north-eastern hemisphere. Close to the stars, we are observing a mass of gas surrounding Mira B, with a size of a few tens of AU, and having Doppler velocities with respect to Mira B reaching 1.5 km\,s$^{-1}$ in each of the red and blue directions, which we interpret as gas flowing from Mira A toward Mira B, being eventually focused or even trapped by Mira B, with a possible small contribution of gas gravitationally bound to Mira B.

Throughout the present article, emphasis has been placed on giving as simple and yet as reliable a picture of what has 
been observed, with the aim of helping with the conception of models that would allow for a deeper understanding of the 
physics at play, but the conception of such models is clearly beyond the scope of the present work.

\section*{Acknowledgements}
This paper makes use of the following ALMA data: ADS/JAO.ALMA\#2013.1.00047.S. ALMA is a partnership of ESO 
(representing its member states), NSF (USA) and NINS (Japan), together with NRC (Canada) , NSC and ASIAA (Taiwan), 
and KASI (Republic of Korea), in cooperation with the Republic of Chile. The Joint ALMA Observatory is operated by ESO, AUI/NRAO and NAOJ.
The data are retrieved from the JVO portal (http://jvo.nao.ac.jp/portal) operated by the NAOJ. 
We are indebted and very grateful to the ALMA partnership, who are making their data available to the public after a 
one year period of exclusive property, an initiative that means invaluable support and encouragement for Vietnamese astrophysics. 
We particularly acknowledge friendly and patient support from the staff of the ALMA Helpdesk. We are 
very grateful to Professors Anne Dutrey, Stephane Guilloteau and Thibaut Le Bertre for their interest in this work 
and the many helpful comments and information they have provided. Financial support is acknowledged from the Vietnam 
National Satellite Centre (VNSC/VAST), the NAFOSTED funding agency under contract xx, the World Laboratory, the Odon 
Vallet Foundation and the Rencontres du Viet Nam.









\appendix
  \section{}
  \begin{figure*}
    \centering
    \includegraphics[height=22 cm]{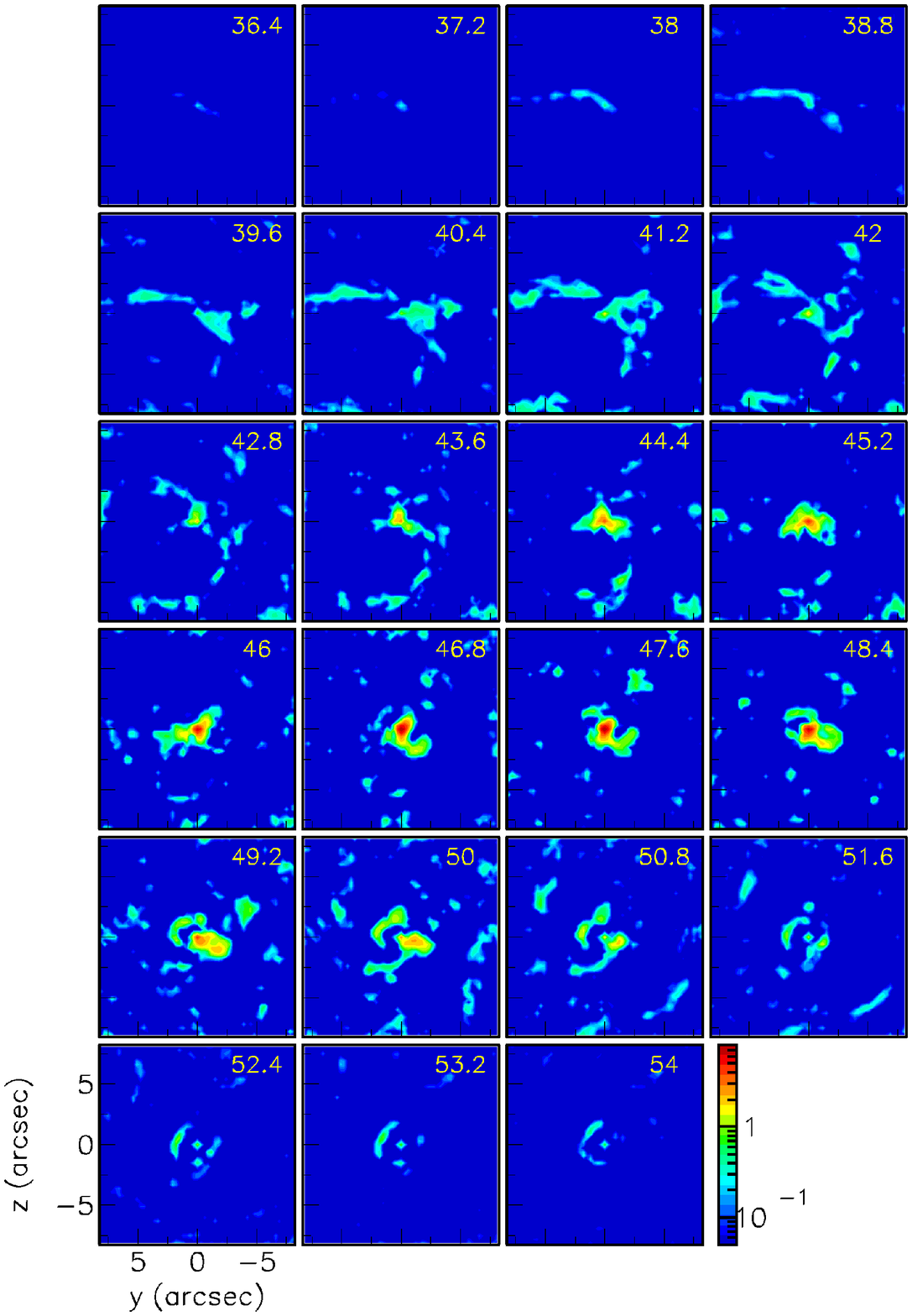}
    \caption{Channel maps covering $8.2''\times8.2''$ obtained by grouping two adjacent
        velocity channels (resulting in a width of 0.8 km\,s$^{-1}$) between 36.0 and 54.4 km\,s$^{-1}$. 
        Each panel is labelled with the central velocity (km\,s$^{-1}$). The colour scale is in units of Jy\,km\,s$^{-1}$ arcsec$^{-2}$.}
    \end{figure*}

\bsp	
\label{lastpage}
\end{document}